\documentclass[12pt]{article}

\usepackage[all]{xy}
\usepackage{amsmath}
\usepackage{amsfonts}
\usepackage{amsthm}
\usepackage{booktabs}
\usepackage{graphicx}
\usepackage{makeidx}
\usepackage{tocloft}
\usepackage{parskip}
\usepackage{upquote}
\usepackage{verbatim}
\usepackage{float}
\makeindex
\usepackage{listings}
\usepackage{url}
\usepackage[utf8x]{inputenc}
\usepackage[usenames]{color}
\usepackage{grffile}

\definecolor{lg}{rgb}{0.9,0.9,0.9}
\definecolor{dg}{rgb}{0.3,0.3,0.3}
\def\ft{\small\tt}
\def\floor#1{\lfloor #1 \rfloor}
\title{QCDUTILS}
\author{Massimo Di Pierro}
\begin{document}
\maketitle

\begin{abstract}
This manual describes a set of utilities developed for Lattice QCD computations. They are collectively called {\ft qcdutils}. They are comprised of a set of Python programs each of them with a specific function: download gauge ensembles from the public NERSC repository, convert between formats, split files by time-slices, compile and run physics algorithms, generate visualizations in the form of VTK files, convert the visualizations into images, perform bootstrap analysis of results, fit the results of the analysis, and plot those results. These tools implement the typical workflow of most Lattice QCD computations and automate it by enforcing filename conventions: the output of one tool is understood by the next tool in the workflow. This manual is organized as a series of autonomous recipes which can be combined together.
\end{abstract}
\newpage
\tableofcontents
\newpage

\goodbreak\section{Introduction}

In this manual we provide a description of the following tools:

\begin{itemize}
\item {\ft qcdutils\_get.py}: a program to download gauge configurations form the NERSC {\it Gauge Connection} archive~\cite{nersc} and convert them from one format to another, including to ILDG~\cite{ildg} and FermiQCD formats~\cite{mdp,fermiqcd}.

\item {\ft qcdutils\_run.py}: a program to download, compile and run various parallel Physics algorithms (for example compute the average plaquette, the topological charge density, two and three points correlation functions). {\ft qcdutils\_run} is a proxy for FermiQCD. Most of the FermiQCD algorithms and examples generate files that are suitable for visualization (VTK files~\cite{vtk})

\item {\ft qcdutils\_vis.py}: a program to manipulate the VTK files generated by {\ft qcdutils\_run} which can be used to split VTK files into components, interpolate them, and generate 3D contour plots as JPEG images. This program uses metaprogramming to write a VisIt~\cite{visit} script and runs it in background.

\item {\ft qcdutils\_vtk.py}: a program that converts a VTK file into a web page (HTML) which displays iso-surfaces computed from the VTK file. The generated files can be visualized in any browser and allows interactive rotation of the visualization. This program is a based on the ``processing.js'' library~\cite{processing}.

\item {\ft qcdutils\_boot.py}: a tools for performing bootstrap analysis of the output of {\ft qcdutils\_run} and other QCD Software. It computes autocorrelations, moving averages, and distributions.

\item {\ft qcdutils\_plot.py}: a tool to plot results from {\ft qcdutils\_boot}.

\item {\ft qcdutils\_fit.py}: a tool to fit results from {\ft qcdutils\_boot.py}.
\end{itemize}

As the {\ft .py} extension implied, these programs are written in Python~\cite{python} (2.7 version recommended).

Together these tools allow automation of the workflow of most Lattice QCD computations from downloading data to computing scientific results, plots, and visualizations.

Notice that each of the utilities has its own help page which you an access using the {\ft -h} command line option. The output for each is reported in the Appendix.

The data downloaded by {\ft qcdutils\_get} can be read by {\ft qcdutils\_run} which executes the physics algorithms implemented in C++. The output can be VTK files manipulated by {\ft qcdutils\_vis} and and then transformed into images and movies by VisIt, or they can be tabulated data that require bootstrap analysis. This is done by {\ft qcdutils\_boot}. The output of the latter plotted by {\ft qcdutils\_plot} and can be fitted with {\ft qcdutils\_fit}.

These files enforce a workflow by following the file naming conventions described in the Appendix but, they do not strictly depend on each other. For example {\ft qcdutils\_boot} can be used to analyze the output of any of your own physics simulations even if you do not use {\ft qcdutils\_run}.

Here is an overview of the workflow:

\begin{displaymath}
  \xymatrix{
    {\ft qcdutils\_}get \ar[r] & run \ar[r] \ar[dr] \ar[ddr] & boot \ar[r] \ar[dr] & plot   \\
    &                             & vis                 & fit    \\
    &                             & vtk                 &
  }
\end{displaymath}

This manual is not designed to be complete or exhaustive because our tools are in continuous development and new features are added every day. Yet is designed to provide enough examples to allow you to explore further. Our analysis and visualizations are created on sample data and aimed exclusively at explaining how to use the tools.

Our hope is that these tools will be useful to practitioners in the field and specifically to graduate students new to the field of Lattice QCD and looking to jumpstart their research projects.

These tools can also be used to automate the workflow of analyzing gauge configurations in real time in order to obtain and display preliminary results.

Some of the tools described here find more general application than Lattice QCD and can be utilized in other scientific areas.

\goodbreak\subsection{Resources}

{\ft qcdutils} can be downloaded from:

\url{http://code.google.com/p/qcdutils}

More information FermiQCD code used by {\ft qcdutils\_run} is available from refs.~\cite{mdp, fermiqcd} and the web page:

\url{http://fermiqcd.net}

More examples of visualizations and links do additional code and examples can be found at:

\url{http://latticeqcd.org}

\goodbreak\subsection{Getting the tools}

There are two ways to get the tools described in here.  The easiest way to get {\ft qcdutils} is to use Mercurial:

Install mercurial from:
\begin{lstlisting}[keywords={}]
http://mercurial.selenic.com/
\end{lstlisting}
\noindent and download {\ft qcdutils} from the googlecode repository

\begin{lstlisting}[keywords={}]
http://code.google.com/p/qcdutils/source/browse/
\end{lstlisting}

using the following commands:

\begin{lstlisting}
hg clone https://qcdutils.googlecode.com/hg/ qcdutils
cd qcdutils
\end{lstlisting}

The command creates a folder called ``qcdutils'' and download the latest source files in there.

You can also download individual files using {\ft wget} (default on Linux systems) or {\ft curl} (default on mac systems):

\begin{lstlisting}
wget http://qcdutils.googlecode.com/hg/qcdutils_get.py
wget http://qcdutils.googlecode.com/hg/qcdutils_run.py
wget http://qcdutils.googlecode.com/hg/qcdutils_vis.py
wget http://qcdutils.googlecode.com/hg/qcdutils_vtk.py
wget http://qcdutils.googlecode.com/hg/qcdutils_boot.py
wget http://qcdutils.googlecode.com/hg/qcdutils_plot.py
wget http://qcdutils.googlecode.com/hg/qcdutils_fit.py
\end{lstlisting}

\goodbreak\subsection{Dependencies}

These files do not depend on each other so you can download only those that you need. {\ft qcdutils\_run} is special because it is a Python interface to the {\ft FermiQCD} library. As it is explained later, when executed, it downloads and compiles {\ft FermiQCD}. It assumes you have {\ft g++} installed.

{\ft qcdutils\_fit.py} and {\ft qcdutils\_plot.py} requires the Python {\ft numpy} and {\ft matplotlib} installed.

All the file require Python 2.x (possibly 2.7) and do not work with Python 3.x.

\goodbreak\subsection{License}

{\ft qcdutils} are released under the GPLv2 license.

\goodbreak\subsection{Acknowledgments}

We thank all members of the USQCD collaborations for making most of their their data and code available to the public, and for a long-lasting collaboration. We thank David Skinner, Schreyas Cholia, and Jim Hetrick for their collaboration in improving and running the NERSC gauge connection. We particularly thank Jim Hetrick for sharing his code for t'Hooft instantons. We thank Chris Maynard for useful discussions about ILDG.  We thanks Simon Catterall, Yannick Meurice, Jonathan Flynn, and all those that over time have submitted patches for FermiQCD thus contributing to make it better. We also thank all of those who have used and who still use FermiQCD, thus providing the motivation for continuing this work. We thank the graduate students that over time have helped with coding, testing, and documentation: Yaoqian Zhong, Brian Schinazi, Nate Wilson, Vincent Harvey, and Chris Baron.

This work was funded by Department of Energy grant DEFC02-06ER41441 and by National Science Foundation grant 0970137.  

\goodbreak\section{Accessing public data with {\ft qcdutils\_get.py}}

\goodbreak\subsection{Searching data on the NERSC {\it Gauge Connection}}

The {\it Gauge Connection}~\cite{nersc} is a repository of Lattice QCD data, primarily but not limited to gauge ensembles, hosted by the National Energy Research Science Center (NERSC) on their High Performance Storage System (HPSS). At the time of writing the Gauge Connection hosts 16 Terabytes of data and makes it publicly accessible to researchers worldwide.

The new Gauge Connection site consists of a set of dynamic web pages in hierarchical structures that closely mimics the folder structure in the HPSS FTP server. Each folder corresponds to a web page. The web page provide a description of the folder content, in the form of an editable wiki, comments about the content, links to sub-folder and links to files contained in the folder. Since folders may contain thousands of files, files with similar filenames are grouped together into filename patterns. For example all files with the same name but different extension or similar names differing only for a numerical value are grouped together. Pages are tagged and can be searched by tag. Users can search for files by browsing the folder structure, searching by tags and can download individual files or all files matching a pattern.

You do not need an account to login and you can use your OpenID account, for example a Google email account. You do not need to login to search but you need to login to download. From now on we assume you are logged in into the Gauge Connection.

\begin{figure}[ht!]
\begin{center}
\begin{tabular}{cc}
\includegraphics[width=8cm]{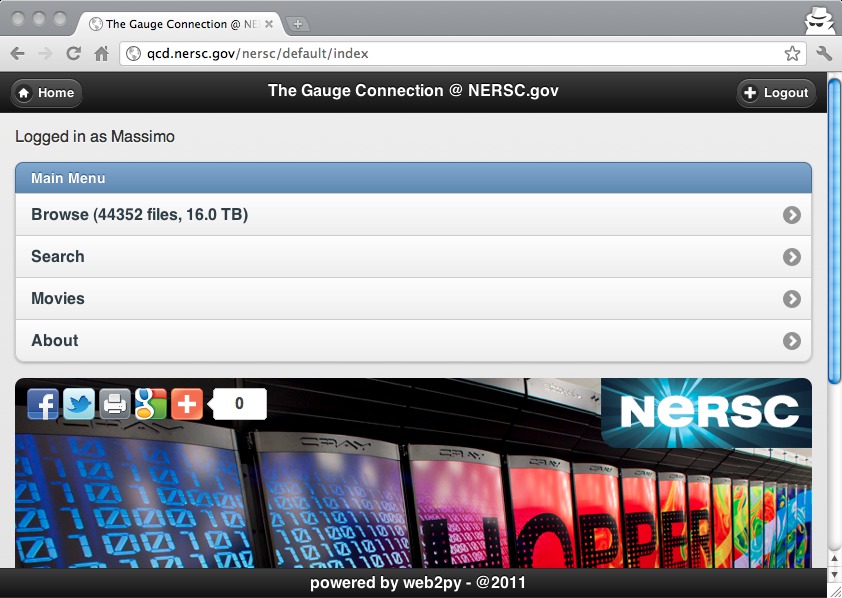} &
\includegraphics[width=8cm]{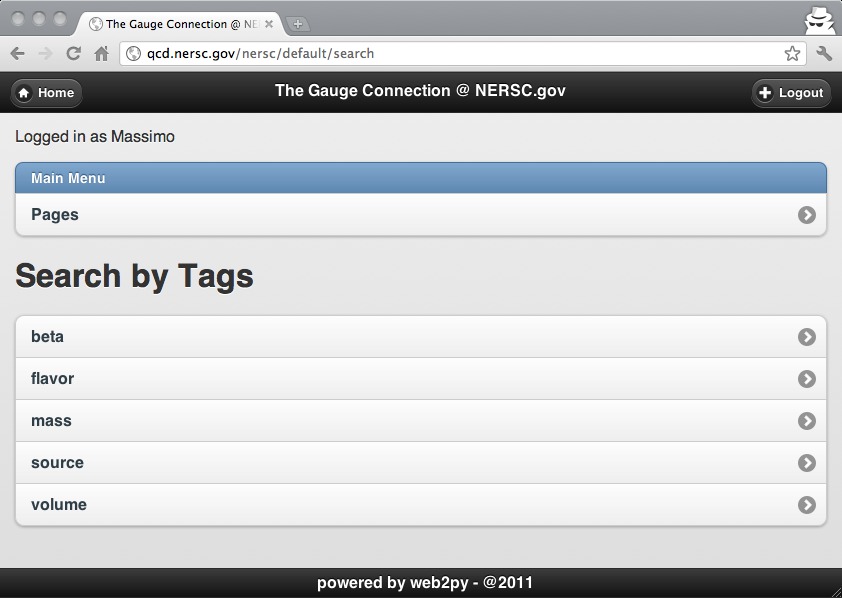}
\end{tabular}
\end{center}
\caption{Main NERSC Gauge Connection web site (left) and search by tag feature (right).\label{nersc1}}\end{figure}

Fig~\ref{nersc1} (left) is a screenshot of the main Gauge Connection site.
Each gauge ensemble is stored in a folder which is represented by a dynamic web page and tagged. You can search these pages by tag, as shown in Fig.~\ref{nersc1} (right).

\begin{figure}[ht!]
\begin{center}
\begin{tabular}{cc}
\includegraphics[width=8cm]{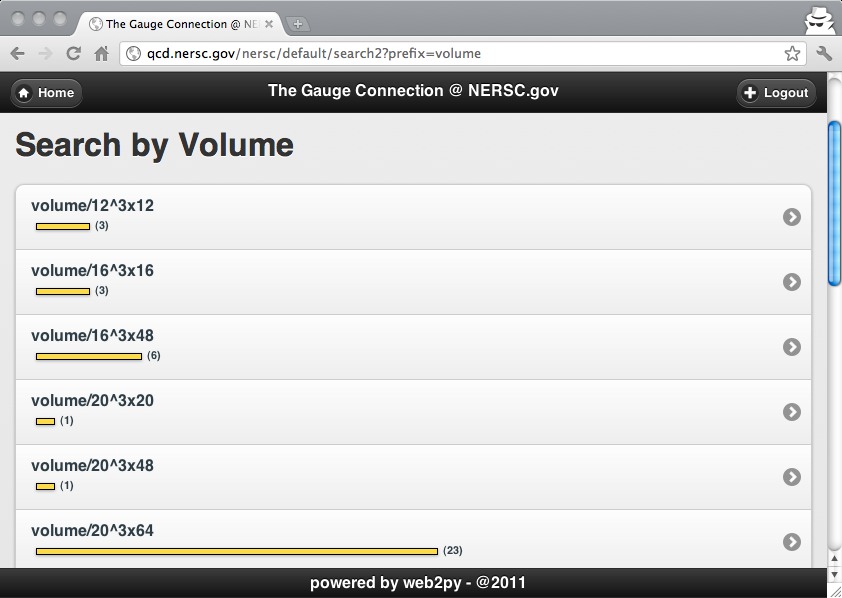} &
\includegraphics[width=8cm]{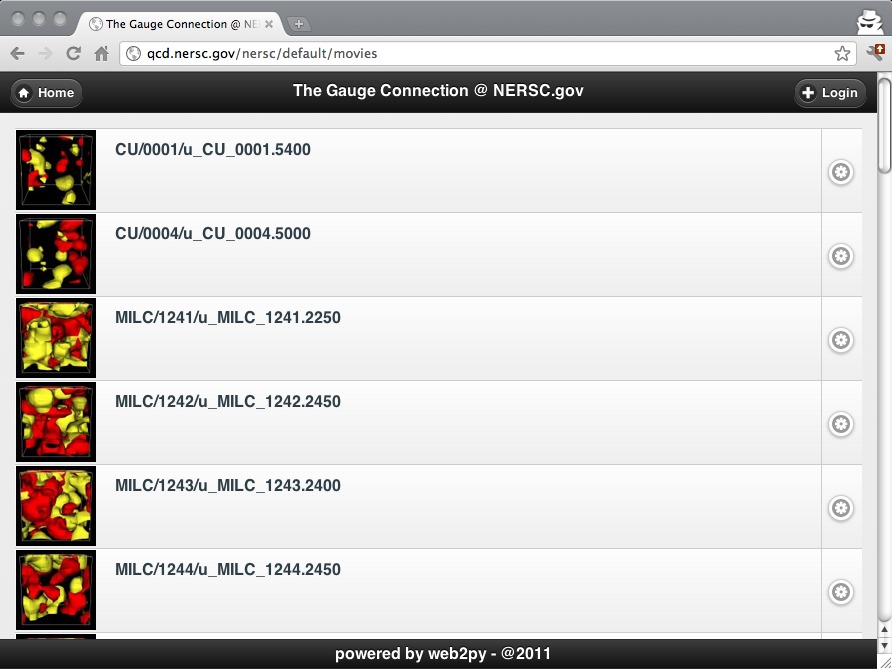}
\end{tabular}
\end{center}
\caption{A page showing statitical information (left) and list of visualzations (right).\label{nersc2}}\end{figure}

Fig.~\ref{nersc2} (left) shows statistical information about tags.

Each tag has the form ``type/value'' where the tag type can be:

\begin{itemize}
\item {\it source}: the name of the organization who donated the data, value can be, for example {\ft MILC}~\cite{milc}.

\item {\it flavor}, the flavor content of the data, value can be ``0'' for quenched data, ``2'' for two flavor unquenched data, ``2+1'' for three flavor unquenched where two quarks have one mass and 1 quark has another mass.

\item {\it kappa}: the $\kappa$  value

\item {\it mass}: the quark mass
\end{itemize}

We have also processed many of the ensambes using some of the tools described here and generated animations of the topological charge densities. This is shown in fig.~\ref{nersc2} (right).

\begin{figure}[ht!]
\begin{center}
\begin{tabular}{cc}
\includegraphics[width=8cm]{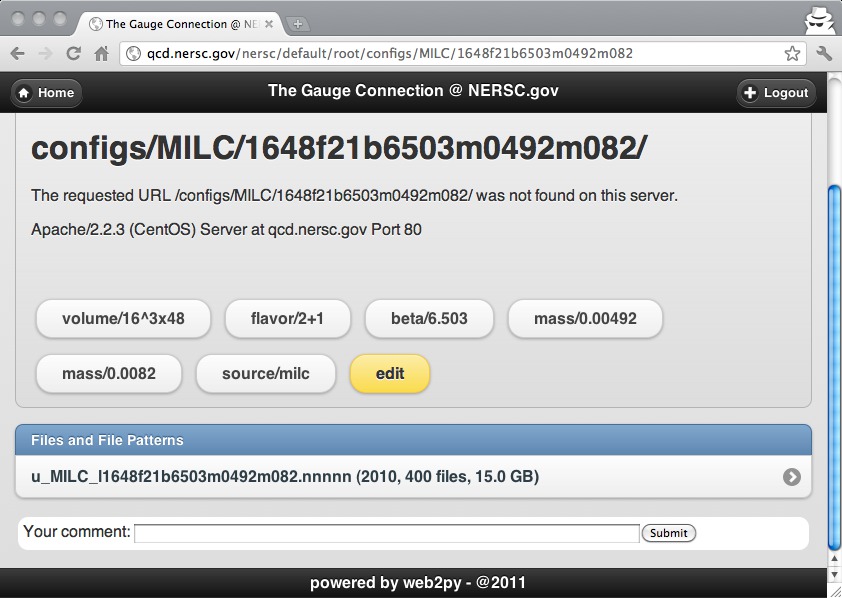} &
\includegraphics[width=8cm]{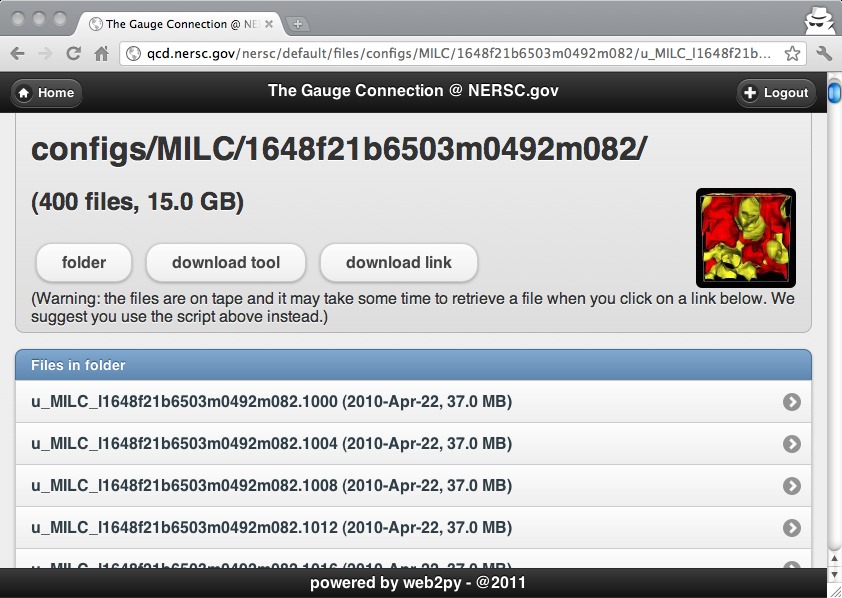}
\end{tabular}
\end{center}
\caption{Folder page (left) and page listing all files in an ensamble (right).\label{nersc3}}
\end{figure}

A screenshot of a folder page is shown in fig.~\ref{nersc3} (left)

You can see a description, a list of tags, list of file patterns in the folder, and comments. The comments are only visible to logged in users. The login link is at top left of the page. 

A screenshot of a page listing the files in an ensemble is shown in fig.~\ref{nersc3} (right)

You can download an individual files by clicking on the file.

To download files in batch you need to first download {\ft qcdutils\_get.py}. The web page page above includes a link [{\it download tool}] that explains where to get and how to use {\ft qcdutils\_get.py}. We suggest you first read the rest of this section but also read the linked instructions which may be more updated. The page also contains a link [{\it download link}] which is used to reference the data for later download.

\goodbreak\subsection{Downloading data from NERSC}

Here we assume you want to download the 400 MILC gauge configurations of size $16^3 \times 48$  and $\beta=6.503$  computed using 2+1 quarks of mass respectively $0.00492$  (light) and $0.082$  (heavy). These files can be found at

\begin{lstlisting}[keywords={}]
http://qcd.nersc.gov/nersc/default/root/configs/MILC/1648f21b6503m0492m082
\end{lstlisting}
\noindent where you should notice the folder name

\begin{lstlisting}
1648f21b6503m0492m082
\end{lstlisting}

It follows the MILC filename convention
\begin{lstlisting}
[time][space]b[beta]m[mass]m[mass]
\end{lstlisting}
and the values of [beta] and [mass] omit the decimal point.

The above page links the pattern page:
\begin{lstlisting}[keywords={}]
http://qcd.nersc.gov/nersc/default/files/configs/MILC/1648f21b6503m0492m082/u_MILC_l1648f21b6503m0492m082.nnnnn
\end{lstlisting}
\noindent where
\begin{lstlisting}
u_MILC_l1648f21b6503m0492m082.nnnnn
\end{lstlisting}
\noindent is the filename pattern and {\ft nnnnn} is just a wildcard for the gauge configuration numbers in the ensemble.

The page contains a [{\it download link}] to a document in JSON format listing all files in the folder matching the pattern and additional meta-data about each file. You do not need to open this document. All you need to do is copy the link address and pass it to {\ft qcdutils\_get.py} as a command line argument. The program opens the URL, download the list, loop over the files in the ensemble, and download them one by one.

Copy the {\it download link} to clipboard and it looks something like this:
\begin{lstlisting}[keywords={}]
http://qcd.nersc.gov/nersc/api/files.json/.../u_MILC_l1648f21b6503m0492m082.nnnnn
\end{lstlisting}
We have shortened the full actual path using {\ft ...}. This URL is a personal token and different users get different URLs for the same data. This allows the server to monitor usage and expire an URL in case of indiscriminate downloads from one user without affecting other users.

To download all data referenced by this link you simply paste the download link after a call to {\ft qcdutils\_get}:

\begin{lstlisting}
python qcdutils_get.py [download link]
\end{lstlisting}

{\ft qcdutils\_get.py} performs the following operations:

Before downlaod, {\ft qcdutils} creates a folder with the same name as the ensemble:
\begin{lstlisting}
u_MILC_l1648f21b6503m0492m082.nnnnn/
\end{lstlisting}
and then download all files in there. The files retain the original file name.

{\ft qcdutils} also creates a file called ``qcdutils.catalog'' where it keeps track of successful downloads. This allows automatic resume on restart: if your download is interrupted, for any reason (for example network problem or server crash), you can re-issue the download command and it resumes where it stopped. {\ft qcdutils} does not download again files that were already downloaded and are currently present on your system.

{\ft qcdutils} can check if a file is complete by checking its size. Data integrity during transmission is guaranteed by the TCP protocol. It is still possible that data is corrupted at the source or locally after download (for example due to a bad disk sector). If a file is found to be corrupted simply delete it, run {\ft qcdutils\_get} again, and it downloads it again. 

Notice that most of the files stored and served by the Gauge Connection are in either the NERSC 3x3 or the NERSC 3x2 file format, described later. If your program can read them, you do not need any conversion. Yet it is likely you need to convert them and this is the subject of the rest of the section.

\goodbreak\subsection{Testing download}

If you encounter any problem downloading real data you can try download a single small demo gauge configuration:

\begin{lstlisting}
python qcdutils_get.py http://qcd.nersc.org/nersc/api/files/demo
\end{lstlisting}

It creates a folder called {\ft demo} and download a single file
\begin{lstlisting}
demo/demo.nersc
\end{lstlisting}

\goodbreak\subsection{Converting to ILDG format (.ildg)}

The {\ft qcdutils\_get.py} can also auto-detect and convert file formats. 
It can input NERSC3x2, NERSC3x3, MILC, UKQCD, ILDG, SciDAC, FermiQCD and it can output ILDG and FermiQCD formats. Other output formats may be supported in the future if this becomes necessary.
\noindent {\ft qcdutils} converts files using the following syntax
\begin{lstlisting}
python qcdutils_get.py -c [target-format] [source]
\end{lstlisting}
\noindent Here [source] can be a {\it download link}, a {\it glob} pattern such as ``demo/*'', or an individual file. [target-format] is one of the following:

\begin{itemize}
\item {\it ildg} converts a gauge configuration to ILDG

\item {\it mdp} converts a gauge configuration to the FermiQCD format

\item {\it slice.mdp} converts a gauge configuration (for example $12^3\times 48$ ) into multiple configuration files, one for each time slice (for example $1\times12^3$ ), in the FermiQCD file format.

\item {\it prop} like {\it mdp} but converts file propagators from {\it Scidac-ILDG format} into the FermiQCD format.

\item {\it slice.prop.mdp} like {\it prop.mdp} but converts a {\it Scidac-ILDG} propagator into FermiQCD time-slice files.
\end{itemize}

Most gauge configuration files are very large and require physics algorithms to run in parallel. Yet some algorithms, specifically some visualization ones, can work on individual time-slices. {\it slice.mdp} and {\it slice.prop.mdp} allow you to break large files into time-slices for this purpose.

Here is an example to convert to ILDG format:

\begin{lstlisting}
python qcdutils_get.py http://qcd.nersc.gov/nersc/api/files/demo
python qcdutils_get.py -c ildg demo/*
\end{lstlisting}

Or in one single line:

\begin{lstlisting}
python qcdutils_get.py -c ildg http://qcd.nersc.gov/nersc/api/files/demo
\end{lstlisting}

If the source file is ``demo/demo.nersc'', the converted file has the ``.ildg'' postfix appended and be called ``demo/demo.nersc.ildg''.
The original file is not deleted. These are the output folder/files:

\begin{lstlisting}
demo/
demo/demo.nerc
demo/demo.nersc.ildg
demo/qcdutils.catalog
\end{lstlisting}

By default {\ft qcdutils} preserves the precision of the input data, but can specify the precision of the target gauge configuration using the {\ft -4} flag for single precision and the {\ft -8} flag for double precision. The input precision is automatically detected. For example:

\begin{lstlisting}
python qcdutils_get.py -c ildg -4 demo/*
\end{lstlisting}

\goodbreak\subsection{Using the catalog file}

The file ``qcdutils.catalog'' is only used internally by qcdutils and should not be deleted or else it loses track of completed downloads and may perform them again unnecessarily.

If can pass a {\ft qcdutils.catalog} to {\ft qcdutils\_get} you get a report about the downloaded files.

\begin{lstlisting}
python qcdutils_get.py demo/qcdutils.catalog
\end{lstlisting}

Notice that output files are never overwritten so make sure you delete the old one if you want to create new ones.

\goodbreak\subsection{Converting gauge configurations to the FermiQCD format (.mdp)}

This option works similarly to the previous section:

\begin{lstlisting}
python qcdutils_get.py http://qcd.nersc.gov/nersc/api/files/demo
python qcdutils_get.py -c mdp demo/*
\end{lstlisting}
\noindent or in one line

\begin{lstlisting}
python qcdutils_get.py -c mdp http://qcd.nersc.gov/nersc/api/files/demo
\end{lstlisting}

It creates

\begin{lstlisting}
demo
demo/demo.nersc
demo/demo.nersc.mdp
demo/qcdutils.catalog.db
\end{lstlisting}

As in the previous case you can specify the precision of the converted file using {\ft -4} or {\ft -8}.

Notice you cannot specify the endianness. The FermiQCD format (.mdp) uses LITTLE endianness by convention because that is the format used internally by x386 compatible architectures.

\goodbreak\subsection{Splitting gauge configurations into time-slices}

Often we need to break a single gauge configuration with $T$  time-slices into $T$  gauge configurations with 1 time-slice each. You can do it using the {\it slice.mdp} output file format:

\begin{lstlisting}
python qcdutils_get.py http://qcd.nersc.gov/nersc/api/files/demo
python qcdutils_get.py -c slice.mdp demo/demo.mdp
\end{lstlisting}

The first line creates the following files:
\begin{lstlisting}
demo/
demo/demo.nersc
demo/qcdutils.catalog.db
\end{lstlisting}
\noindent while the second line creates:
\begin{lstlisting}
demo/demo.nersc.t0001.mdp
demo/demo.nersc.t0002.mdp
demo/demo.nersc.t0003.mdp
demo/demo.nersc.t0004.mdp
\end{lstlisting}

Four files because ``demo.nersc'' contains 4 timeslices.

\goodbreak\subsection{Splitting ILDG progators into timeslices}

We can play the same trick with propagators. While for gauge configurations {\ft qcdutils} can read multiple file formats, for input propagators qcd can only read FermiQCD and SciDAC propagators.

Given a file ``propagator.scidac'' we can convert it into FermiQCD format:

\begin{lstlisting}
python qcdutils_get.py -c prop.mdp propagator.scidac
\end{lstlisting}
\noindent which creates

\begin{lstlisting}
propagator.scidac.prop.mdp
\end{lstlisting}
\noindent or split into time-slices

\begin{lstlisting}
python qcdutils_get.py -c slices.prop.mdp propagator.scidac
\end{lstlisting}

This creates

\begin{lstlisting}
propagator.scidac.t0000.prop.mdp
propagator.scidac.t0001.prop.mdp
propagator.scidac.t0002.prop.mdp
...
\end{lstlisting}

In this case you can specify the target precision.

\goodbreak\section{Details about file formats}

In this section we show simplified code snippets that should help you understand the different file formats used in Lattice QCD. They are very similar to the actual code implemented in {\ft qcdutils} but simplified for readability.

\goodbreak\subsection{NERSC file format (3x3)}

To better illustrate each data format we present a minimalist program to store data in the corresponding format.

We assume the input is available through an instance of the following class called {\ft data}:

\begin{lstlisting}
class GenericGauge(object):
    def u(x,y,z,t,mu):
        # u_ij below are complex numbers
        return [[u_00,u_01,u_02],
                [u_10,u_11,u_12],
                [u_20,u_21,u_22]]
\end{lstlisting}

In this section (and only in this section) we follow the convention that
$\mu=0$ is $X$, $1$ is $Y$, $2$ is $Z$ and $3$ is $T$. Everywhere else, in particular in the input parameters of {\ft qcdutils\_run} $\mu=0$ is $T$, $1$ is $X$, $2$ is $Y$ and $3$ is $Z$, which is the FermiQCD convention.

The following code shows how to read {\ft data} and write it in the NERSC3x3 format:

\begin{lstlisting}
NERSC_3x3_HEADER = """BEGIN_HEADER
HDR_VERSION = 1.0
DATATYPE = 4D_SU3_GAUGE_3x3
DIMENSION_1 = %(NX)i
DIMENSION_2 = %(NY)i
DIMENSION_3 = %(NZ)i
DIMENSION_4 = %(NT)i
CHECKSUM = %(checksum)s
LINK_TRACE = %(linktrace)f
PLAQUETTE = %(plaquette)f
CREATOR = %(creator)s
ARCHIVE_DATE = %(archive_date)s
ENSEMBLE_LABEL = %(label)s
FLOATING_POINT = %(precision)s
ENSEMBLE_ID = %(ensemble_id)s
SEQUENCE_NUMBER = %(sequence_number)i
BETA = %(beta)f
MASS = %(mass)f
END_HEADER
"""
def save_3x3_nersc(filename,metadata,data):
  f = open(filename,'wb')
  f.write(NERSC_3x3_HEADER % metadata)
  nt = metadata['NT']
  nx = metadata['NX']
  ny = metadata['NY']
  nz = metadata['NX']
  if metadata['PLOATING_POINT']=='IEEE32':
     couple = '>2f'
  elif metadata['PLOATING_POINT']=='IEEE64':
     couple = '>2d'
  else:
     raise RuntimeError, "Unknown precision"
  for t in range(nt):
    for z in range(nz):
      for y in range(ny):
        for x in range(nz):
          for mu in range(0,1,2,3):
            u = data.u(x,y,z,t,mu)
            for i in range(3):
              for j in range(3)
                c = u[i][j]
                re,im = real(c),imag(c)
                f.write(struct.pack(couple,re,im))
\end{lstlisting}

The variable {\ft couple} determines how to pack in binary the 2 variables  {\ft re,im} using big endianness (``$>$'') in single (``f'') or double precision (``d''). For more info read the documentation on the Python ``struct'' package.

All common file formats used by the community to store QCD gauge configuration require two loops: one loop over the lattice sites and one loop over the link directions at each lattice site.

In the NERSC, ILDG and MILC case, the first loop is:
\begin{lstlisting}
for t ...
  for z ...
    for y ...
      for x ...
\end{lstlisting}
and the second loop is:
\begin{lstlisting}
for mu in (X,Y,Z,T) # (0,1,2,3)
\end{lstlisting}

\goodbreak\subsection{NERSC file format (3x2)}

The NERSC 3x2 format is more common than NERSC 3x3 and here is how to write it:

\begin{lstlisting}
NERSC_3x2_HEADER = """BEGIN_HEADER
HDR_VERSION = 1.0
DATATYPE = 4D_SU3_GAUGE
DIMENSION_1 = %(NX)i
DIMENSION_2 = %(NY)i
DIMENSION_3 = %(NZ)i
DIMENSION_4 = %(NT)i
CHECKSUM = %(checksum)s
LINK_TRACE = %(linktrace)f
PLAQUETTE = %(plaquette)f
CREATOR = %(creator)s
ARCHIVE_DATE = %(archive_date)s
ENSEMBLE_LABEL = %(label)s
FLOATING_POINT = %(precision)s
ENSEMBLE_ID = %(ensemble_id)s
SEQUENCE_NUMBER = %(sequence_number)i
BETA = %(beta)f
MASS = %(mass)f
END_HEADER
"""
def save_3x2_nersc(filename,metadata,data):
  f = open(filename,'wb')
  f.write(NERSC_3x2_HEADER % metadata)
  nt = metadata['NT']
  nx = metadata['NX']
  ny = metadata['NY']
  nz = metadata['NZ']
  if metadata['PLOATING_POINT']=='IEEE32':
     couple = '>2f'
  elif metadata['PLOATING_POINT']=='IEEE64':
     couple = '>2d'
  else:
     raise RuntimeError, "Unknown precision"
  for t in range(nt):
    for z in range(nz):
      for y in range(ny):
        for x in range(nz):
          for mu in range(0,1,2,3):
            u = data.u(x,y,z,t,mu)
            for i in range(3):
              for j in range(2) # here
                c = u[i][j]
                re,im = real(c),imag(c)
                f.write(struct.pack(couple,re,im))
\end{lstlisting}
Notice it differs from NERSC 3x3 by only two lines. One line is in the header:
\begin{lstlisting}
DATATYPE = 4D_SU3_GAUGE
\end{lstlisting}
instead of
\begin{lstlisting}
DATATYPE = 4D_SU3_GAUGE_3x3
\end{lstlisting}
and in the line marked by a {\ft here}.

This file format does not store all the 3x3 matrices but only the first two rows. The third row can be reconstructed when reading the file using the condition that the third row is the (complex) vector product of the first two.

{\ft qcdutils} can reads and rebuilds the missing rows using this code:

\begin{lstlisting}
def reunitarize(s):
  (a1re, a1im, a2re, a2im, a3re, a3im,
   b1re, b1im, b2re, b2im, b3re, b3im) = s
  c1re = a2re*b3re - a2im*b3im - a3re*b2re + a3im*b2im
  c1im = -(a2re*b3im + a2im*b3re - a3re*b2im - a3im*b2re)
  c2re = a3re*b1re - a3im*b1im - a1re*b3re + a1im*b3im
  c2im = -(a3re*b1im + a3im*b1re - a1re*b3im - a1im*b3re)
  c3re = a1re*b2re - a1im*b2im - a2re*b1re + a2im*b1im
  c3im = -(a1re*b2im + a1im*b2re - a2re*b1im - a2im*b1re)
  return (a1re, a1im, a2re, a2im, a3re, a3im,
          b1re, b1im, b2re, b2im, b3re, b3im,
          c1re, c1im, c2re, c2im, c3re, c3im)
\end{lstlisting}

\goodbreak\subsection{MILC file format}

The MILC file format is the same as the NERSC 3x3 but the header contains different information, it uses a binary format, and the endianness is not spedified (although generally large endianness is used). The binary data after the header is the same as NERSC3x3.

Here is an example of code to write a MILC gauge configuration in Python:

\begin{lstlisting}
def save_milc(filename,metadata,data):
  f = open(filename,'wb')
  milc_header = '>i4i64siii'
  milc_magic = 20103
  f.write(struct.pack(milc_header,
            milc_magic,
            metadata['NX'],
            metadata['NY'],
            metadata['NZ'],
            metadata['NT'],
            metadata['ARCHIVE_DATE'][:64],
            metadata['ORDER'],
            metadata['CHECKSUM1'],
            metadata['CHECKSUM2']))
  nt = metadata['nt']
  nx = metadata['nx']
  ny = metadata['ny']
  nz = metadata['nz']
  if metadata['PLOATING_POINT']=='IEEE32':
     couple = '>2f'
  elif metadata['PLOATING_POINT']=='IEEE64':
     couple = '>2d'
  else:
     raise RuntimeError, "Unknown precision"
  for t in range(nt):
    for z in range(nz):
      for y in range(ny):
        for x in range(nz):
          for mu in range(0,1,2,3):
            u = data.u(x,y,z,t,mu)
            for i in range(3):
              for j in range(3)
                c = u[i][j]
                re,im = real(c),imag(c)
                f.write(struct.pack(couple,re,im))
\end{lstlisting}

When reading a MILC gauge configuration, {\ft qcdutils\_get} checks for the magic number and determines the endianness from the first 4bytes of the header. {\ft qcdutils} also determines the precision from the total file size.

\goodbreak\subsection{FermiQCD file format}

The file format used by FermiQCD is called MDP and files have a ".mdp" extensions. They are very similar to the MILC format with these differences:
\begin{itemize}
\item the header has a different format and stores slightly different information

\item the endianness is always little-endian.

\item the inner loop over mu has the same order as the outer loop

\item it is designed to work for an arbitrary number of dimensions (from 1D lattices to 10D lattices and have an arbitrary site structure) and the FermiQCD code deal with this aspect in an automated way that is explored later.
\end{itemize}

For regular QCD ($SU(3)$  matrices per link and 4D lattice) a FermiQCD gauge configuration can be generated using the following code:

\begin{lstlisting}
def save_fermiqcd_4d_su3(filename,metadata,data):
  f = open(filename,'wb')
  nt = metadata['nt']
  nx = metadata['nx']
  ny = metadata['ny']
  nz = metadata['nz']
  header_format = '<60s60s60sLi10iii'
  maginc_number = 1325884739
  ndim,
  if metadata['PLOATING_POINT']=='IEEE32':
     couple = '>2f'
     metadata['SITE_SIZE'] = 4*9*2*4
  elif metadata['PLOATING_POINT']=='IEEE64':
     couple = '>2d'
     metadata['SITE_SIZE'] = 4*9*2*4
  else:
     raise RuntimeError, "Unknown precision"
  f.write(struct.pack(header_format,
            'File Type: MDP FIELD',
            metadata['FILENAME'],
            metadata['ARCHIVE_DATE'][:60],
            magic_number,ndim,nt,nx,ny,nz,0,0,0,0,0,0,
            metadata['SITE_SIZE'],nt*nx*ny*nz))
  for t in range(nt):
    for z in range(nz):
      for y in range(ny):
        for x in range(nz):
          for mu in range(3,2,1,0):
            u = data.u(x,y,z,t,mu)
            for i in range(3):
              for j in range(3)
                c = u[i][j]
                re,im = real(c),imag(c)
                f.write(struct.pack(couple,re,im))
\end{lstlisting}

Notice the following:
\begin{itemize}
\item The header is binary but uses a string ``File Type: MDP FIELD'' to identify the file format and version. This allows you to identify the file using an ordinary editor, like in the NERSC format.

\item It still uses an integer magic number to allow the reader to check the endianness.

\item It requires an {\ft ndim} variable which is set to 4 because this manual mostly deals with 4D fields.

\item It stores in {\ft metadata['SITE\_SIZE']} the number of bytes for each lattice site. For single precision this is 4 directions times 9 SU(3) matrix elements times 2 (real+complex) times 4 (4 bytes for IEEE32, single precision, float) = 288 bytes. It is 576 bytes for double precision.
\end{itemize}

\goodbreak\subsection{LIME file format}

The file formats described so far store metadata in a header which precedes the binary data.

The LIME data format is different. It is similar to TAR or MIME as scope. It is designed to package multiple files into one file.

A LIME file is divided into segments (sometime called records in the literature although it does not strictly conform to the definition of a record because LIME has nothing to do with databases). A segment is comprised of five parts: a magic number, a version number, an integer storing the size of the binary data, a segment name, and the binary data.

The magic number identifies the file as a LIME file and the version number identifies the LIME version. This information is repeated for each segment.

Notice that LIME records do not declare the type of the segments and this has to be inferred from the name of the segments. 
One important caveat of LIME is that some segments contain binary data, while some contain ASCII strings such as XML.
Segments that contain ASCII strings are null-terminated and the terminating zero is counted in the size. 
Binary segments are not null-terminated. This is an important detail when reading the data.

{\ft qcdutils} contains a class LIME that can be used to open LIME files and read/write segments in or out of order. 

A minimalist implementation of the LIME file format is the following:

\begin{lstlisting}
class Lime(object):
  def __init__(self,filename,mode,version = 1):
    self.magic = 1164413355
    self.version = version
    self.filename = filename
    self.mode = mode
    self.file = open(filename,mode)
    self.records = [] # [(name,position,size)]
    if mode == 'r' or mode == 'rb':
      while True:
        header = self.file.read(144)
        if not header: break
        magic, null,size, name = struct.unpack('!iiq128s',header)
        if magic != 1164413355:
          raise IOError, "not in LIME format"
        name = name[:name.find('\0')]
        position = self.file.tell()
        self.records.append((name,position,size)) # in bytes
        padding = (8 - (size % 8)) % 8
        self.file.seek(size+padding,1)
  def read(self,record):
    (name,position,size) = self.records[record]
    self.file.seek(position)
    return (name, self.file, size)
  def __iter__(self):
    for record in range(len(self)):
      yield self.read(record)
   def write(self,name,data,size = None,chunk = MAXBYTES):
    position = self.file.tell()
    header = struct.pack('!iiq128s',self.magic,self.version,size,name)
    self.file.write(header)
    self.file.write(data)
    self.file.write('\0'*(8 - (size % 8)) % 8)
    self.records.append((name,size,position))
  def close(self):
    self.file.close()
\end{lstlisting}

The actual implementation in {\ft qcdutils} is more complex because it performs more checks and because it can read and write segments even if they do not fit in RAM, which is not the case in the example above.

Here is an example of usage from Python:

Open a LIME file for writing

\begin{lstlisting}
>>> from qcdutils_get import Lime
>>> lime = Lime('test.lime','w')
\end{lstlisting}

Write two records in it

\begin{lstlisting}
>>> lime.write('record1','01234567')
>>> lime.write('record2','other binary data')
\end{lstlisting}
Close it

\begin{lstlisting}
>>> lime.close()
\end{lstlisting}

Open the file again for reading:
\begin{lstlisting}
>>> lime = Lime('test.lime','r')
\end{lstlisting}
Loop over the segments and print, name size, content:
\begin{lstlisting}
>>> for name,reader,size in lime:
...     print (name, size, reader.read(size))
\end{lstlisting}

\goodbreak\subsection{ILDG file format}

The ILDG file format uses LIME to package two segments:
\begin{itemize}
\item One segment contains the metadata marked up in XML.

\item One segment contains the binary data, in the same format as in MILC and NERSC 3x3.
\end{itemize}

The XML markup is specified by ILDG for 4D gauge files. Notice that because the first segment refers to the second, many programs that read ILDG expect the metadata segment to precede the data segment.

Here is an example of code to write an ILDG file:

\begin{lstlisting}
def save_ildg(filename,metadata,data,lfn):
  lime = Lime(filename,'wb')
  lime.write('ildg-format',"""
    <?xml version = "1.0" encoding = "UTF-8"?>
      <ildgFormat>
        <version>%(VERSION)s</version>
        <field>su3gauge</field>
        <precision>%(PRECISION)s</precision>
        <lx>%(NX)s</lx>
        <ly>%(NY)s</ly>
        <lz>%(NZ)s</lz>
        <lt>%(NT)s</lt>
      </ildgFormat>
""".strip() % metadata)
  nt = metadata['NT']
  nx = metadata['NX']
  ny = metadata['NY']
  nz = metadata['NX']
  def writer():
     for t in range(nt):
       for z in range(nz):
         for y in range(ny):
           for x in range(nz):
             for mu in range(0,1,2,3):
               u = data.u(x,y,z,t,mu)
               for i in range(3):
                 for j in range(3)
                  c = u[i][j]
                  re,im = real(c),imag(c)
                  yield struct.pack(couple,re,im)
  self.lime.write('ildg-binary-data',writer)
  self.lime.write('ildg-data-LFN',lfn)
\end{lstlisting}

Notice that this file takes the same arguments as save\_3x3\_nersc plus an addition one called {\ft lfn}. {\ft lfn} stands for {\it lattice file name}. 
\begin{lstlisting}
lfn://myCollab/myFilename
\end{lstlisting}
The {\ft lfn} is intended to be a Unique Resource Identifier (URI) but it not a Universal Resource Locator (URL). The prefix {\ft lfn} is not a protocol like {\ft http} or {\ft ftp}.

\goodbreak\subsection{SciDAC file format}

The SciDAC format is used primarily for storing propagators.
It uses LIME and it packages the following segments:
\begin{itemize}
\item {\ft scidac-binary-data}: 
the actual binary data
\item {\ft scidac-private-file-xml}
\item {\ft scidac-private-record-xml}
\end{itemize}

We do not describe it here becuase this file type is not used by the tools which are described in this manual. Yet we observer that {\ft qcdutils\_get} can convert this files into FermiQCD propagators.

\goodbreak\section{Running physics algorithms with {\ft qcdutils\_run.py}}

{\ft qcdutils\_run.py} is a program for downloading, compiling, and running FermiQCD~\cite{mdp, fermiqcd}. FermiQCD is a library for parallel Lattice QCD algorithms. The library has been improved over time and it now includes algorithms for visualization of Lattice QCD data. You can learn more about LatticeQCD from refs.~\cite{latticeqcd1, latticeqcd2, latticeqcd3}. You can learn more about FermiQCD from:

\url{http://fermiqcd.net}

After you download qcdutils, run the following command:
\begin{lstlisting}
python qcdutils_run.py -download
\end{lstlisting}
This creates a local folder called ``fermiqcd'', download the latest FermiQCD source from the google code repository:
\begin{lstlisting}[keywords={}]
http://code.google.com/p/fermiqcd
\end{lstlisting}

The source include a file ``fermiqcd.cpp'' file, which can parse command line arguments and run various physics algorithms, some described in this section. {\ft qcdutils\_run.py} compiles this source file and stores the compiled one in:
\begin{lstlisting}
fermiqcd/fermiqcd.exe
\end{lstlisting}

Notice the {\ft .exe} extension is used on all supported platforms.

{\ft qcdutils\_run} requires {\ft g++} and you need to install it separately.

Now you can run physics algorithms with:
\begin{lstlisting}
python qcdutils_run.py [options]
\end{lstlisting}
{\ft qcdutils\_run.py} internally calls {\ft fermiqcd.exe} and pass its [options] along.

You can learn more about the FermiQCD options with
\begin{lstlisting}
python qcdutils_run.py -h
\end{lstlisting}

The output is reported in the appendix but you are encouraged to run it yourself with the latest code.

{\ft qcdutils\_run.py} simply passes its command line arguments to ``fermiqcd.exe'' which parses  and calls the corresponding algorithms. Some arguments are special ({\ft -download}, {\ft -compile}, {\ft -mpi}, {\ft -options}, {\ft -h}) because they are handled by {\ft qcdutils\_run} directly. In particular a call to {\ft -options} introspects the source of ``fermiqcd.cpp'' and figures out which arguments are supported.

Notice that FermiQCD can do more of what {\ft qcdutils\_run} can access. For example it supports staggered fermions (including asqtad), staggered mesons, and domain wall fermions. It can do visualizations using those fields too, but that is not discussed here.

You can fork ``fermiqcd.cpp'' and force ``qcdutils\_run'' to use your own source code:

\begin{lstlisting}
python qcdutils_run.py -compile -source myownfermiqcd.cpp
\end{lstlisting}

\goodbreak\subsection{Running in parallel}

There are two ways to run FermiQCD in parallel with {\ft qcdutils\_run.py}. On an SMP machine you can simply run with the option {\ft -PSIM\_NPROCS=<number>}. Here is an example that loads a gauge configuration and computes the plaquette in parallel using 4 processes:

\begin{lstlisting}
python qcdutils_run.py  -PSIM_NPROCS=4 \
    -gauge:start=load:load=demo/demo.nersc.mdp -plaquette
\end{lstlisting}

When running in parallel with {\ft -PSIM\_NPROCS}, FermiQCD uses fork to create the parallel processes and uses named pipes for the message passing. Most PCs and workstations do not allow dynamic memory allocation of more then 2GB of contiguous space and this creates problems when processing large lattices, even if there is enough total RAM available. {\ft -PSIM\_NPROCS} is designed to overcome this limitation. 

FermiQCD with {\ft -PSIM\_PROCS} enables you to run parallel processes on one machine even if there is only enough RAM to run one of them at time but not all of them concurrently. This is because only one of the parallel processes needs to be loaded in RAM at once and the OS can automatically switch between processes by swapping to disk. Communications between the parallel processes are also buffered to disk and therefore they work as expected. For example: 

\begin{itemize}
\item {\ft qcdutils\_run.py -PSIM\_NPROCS=2} forks two processes (0 and 1)
\item p1 is put to sleep and p0 is executed
\item If p0 sends data to p1 the data is stored in a named pipe
\item When p0 is completed or attempts to receive data it is put to sleep
\item When p0 is put to sleep, p1 is loaded in RAM and continues execution.
\item p1 can receive the data sent from p0 by reading form the named pipe.
\end{itemize}

While this is not very efficient, it does allow to run most algorithms even when there is not enought RAM available. The communication patters are implemented in ways that avoid deadlocks.

A better option is to use MPI and this is the preferred option for production runs. If you want to use MPI, it must be pre-installed on your system separately. On Debian/Ubuntu Linux machines this is done with:

\begin{lstlisting}
sudo apt-get install mpich2
cd ~
touch .mpd.conf
chmod 600 .mpd.conf
mpd &
\end{lstlisting}

In order to use it from {\ft qcdutils\_run} you need to recompile FermiQCD with MPI:
\begin{lstlisting}
python qcdutils_run.py -compile -mpi
\end{lstlisting}
\noindent This makes an mpi-based executable for FermiQCD:
\begin{lstlisting}
fermiqcd/fermiqcd-mpi.exe
\end{lstlisting}

You can run it with

\begin{lstlisting}
python qcdutils_run.py -mpi=4 \
    -gauge:start=load:load=demo/demo.nersc.mdp -plaquette
\end{lstlisting}

Internally it calls {\ft mpirun}.

\goodbreak\subsection{General syntax}

The main options of ``qcdutils\_run.py'' are:
\begin{itemize}
\item {\ft -gauge}: creates, loads, and saves gauge configurations
\item {\ft -plaquette}: computes the average plaquette 
\item {\ft -plaquette\_vtk}: generates images of the plaquette density
\item {\ft -polyakov}: computes Polyakov lines
\item {\ft -polyakov\_vtk}: computes images from polyakov lines
\item {\ft -topcharge}: computes the total topological charge
\item {\ft -topcharge\_vtk}: generates images of the topological charge density
\item {\ft -cool}: cools the gauge configurations
\item {\ft -cool\_vtk}: cools the gauge configurations and save images of the topological charge at every step
\item {\ft -quark}: computes a quark propagator (different sources are possible)
\item {\ft -pion}: computes a pion propagator (and optionally saves images of the pion propagator)
\item {\ft -meson}: computes a meson propagator (and optionally saves images of the meson propagator)
\item {\ft -current\_static}: computes a three points correlation function by inserting a light-light between two heavy-light meson operators (and optionally saves images of the current density)
\item {\ft -4quark}: computes all possible contractions of a 4-quark operator between two light mesons.
\end{itemize}

Each option takes optional attributes in the form {\ft:name=value}. All attributes have default values. The {\ft -pion}, {\ft -meson} and {\ft current\_static} operators take an optional {\ft :vtk=true} argument needed to save the VTK files for visualization.

Multple options can be listed and executed together in one run. Although we recommend separating the following operations in different runs:
\begin{itemize}
\item Generate gauge configurations,
\item Compute propagators on each gauge configuration.
\item Measure opeartors by reading previously computed gauge configurations and propagators.
\end{itemize}

The code described here should be considered and example and other cases can be dealt with by modifying the provided examples.

\goodbreak\subsection{Creating a cold or hot gauge configuration}

You can create a cold gauge configuration with the following command
\begin{lstlisting}
python qcdutils_run.py -gauge:start=cold:nt=16:nx=4:ny=4:nz=4:nc=3
\end{lstlisting}

The {\ft -gauge} option sets the gauge parameters of FermiQCD. The option is followed by parameters separated by a colon. All parameters have default values. 

{\ft qcdutils\_run.py} creates a cold gauge configuration with volume
{\ft nt=16:nx=4:ny=4:nz=4}, $SU(N_c)$ with {\ft nc=3}, and saves it with the name ``cold.mdp''.

The order of the parameters is not important.
All parameters have default values. The output lists all parameters which are used.

You can also run
\begin{lstlisting}
python qcdutils_run.py -gauge:start=hot:nt=16:nx=4
\end{lstlisting}
to generate a ``hot.mdp'' gauge configuration. Notice {\ft nc=3} is the default.

\goodbreak\subsection{Loading a gauge configuration}

The {\ft start} attribute of the {\ft -gauge} option takes four possible values:
\begin{itemize}
\item {\ft cold}: makes a cold gauge configuration
\item {\ft hot}: makes a hot gauge configuration
\item {\ft instantons}: makes a cold configuration containing one instanton and an, optionally, one anti-instanton at given positions.
\item {\ft load}: loads one or more gauge configurations (if more then one, it loops over them)
\end{itemize}
When not set, {\ft start} defaults to {\ft load}, and FermiQCD expects to load input gauge configurations.

In this case, the {\ft load} attribute of the {\ft -gauge} option specifies the pattern of the filenames to read.

You can specify one single gauge configuration by filename or multiple configurations using a glob pattern (for example ``*.mdp'').

Here is an example that loads all gauge configurations in the ``demo'' folder and computes their average plaquette ({\ft -plaquette}):

\begin{lstlisting}
python qcdutils_run.py -gauge:start=load:load=demo/*.mdp -plaquette
\end{lstlisting}

Similarly if you want to download a stream of NERSC gauge configurations and compute the average plaquette on each of them you can do:

\begin{lstlisting}
python qcdutils_get.py -c mdp -4 http://qcd.nersc.org/nersc/api/files/demo
python qcdutils_run.py -gauge:load=demo/*.mdp -plaquette > run.log
grep plaquette run.log
\end{lstlisting}

When loading gauge configurations there is no need to specify the volume since FermiQCD reads that information from the input files.

If you peek into ``fermiqcd/fermiqcd.cpp'' you can find code like this:

\begin{lstlisting}[language=C++]
if(arguments.have("-plaquette")) {
  mdp << "plaquette = " << average_plaquette(U) << endl;
}
\end{lstlisting}
Here {\ft arguments.have("-plaquette")} checks that the option is present and {\ft average\_plaquette(U)} performs the computation for the input gauge configuration {\ft U}. {\ft mdp} is the parallel output stream and it double as wrapper object for the MPI communicator.

\goodbreak\subsection{Heatbath Monte Carlo}

Whether you start form a cold, hot or loaded gauge configuration you can generate more by using the {\ft n} attribute. In this example:

\begin{lstlisting}
python qcdutils_run.py -gauge:start=cold:beta=4:n=10:therm=100:steps=5
\end{lstlisting}

FermiQCD starts from a cold configuration, and using the Wilson gauge action~\cite{wilsongauge} (default) generates {\ft n=10} gauge configurations. It perform 100 thermalization steps ({\it therm}) starting from the cold one and then 5 {\it steps} separating the one configuration from the next.  

It saves the gauge configuration files with progressive names:
\begin{lstlisting}
cold.mdp
cold.0000.mdp
cold.0001.mdp
...
cold.0099.mdp
\end{lstlisting}
If you want to change ``cold'' prefix of numbered filename you can specify the {\ft prefix} attribute of the {\ft -gauge} option. When this attribute is missing, {\ft prefix} defaults to the name of the starting gauge configuration, i.e. ``cold''.

When you start from hot or cold, FermiQCD generates output files in the current working directory. If you start from a loaded file, it generates output files (gauge configurations, propagators, vtk files) in the same folder as the input files.

You can use the optional {\ft alg} attribute to use an improved action or a SSE2 optimized action. 

Here is the relevant code in ``fermiqcd.cpp'':

\begin{lstlisting}[language=C++]
   int nconfigs = arguments.get("-gauge","n",0);
   ...
   for(int n=-1; n<nconfigs; n++) {
      if(n>=0) {
        int niter =(n==0)?ntherm:nsteps;
        if (gauge_action=="wilson")
          WilsonGaugeAction::heatbath(U,gauge,niter);
        else if (gauge_action=="wilson_improved")
          ImprovedGaugeAction::heatbath(U,gauge,niter);
\end{lstlisting}

Use {\ft -options} to see which algorithms are available. For example you can declare an improved gauge action:

\begin{lstlisting}
-gauge:action=wilson_improved:beta=...:zeta=...:u_s=...:u_t=...
\end{lstlisting}

where $\zeta$, $u_t$, and $u_s$ are the parameters of the improved un-isotropic action defined in ref.~\cite{morningstar}.

\goodbreak\subsection{Computing a pion propagator}

We define a pion propagator as

\begin{eqnarray}
C_2[t_1] &=& \sum_\mathbf{x}
\left<\pi(0,\mathbf{0}) | \pi(+t,\mathbf{x})\right> \\
&=& \sum_\mathbf{x} \sum_{ij,\alpha\beta\delta\rho}
\left<0\right|
\bar q_a^{i\alpha}(0)\gamma^5_{\alpha\beta} q_b^{i\beta}(0)
\bar q_b^{j\delta}(t,\mathbf{x})\gamma^5_{\delta\rho} q_a^{i\rho}(t,\mathbf{x})
\left|0\right> \label{c2eq}\\
&=& \sum_\mathbf{x} \sum_{i,\alpha}\left|S^{ii,\alpha\alpha}(t,\mathbf{x})\right|^2
\end{eqnarray}
where
\begin{equation}
S^{ij,\alpha\beta}(t,\mathbf{x}) \equiv \left<0\right|
\{q^{i\alpha}(0), \bar q^{j\beta}(t,\mathbf{x})\}
\left|0\right>
\end{equation}
is a quark propagator with source at $\mathbf{0}$. Here $a$ and $b$ label quark flavours, $i$ and $j$ label color indexes, $\alpha$, $\beta$, $\delta$, $\rho$ label spin indexes. Notice we used the known identity

\begin{equation}
\left<0\right|\{q^{i\alpha}(t,\mathbf{x}), \bar q^{j\beta}(0)\}\left|0\right> 
= \sum_{\rho\delta}\gamma^5_{\alpha\rho}S^{\ast,ji,\delta\rho}(t,\mathbf{x})\gamma^5_{\delta\beta}
\end{equation}

You can compute C2 using the following syntax:

\begin{lstlisting}
python qcdutils_run.py \
       -gauge:start=cold:beta=4:n=10:steps=5:therm=100 \
       -quark:kappa=0.11:c_sw=0.4:save=false -pion > run.log
\end{lstlisting}

{\ft qcdutils\_run} calls ``fermiqcd/fermiqcd.exe'' which generates 10 gauge configurations and, for each, computes a quark propagators with the given values of $\kappa$  and $c_{SW}$  using a fast implementation of the clover action (another attribute that can be set) and compute the pion progator. 

The {\ft -quark} option loops over the $j,\beta$ indexes and computes the $S^{ij,\alpha\beta}(t,\mathbf{x})$. The {\ft -pion} options loops over the $i,\alpha$ indexes and for every $t$ computes the zero momentum Fourier transform in $\mathbf{x}$ of eq.~\ref{c2eq}.

Notice that by default qcdutils saves all the $S$ components. We can avoid it with {\ft save=false}.

The pion propagator for each gauge configuration can be found in the output log file.

\begin{lstlisting}
grep C2 run.log
\end{lstlisting}

The output of {\ft qcdutils\_run} in this case is looks like the following.

\begin{lstlisting}
C2[0] = 14.4746
...
C2[15] = 0.794981
C2[0] = 14.4746
...
C2[15] = 0.794981
...
\end{lstlisting}
For each $t$, {\ft C[t]} takes a different value on each gauge configuration.

In some of the following example we rely on the output pattern:
\begin{lstlisting}
C2[...] = ...
\end{lstlisting}

Later we show how to use {\ft vtk=true} option to save the progagator as function of $x$ and visualize it. We also show tools to automate the analysis of logfiles like ``run.log''.

If you peek into ``fermiqcd.cpp'' you find the following code that computes the pion propagator:

\begin{lstlisting}[language=C++]
for(int a=0; a<4; a++)
  for(int i=0; i<nc; i++) {
    psi = 0;
    if (on_which_process(U.lattice(),0,0,0,0)==ME) x.set(0,0,0,0);
    psi(x,a,i)=1;
    psi.update();
    [...]
    mul_invQ(phi,psi,U,quark,abs_precision,rel_precision);
    [...]
    if (arguments.have("-pion")) {
       [...]
       forallsitesandcopies(x) {
          for(int b=0; b<4; b++)
            for(int j=0; j<nc; j++) {
              tmp = real(phi(x,b,j)*conj(phi(x,b,j)));
              pion[(x(TIME)-t0+NT)%NT] += tmp;
              Q(x) += tmp;
            }
        }
    }
 }
\end{lstlisting}

Notice the field $Q$ which is used in the next section.
It is used for 3D visualizations of the propagator. 

\goodbreak\subsection{Action and inverters}

You can change the action by setting the {\ft action} attribute of the {\ft -quark} option to one of the following: {\ft clover\_fast}, {\ft clover\_slow}, {\ft clover\_sse2}. The first of them is the fastest portable implementation. The second is a slower but more readable one. The first two support arbitrary $SU(N_c)$ gauge groups while the latter is optimized in assembler for $N_c=3$. All of them support clover, and un-isotropic actions. The attributes are 

\begin{center}
\begin{tabular}{|c|c|} \hline
{\ft kappa} & $\kappa$ \\
{\ft kappa\_s} & $\kappa_s$ \\
{\ft kappa\_t} & $\kappa_t$ \\
{\ft c\_sw} & $c_{SW}$ \\
{\ft c\_E} &$c_E$ \\
{\ft c\_B} &$c_B$ \\ \hline
\end{tabular}
\end{center}

If separate values for $\kappa_{s,t}$ are not specified, $\kappa$ is used for both. $c_E$ is the coefficient that multiplies the electric part of the SW term, $c_B$ multiplies the magnetic part. $c_{SW}$ defaults to 0.

The inverter can be specified using the {\ft alg} attribute of the {\ft -quark option} and it can be one of the following: {\ft bicgstab}, {\ft minres}, {\ft bicgstabvtk}, {\ft minresvtk}. The meaning of the first two is obvious. The second two perform the extra task of saving the field components and the residue at every step of the inversion as a VTK file.

The {\ft -quark} option also takes an optional {\ft source\_type} attribute which can be {\ft point} or {\ft wall} and, if point, a {\ft source\_point} attribute to position the source at {\ft zero} or the {\ft center} of the lattice. It also takes the optional {\ft smear\_steps} and {\ft smear\_alpha} which are used to smear the sink.

The relevant code in ``fermiqcd.cpp'' is:

\begin{lstlisting}
  for(int a=0; a<4; a++)
    for(int i=0; i<nc; i++) {
      if(source_type==''point'') {
        psi = 0;
        if (on_which_process(U.lattice(),t0,x0,y0,z0)==ME) {
          x.set(t0,x0,y0,z0);
          psi(x,a,i)=1;
        }
      }
      [...]
      psi.update();
      [...]
      if (arguments.get(``-quark'',''load'',''false|true'')==''true'') {
        phi.load(quarkfilename);
      } else {
        mul_invQ(phi,psi,U,quark,abs_precision,rel_precision);
        phi.save(quarkfilename);
      }
      [...]
      if(use_propagator) {
        forallsites(x) {
          forspincolor(b,j,nc) {
            S(x,a,b,i,j) = phi(x,b,j);
          }
        }
      }
    }
\end{lstlisting}

Notice that in FermiQCD inverters are action agnostic. A call to {\ft mul\_Q(phi,psi,U,...)} computes $\phi=Q[U]\psi$ where $Q$ is the selected action for the type of fermion $\psi$ (in this document we deal only with wilson type fermions but it works with staggered and domain wall too). A call to {\ft mul\_invQ(phi,psi,U,...)} computes $\phi=Q^{-1}[U]\psi$ using the same $Q$ and the selected inverter. There is no code in the inverter which is action specific.

\goodbreak\subsection{Meson propagators}

Given a meson created by $\bar q \Gamma q \left|0\right>$, a meson propagator can be defined as follows:

\begin{eqnarray}
C_2[t_1] &=& \sum_\mathbf{x}
\left<\Gamma^{source}(0,\mathbf{0}) | \Gamma^{sink}(+t,\mathbf{x})\right> \\
&=& \sum_\mathbf{x} \sum_{ij,\alpha\beta\delta\rho}
\left<0\right|
\bar q_a^{i\alpha}(0)\Gamma^{source}_{\alpha\beta} q_b^{i\beta}(0)
\bar q_b^{j\delta}(t,\mathbf{x})\Gamma^{sink}_{\delta\rho} q_a^{i\rho}(t,\mathbf{x})
\left|0\right> \\
&=& \sum_\mathbf{x} \sum_{...}
S^{ij,\beta\delta}(t,\mathbf{x})(\Gamma^{sink}\gamma^5)_{\delta\rho}
S^{\ast ij,\alpha\rho}(t,\mathbf{x})(\gamma^5\Gamma^{source})_{\alpha\beta}
\end{eqnarray}

The command to compute an arbitrary meson propagator and reuse the previously computed propagators (the code assumes different flavours of degenerate quarks, i.e. same mass):

\begin{lstlisting}
python qcdutils_run.py \
       -gauge:start=cold:beta=4:n=10:steps=5:therm=100 \
       -quark:kappa=0.11:c_sw=0.4:save=false \
       -meson:source_gamma=1:sink_gamma=1 > run.log
\end{lstlisting}

The {\ft source\_gamma} and {\ft sink\_gamma} attributes can be specified according to the following table:

\begin{center}
\begin{tabular}{|c|c|}\hline
{\ft source\_gamma}/{\ft sink\_gamma} & $\Gamma^{source}$/$\Gamma^{sink}$ \\ \hline
I & $1$ \\
5 & $\gamma^5$ \\
0 & $\gamma^0$ \\
1 & $\gamma^1$ \\
2 & $\gamma^2$ \\
3 & $\gamma^3$ \\
05 & $\gamma^0\gamma^5$ \\
15 & $\gamma^1\gamma^5$ \\
25 & $\gamma^2\gamma^5$ \\
35 & $\gamma^3\gamma^5$ \\
01 & $\gamma^0\gamma^1$ \\
02 & $\gamma^0\gamma^2$ \\
03 & $\gamma^0\gamma^3$ \\
12 & $\gamma^1\gamma^2$ \\
13 & $\gamma^1\gamma^3$ \\
23 & $\gamma^2\gamma^3$ \\ \hline
\end{tabular}
\end{center}

The relevant code in ``fermiqcd.cpp'' is described here:

\begin{lstlisting}[language=C++]
  if(arguments.have("-meson")) {
    [...]
    G1 = Gamma5*parse_gamma(arguments.get("-meson","source_gamma",...)
    G2 = parse_gamma(arguments.get("-meson","sink_gamma",...))*Gamma5
    forspincolor(a,i,U.nc) {
      forspincolor(b,j,U.nc) {
        forallsites(x) {
          s1=s2=0;
          for(int c=0;c<4;c++) {
            s1 += S(x,a,c,i,j)*G2(c,b);
            s2 += conj(S(x,c,b,i,j))*G1(c,a);
          }
          tmp = abs(s1*s2);
          meson[(x(TIME)-t0+NT)%NT] += tmp;
          Q(x) += tmp;
        }
      }
    }
\end{lstlisting}

As before we use a scalar field $Q$ for data visualization.

In this and the other examples the two quarks are degenerate but it is possible to change one of the quark propagators by simply replacing it in the code for a different one. We leave it to the reader as an exercise. A next version of ``fermiqcd.cpp'' will have an option {\ft -quark2} for doing this automatically.

\goodbreak\subsection{Current insertion}

We define it as follows (for two light quarks $a$, $b$ and one static quark $h$):

\begin{eqnarray}
C_{current}[t] = &=& \sum_\mathbf{x}
\left<\Gamma^{source}_{ha}(-t,\mathbf{x})\right|
\bar q_{a} \Gamma^{current} q_{b}(0)
\left|\Gamma^{sink}_{bh}(+t,\mathbf{x})\right> \nonumber \\
&=& \sum_\mathbf{x} \sum_{...}
\left<0\right|
\bar h^{i\alpha}(-t,\mathbf{x})\Gamma_{\alpha\beta} q_a^{i\beta}(-t,\mathbf{x})
\bar q_a^{r,\zeta} \Gamma^{current}_{\zeta\theta} q_b^{s\theta}
\bar q_b^{j\delta}(t,\mathbf{x})\Gamma_{\delta\rho} h^{i\rho}(t,\mathbf{x})
\left|0\right> \nonumber \\
&=& \sum_\mathbf{x} \mathrm{tr}(\Gamma^{source} \gamma^5 S^\dagger(-t,\mathbf{x}) \gamma^5 \Gamma^{current} S(t,\mathbf{x}) \Gamma^{sink} H^\dagger(-t,t,\mathbf{x}))
\end{eqnarray}

Here $H$ is the heavy quark propagator according to Heavy Quark Effective Theory~\cite{hqet} (from $(-t,\mathbf{x})$ to $(t,\mathbf{x})$):

\begin{equation}
H(-t,t,x) = \frac12(1+\gamma^0)U_0(-t,x)U_0(-t+1,x)...U_0(t-1,x)
\end{equation}

You can compute it with

\begin{lstlisting}
python qcdutils_run.py \
   -gauge:start=cold:beta=4:n=10:steps=5:therm=100 \
   -quark:kappa=0.11:c_sw=0.4:save=false \
   -current_static:source_gamma=1:sink_gamma=1:current_gamma=I > run.log
\end{lstlisting}

The relevant code in ``fermiqcd.cpp'' is:

\begin{lstlisting}[language=C++]
G1 = parse_gamma(arguments.get("-current_static","source_gamma",...))*Gamma5;
G2 = parse_gamma(arguments.get("-current_static","sink_gamma",...));
G3 = Gamma5*parse_gamma(arguments.get("-current_static","current_gamma",...));
G4 = G2*(1-Gamma[0])/2*G1;
forallsites(x)
     if(x(TIME)>=0) {
        z.set((NT+2*t0-x(TIME))%NT,x(1),x(2),x(3));
        forspincolor(a,i,U.nc) {
          forspincolor(b,j,U.nc) {
            s1 = s2 = 0;
            for(int c=0; c<4; c++) {
              s1 += conj(S(z,c,a,j,i))*G3(c,b);
              for(int k=0; k<U.nc; k++)
                s2 += S(x,b,c,j,k)*G4(c,a)*conj(Sh(x,i,k));
            }
            tmp = abs(s1*s2);
            current[(x(TIME)-t0+NT)%NT] += tmp;
            Q(x) += tmp;
          }
        }
      }
\end{lstlisting}

Here $Sh$ is the product of links from $-t$ to $t$ along the time direction.

\goodbreak\subsection{Four quark operators}

Instead of inserting a current we can insert a 4-quark operator between two meson operators (light-light):

\begin{eqnarray}
C_3[t_1][t_2] &=& \sum_\mathbf{x_1}\sum_\mathbf{x_2}
\left<\Gamma^{source}(-t_1,\mathbf{x_1})\right|
\bar q_a \Gamma_A q_b \otimes \bar q_c \Gamma_B q_d
\left|\Gamma^{sink}(+t_2,\mathbf{x_2})\right> \\
&=& \textrm{tr}(\Gamma^{source}\gamma^5 S^\dagger(-t_1, \mathbf{x_1}) \gamma^5 \Gamma_A S(-t_1,\mathbf{x_1})) \textrm{tr}(\Gamma^{sink}\gamma^5 S^\dagger(t_2, \mathbf{x_2}) \gamma^5 \Gamma_B S(t_2,\mathbf{x_1})) \nonumber \\
&or& \textrm{tr}(\Gamma^{source}\gamma^5 S^\dagger(-t_1, \mathbf{x_1}) \gamma^5 \Gamma_A S(t_2,\mathbf{x_2}) \Gamma^{sink}\gamma^5 S^\dagger(t_2, \mathbf{x_2}) \gamma^5 \Gamma_B S(-t_1,\mathbf{x_1})) \nonumber
\end{eqnarray}

The {\it or} indicates that there are two possible contrations. FermiQCD computes both of them and writes them seperately in the output.

Here $\Gamma_A\otimes\Gamma_B$  is the spin/color structure of the 4-quark operator. We are also ignoring the contractions that corresponds to disconnected diagrams.

We can compute $\Gamma_A\otimes\Gamma_B$  for $\gamma_5\otimes\gamma_5$  in spin and $\mathbf{1}\otimes\mathbf{1}$ in color ({\ft 5Ix5I}) with:

\begin{lstlisting}
python qcdutils_run.py \
     -gauge:start=cold:beta=4:n=10:steps=5:therm=100 \
     -quark:kappa=0.11 -4quark:source=1:operator=5Ix5I > run.log
\end{lstlisting}

In this example, {\ft source=1} indicates that $\Gamma^{source}=\Gamma^{sink}=\gamma^1$.

This generates the following output, repeated for each of the 10 gauge configurations:

\begin{lstlisting}
C3[0][0] = 9.12242
C3[0][1] = 0.485189
...
C3[15][15] = 9.12242
\end{lstlisting}

Notice the program computes the two contractions of the operator and writes one in {\ft C3} and one in {\ft C3x}. 

Instead of {\ft source=1} you can use any of the operators defined for mesons.

Instead of {\ft 5Ix5I} 4-quark operator you can use any the following other operators:
{\ft 5Ix5I}, {\ft 0Ix0I}, {\ft 1Ix1I}, {\ft 2Ix2I}, {\ft 3Ix3I}, {\ft 05Ix05I}, {\ft 15Ix15I}, {\ft 25Ix25I}, {\ft 35Ix35I}, {\ft 01Ix01I}, {\ft 02Ix02I}, {\ft 03Ix03I}, {\ft 12Ix12I}, {\ft 13Ix13I}, {\ft 23Ix23I}, {\ft 5Tx5T}, {\ft 0Tx0T}, {\ft 1Tx1T}, {\ft 2Tx2T}, {\ft 3Tx3T}, {\ft 05Tx05T}, {\ft 15Tx15T}, {\ft 25Tx25T}, {\ft 35Tx35T}, {\ft 01Tx01T}, {\ft 02Tx02T}, {\ft 03Tx03T}, {\ft 12Tx12T}, {\ft 13Tx13T}, {\ft 23Tx23T}. Here the numerical part represents the $\Gamma\otimes\Gamma$ stucture of the 4-quark operator, the $I$ or $T$ represents its color structure. {\ft TxT} stands for $\sum_a T^a \otimes T^a$ with $T^a = \lambda^a/2$, and $\lambda^a$ is the $SU(3)$ generator.

Here is the relevant source code in ``fermiqcd.cpp":

\begin{lstlisting}[language=C++]
    mdp_matrix G = parse_gamma(arguments.get("-4quark","source",...);
    forspincolor(a,i,U.nc) {
      for(int c=0; c<4; c++)
        for(int d=0; d<4; d++)
          if(G(c,d)!=0)
            forallsites(x)
              for(int k=0; k<U.nc; k++)
                 open_prop[a][b][i][j][(x(TIME)-t0+NT)%NT] +=
                    S(x,a,c,i,k)*conj(S(x,b,d,j,k))*G(c,d);
    string op4q = arguments.get("-4quark","operator",...);
    if(arguments.have("-4quark")) {
        for(int a=0; a<4; a++)
          for(int b=0; b<4; b++)
            for(int c=0; c<4; c++)
              for(int d=0; d<4; d++) {
                mdp_complex g1 = G1(b,a);
                mdp_complex g2 = G2(d,c);
                if(g1!=0 && g2!=0)
                  for(int i=0; i<U.nc; i++)
                    for(int j=0; j<U.nc; j++)
                      if(!rotate) {
                        c3a+=abs(open_prop[a][b][i][i][t1s]*g1*
                                 open_prop[c][d][j][j][t2s]*g2);
                        c3b+=abs(open_prop[c][b][j][i][t1s]*g1*
                                 open_prop[a][d][i][j][t2s]*g2);
                      } else
                        [...]
                      }
              }
    }
  }
\end{lstlisting}

Notice the two contractions are computed separately.  The case {\ft rotate==true} corresponds to the {\ft TxT} color stucture.

\goodbreak\section{Images and movies with {\ft qcdutils\_vis.py} and {\ft qcdutils\_vtk.py}}

In this section we describe how to make 3D visualizations using VisIt~\cite{visit} and how to embed visualizations into web pages using ``processing.js``~\cite{processing}.

VisIt is a visualization software developed at Lawrence Livermore National Lab based on the VTK toolkit. It provides a GUI which can be used to open the VTK files created by FermiQCD (or other scientific program) in interactive mode, but it can also be scripted using the Python language.

``processing.js'' is a lightweight javascript library that allows drawing on an HTML canvas using the {\it processing} language or the javascript language.

{\ft qcdutils} uses meta-programming to generate VisIt scripts ({\ft qcdutils\_vis}) or processing.js scripts ({\ft qcdutils\_vtk}). The former is more flexible and is more appropriate for making high resolution images. The latter makes it easy to embed 3D visualizations into web pages.

Using VisIt is intuitive but there are certain tasks which can be repetitive. For example if you have multiple VTK files containing topological change density (or any other scalar field), you have to determine the optimal threshold values for the contour plots. If you have many files you may want to interpolate between them for a smoother visualization.
{\ft qcdutils\_vis} helps with these tasks. In particular it can:

\begin{itemize}
\item Split VTK files containing multiple time-slices into separate VTK files, one for each slice.
\item Interpolate each couple of consecutive VTK files and make new ones in between. This is necessary for smoother visualizations.

\item Compute automatic thresholds values for contour plots.

\item Resample the points by interpolating between the.

\item Generate VisIt scripts which converts VTK files to JPEG format (these script can be saved, edited, and reused).

\item Pipe the above operations and run them for multiple files.
\end{itemize}

Images generate in this way can be assembled into mpeg4 (or quicktime or avi) movies using ffmpeg (an open source tool that is distributed with VisIt) but there are other and better tools available. We strongly recommend ``MPEG Streamclip''. It is much faster, robust, and much easier to properly confgure than ffmpeg.

\goodbreak\subsection{About VTK file format}

There are many VTK file formats. {\ft qcdutils} uses the binary VTK file format described below to store scalar fields, usually by timeslices.

A typical file has the following content:

\begin{lstlisting}
# vtk DataFile Version 2.0
filename.vtl
BINARY
DATASET STRUCTURED_POINTS
DIMENSIONS 4 4 4
ORIGIN     0   0   0
SPACING    1   1   1
POINT_DATA 64
SCALARS scalars_t0 float
LOOKUP_TABLE default
[binary data]
SCALARS scalars_t1 float
LOOKUP_TABLE default
[binary data]
...
\end{lstlisting}

It consists of an ASCII header declaring the 3D dimensions (4 4 4) and the total number of points ($4\times 4 \times 4 = 64$ ). This is following by blocks representing the time-slices. Each block as its own ASCII header:

\begin{lstlisting}
SCALARS scalars_t0 float
LOOKUP_TABLE default
\end{lstlisting}
\noindent followed by binary data (64 floating point numbers).

{\ft scalars\_t0}, {\ft scalars\_t1}, etc. are the names of the fields as stored by FermiQCD. When time-slices are extracted by {\ft qcdutils\_vis} the slices are renamed as {\ft slice}.

Given any VTK file, for example {\ft demo.vtk} we can visualize it using {\ft qcdutils\_vis.py} using the following syntax:

\begin{lstlisting}
python qcdutils_vis.py -r 'scalars_t0' -p default demo.vtk
\end{lstlisting}

{\ft qcdutils\_vis.py} generates images in JPEG format.

Similarly we can visualize by creating an interactive 3D web page:

\begin{lstlisting}
python qcdutils_vtk.py -u 0.10 -l 0.90 demo.vtk
\end{lstlisting}

If the filename is a glob pattern (*.vtk), both tools loop and process and files matching the pattern.

{\ft qcdutils\_vtk} computes the range of values in the scalar field from the maximum to the minimum. {\ft -u 0.10} indicates we want an isosurface at 10\% form the max and {\ft -l 0.90} indicates we want another isosurface at 90\% from the max (10\% from the min). It is also possible to specify the colors of the iso-surfaces.

{\ft qcdutils\_vtk} generates HTML files with the same as the input VTK files followed by the {\ft .html} postfix. The isosurfaces are computed by the Python program itself but the representation of the isosurfaces is embedded in the html file, together with the ``processing.js'' library, and with custom JS code. These files are not static images. You can rotate them in the browser using the mouse.

\goodbreak\subsection{Plaquette}

As an example, we want to make a movie of the plaquette as function of the time-slice. We follow this workflow:
\begin{itemize}
\item Load a gauge configuration.
\item Compute the plaquette at each lattice site.
\item Save the plaquette as a VTK file.
\item Split the VTK file into one file per time-slice.
\item Interpolate the timeslices to generate more frames.
\item Generate contour plots for each frame and save them as JPEG files.
\end{itemize}

This can be done in two steps. Step one:

\begin{lstlisting}
python qcdutils_run.py \
       -gauge:load=demo/demo.nersc.mdp \
       -plaquette_vtk
\end{lstlisting}

This command uses FermiQCD to load the gauge configurations. For each of them it computes the trace of the average plaquette at each lattice site, and generates one VTK file contain the 4D scalar for the plaquette. This file is saved in a new file with the same prefix as the input but ending in ``.plaquette.vtk''.

Step two:

\begin{lstlisting}
python qcdutils_vis.py -s '*' -i 9 -p default 'demo/*.plaquette.vtk'
\end{lstlisting}

It reads all files matching the pattern ``demo/*.plaquette.vtk'', extracts all time-slices with names matching ``*'' (all time slices), and interpolates each couple of VTK files by adding 9 more frames ({\ft -i 9}), then generates a VTK script that reads each VTK file, resamples it, and stores contour plots in JPEG files with consecutive file filenames..

The generated script has a unique name which looks like this:
\begin{lstlisting}
qcdutils_vis_2fac1b86-5b86-42ee-8552.py
\end{lstlisting}

{\ft qcdutils\_vis} writes and runs the script. It saves it for you in case you want to read and modify it.

When it runs, it loops over all the frames, resamples them, computes the contour plots and saves each frame into one JPEG image:
\begin{lstlisting}
qcdutils_vis_2fac1b86-5b86-42ee-8552_0000.00.jpeg
qcdutils_vis_2fac1b86-5b86-42ee-8552_0001.01.jpeg
...
qcdutils_vis_2fac1b86-5b86-42ee-8552_0003.00.jpeg
\end{lstlisting}
Here 0000, 0001, 0002, 0003 are the original frames (timeslices) and the. {\ft .01}, {\ft.02}, ..., {\ft .09} are the interpolated ones.

Notice that the {\ft -i 9} option is very important to obtain smooth sequences of images to be assembled into movies.

The option
\begin{lstlisting}
-p default
\end{lstlisting}
is equivalent to
\begin{lstlisting}
-p 'AnnotationAttributes[];ResampleAttributes[];ContourAttributes[]'
\end{lstlisting}
Here {\ft Annotation}, {\ft Resample} and {\ft Contour} are VisIt functions. Using {\ft -p} you can set the attributes for each functions.

For example, to remove the bounding box you would replace
\begin{lstlisting}
AnnotationAttributes[]
\end{lstlisting}
with
\begin{lstlisting}
AnnotationAttributes[axes3D.bboxFlag=0]
\end{lstlisting}

To increase the re-sampling points from 100 to 160 you would replace:
\begin{lstlisting}
ResampleAttributes[]
\end{lstlisting}
with
\begin{lstlisting}
ResampleAttributes[samplesX=160;samplesY=160;samplesZ=160]
\end{lstlisting}

To change the color of the 9th contour to Orange, you would replace:
\begin{lstlisting}
ContourAttributes[]
\end{lstlisting}
with
\begin{lstlisting}
ContourAttributes[SetMultiColor(9,orange)]
\end{lstlisting}

The argument of the {\ft <function>Attributes[...]} are VisIt attributes and they are described in the VisIt documentation.

\begin{figure}[ht!]
\begin{center}
\begin{tabular}{cc}
\includegraphics[width=8cm]{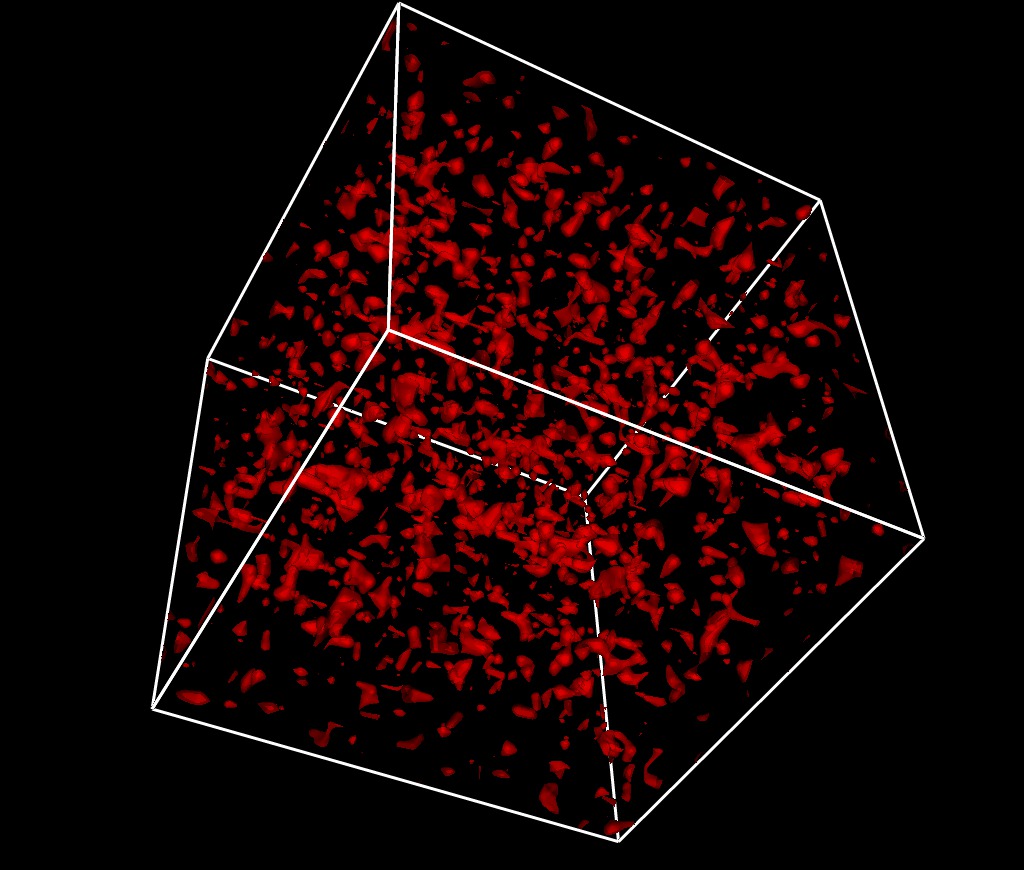} &
\includegraphics[width=8cm]{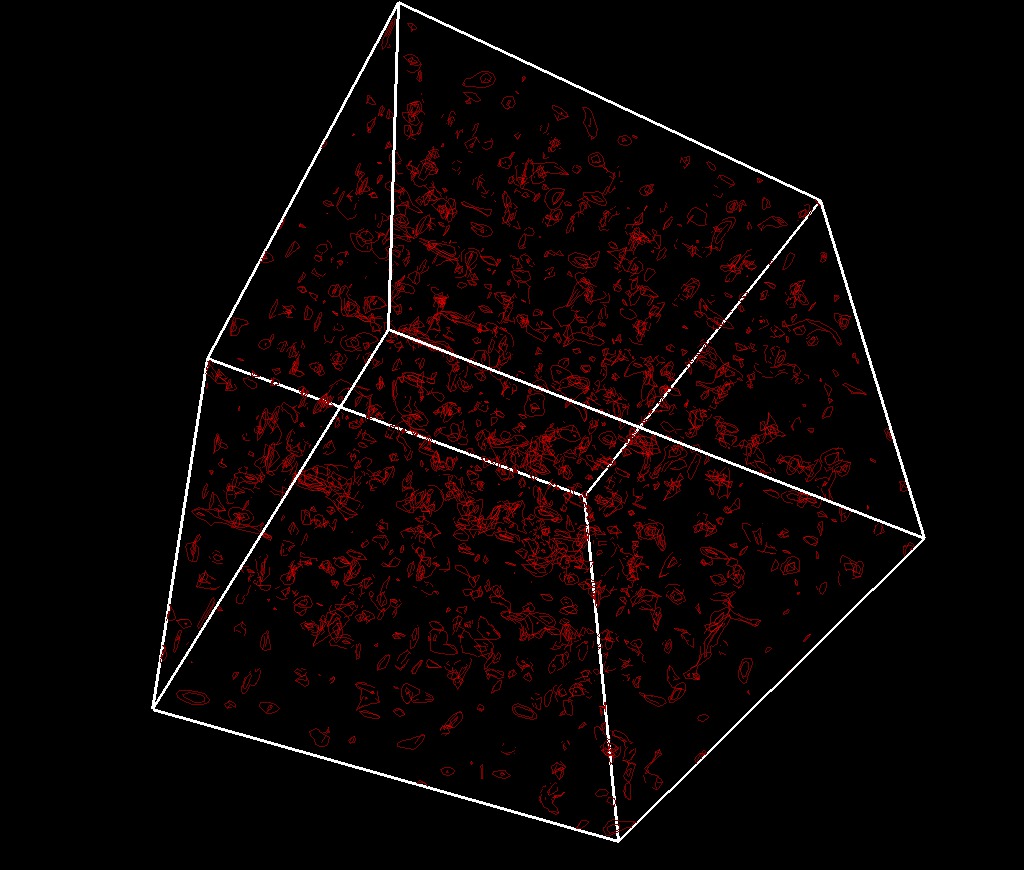}
\end{tabular}
\end{center}
\caption{Visualization of a contour plot for the average plaquette (left) and the intersection of the contours with the bounding box (right)\label{plaquette}}\end{figure}

The relevant page of code in ``fermiqcd.cpp'' that computes the VTK plaquette is here:

\begin{lstlisting}[language=C++]
void plaquette_vtk(gauge_field& U, string filename) {
  mdp_field<mdp_real> Q(U.lattice());
  mdp_site x(U.lattice());
  forallsites(x) if(x(0)==0) {
  Q(x)=0;
  for(int mu=0; mu<4; mu++)
    for(int nu=mu+1; nu<4; nu++)
    Q(x)+=real(trace(plaquette(U,x,mu,nu)));
  }
  Q.save_vtk(filename,-1);
}
[...]
if (arguments.have("-plaquette_vtk")) {
  plaquette_vtk(U,newfilename+".plaquette.vtk");
}
\end{lstlisting}

Notice how the plaquette is computed for each {\ft x}, summed over {\ft mu,nu}, stored in a scalar field {\ft Q(x)}, and then saved in a file. This strategy can be used to visualize any FermiQCD scalar field with minor modifications of the source.

\goodbreak\subsection{Topological charge density}

Similarly to the average plaquette we can make images corresponding to the topological charge density.

To generate the topological change density we need to cool the gauge configurations ({\ft -cool}) and then compute the topological charge ({\ft -topcharge\_vtk}):

\begin{lstlisting}
python qcdutils_run.py \
       -gauge:load=demo/demo.nersc.mdp \
       -cool:steps=20 -topcharge_vtk
\end{lstlisting}

\begin{lstlisting}
python qcdutils_vis.py -s '*' -i 9 -p default 'demo/*.topcharge.vtk'
\end{lstlisting}

\begin{figure}[ht!]
\begin{center}
\includegraphics[width=8cm]{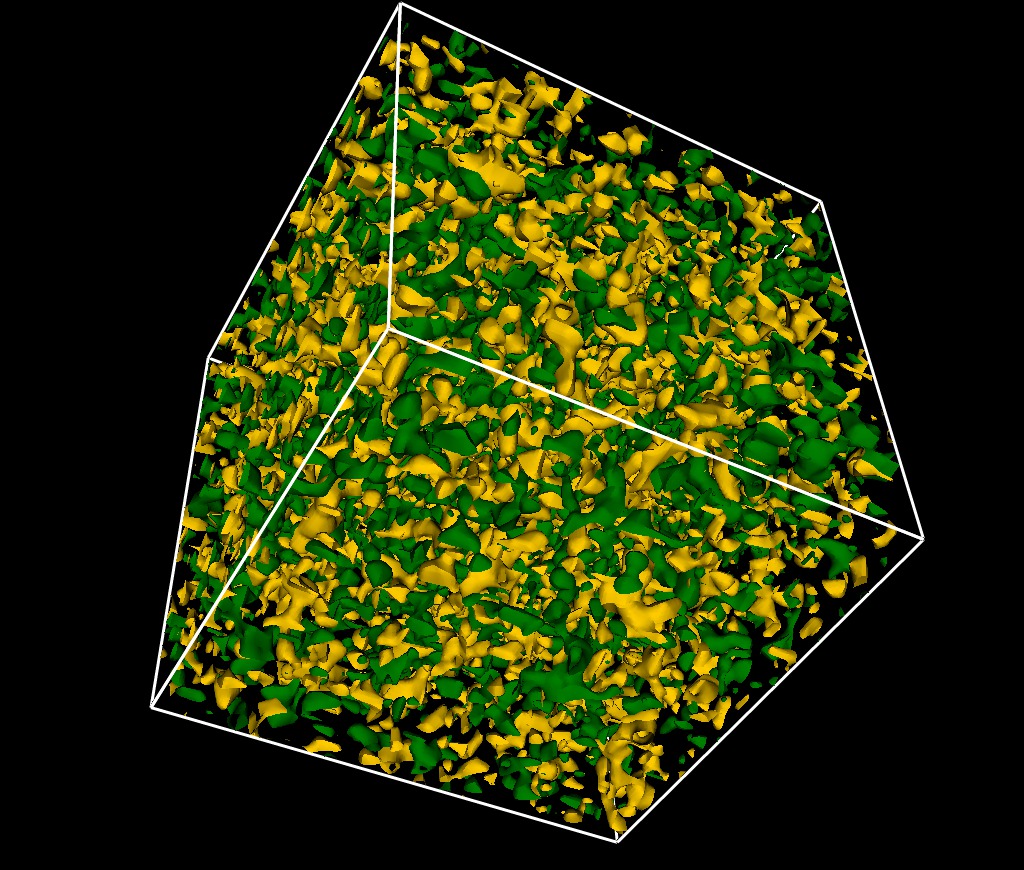}
\end{center}
\caption{Visualization of the topological charge density.\label{topcharge}}\end{figure}

The relavent code in ``fermiqcd.cpp'' is below:

\begin{lstlisting}[language=C++]
if (arguments.have("-topcharge_vtk")) {
   float tc = topological_charge_vtk(U,newfilename+".topcharge.vtk",-1);
   mdp << "topcharge = " << tc << endl;
}
\end{lstlisting}

{\ft topological\_charge\_vtk} is defined in ``fermiqcd\_topological\_charge.h''. The {\ft -1} arguments indicates we want to save all time slices. The actual code to compute the topological charge density is:

\begin{lstlisting}[language=C++]
void topological_charge(mdp_field<float> &Q, gauge_field &U) {
  compute_em_notrace_field(U);
  mdp_site x(U.lattice());
  forallsitesandcopies(x) {
    Q(x)=0;
    for(int i=0; i<U.nc; i++)
      for(int j=0; j<U.nc; j++)
        Q(x)+=real(U.em(x,0,1,i,j)*U.em(x,2,3,j,i)-
                   U.em(x,0,2,i,j)*U.em(x,1,3,j,i)+
                   U.em(x,0,3,i,j)*U.em(x,1,2,j,i));
  }
  Q.update();
}
\end{lstlisting}

Here {\ft U.em} is the eletro-magnetic field computed from {\ft U}.

\goodbreak\subsection{Cooling}

Sometimes we may want to see the changes in the topological charge density as the configuration is cooled down. This requires computing the topological charge density at every cooling step. This can be done with the {\ft -cool\_vtk} option:

\begin{lstlisting}
python qcdutils_run.py  \
       -gauge:start=load:load=demo/demo.nersc.mdp \
       -cool_vtk:cooling=10  > run.log
\end{lstlisting}

\begin{lstlisting}
python qcdutils_vis.py -r 'scalars_t0' -i 9 -p default 'demo/*.cool??.vtk'
\end{lstlisting}

The {\ft -cool\_vtk} option creates VTK files ending in ``cool00.vtk'', ``cool01.vtk'',..., ``cool49.vtk''. To make a smooth movie we do not break files into time-slices (no {\ft -s} option) but instead we extract the same slice for every file ({\ft -r 'scalars\_t0}). Then we interpolate the frames ({\ft -i 9}).

\begin{figure}[ht!]
\begin{center}
\begin{tabular}{cc}
\includegraphics[width=8cm]{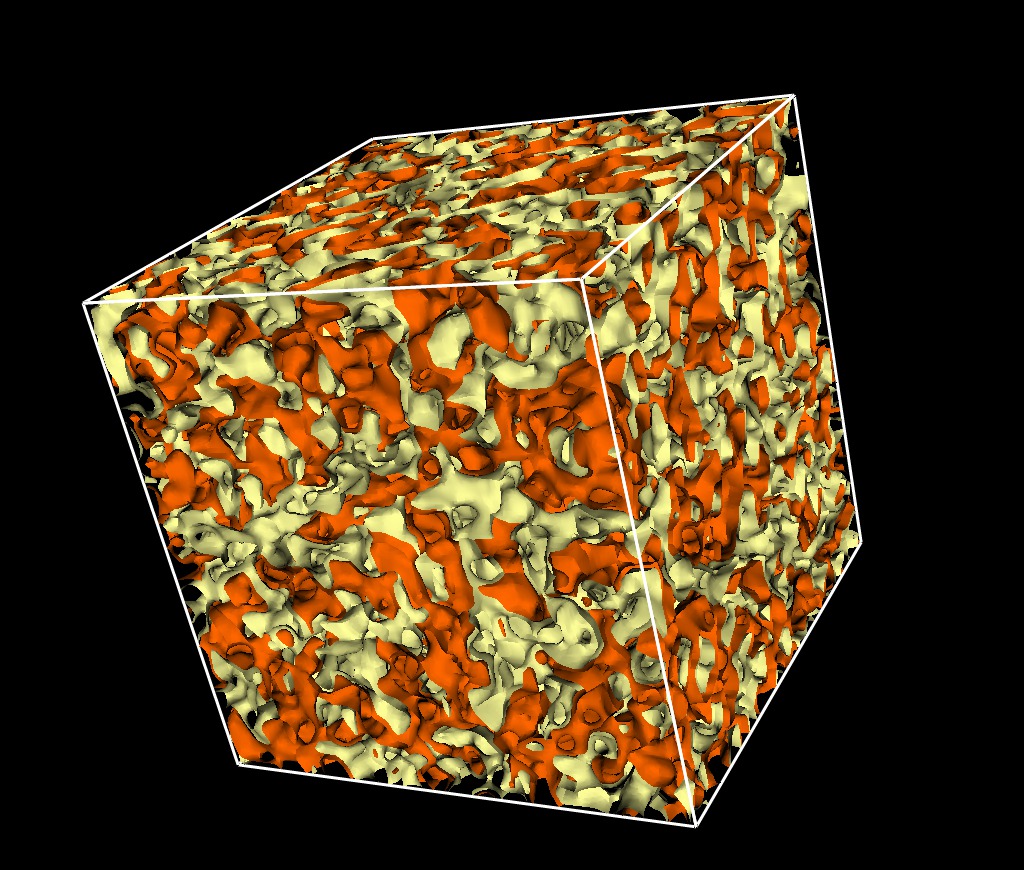} &
\includegraphics[width=8cm]{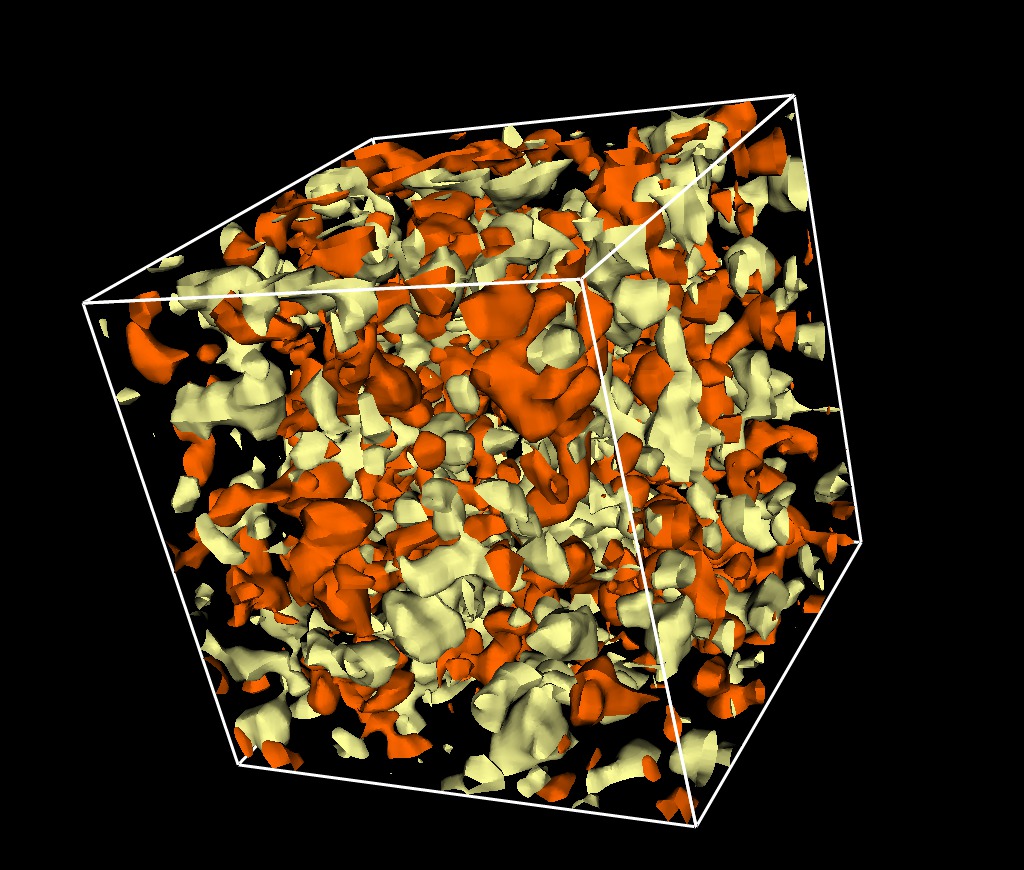} \\
\includegraphics[width=8cm]{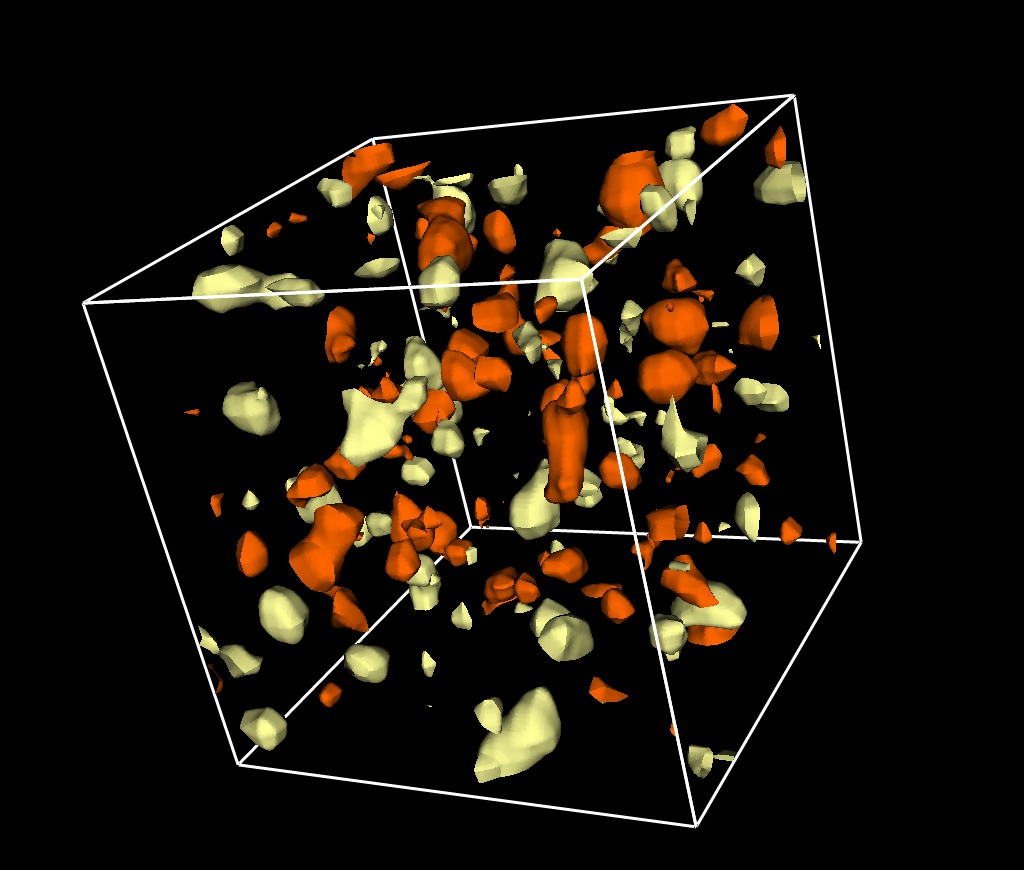} &
\includegraphics[width=8cm]{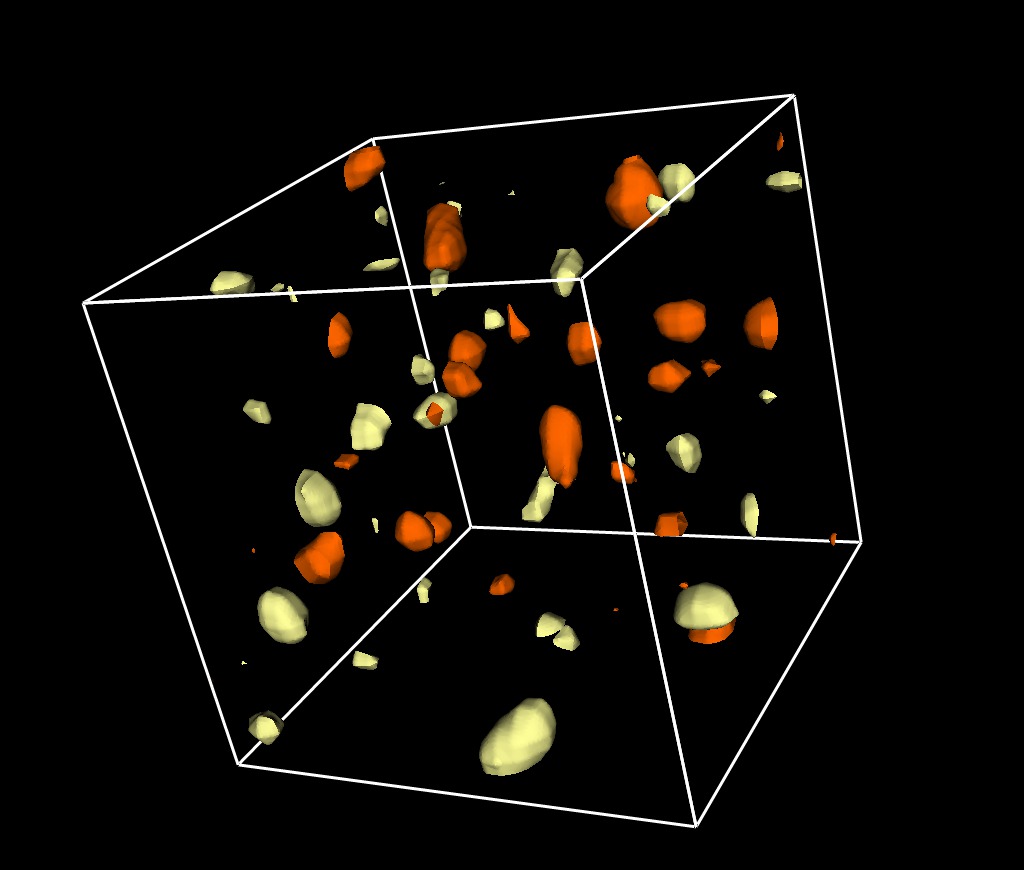}
\end{tabular}
\end{center}
\caption{Visualization of the topological charge density at different cooling stages.\label{cool}}\end{figure}

The above code generates JPEG images showing different stages of cooling of the data. You can see some of the images in fig.~\ref{cool}

The relevant code in ``fermiqcd.cpp'' is here:

\begin{lstlisting}[language=C++]
void cool_vtk(gauge_field& U, mdp_args& arguments, string filename) {
  if (arguments.get("-cool","alg","ape")=="ape")
    for(int k=0; k<arguments.get("-cool_vtk","n",20); k++) {
      ApeSmearing::smear(U,
        arguments.get("-cool_vtk","alpha",0.7),
        arguments.get("-cool_vtk","steps",1),
        arguments.get("-cool_vtk","cooling",10));
      topological_charge_vtk(U,filename+".cool"+tostring(k,2)+".vtk",0);
    }
  else
    mdp.error_message("cooling algorithm not supported");
}
\end{lstlisting}

The smearing algorithm is in the ``topological\_charge\_vtk'' file:

\begin{lstlisting}[language=C++]
class ApeSmearing {
  public: static void smear(gauge_field &U,
                            mdp_real alpha=0.7,
                            int iterations=20,
                            int cooling_steps=10) {
    gauge_field V(U.lattice(),U.nc);
    mdp_site x(U.lattice());
    for(int iter=0; iter<iterations; iter++) {
      cout << "smearing step " << iter << "/" << iterations << endl;
      V=U;
      for(int mu=0; mu<4; mu++) {
        forallsites(x) {
          U(x,mu)=(1.0-alpha)*V(x,mu);
          for(int nu=0; nu<U.ndim; nu++)
            if(nu!=mu)
              U(x,mu)+=(1.0-alpha)/6*
                (V(x,nu)*V(x+nu,mu)*hermitian(V(x+mu,nu))+
                 hermitian(V(x-nu,nu))*V(x-nu,mu)*V((x-nu)+mu,nu));
          U(x,mu)=project_SU(U(x,mu),cooling_steps);
        }
      }
      U.update();
    }
  }
};
\end{lstlisting}

\goodbreak\subsection{Polyakov lines}

A Polyakov line is the trace of the product of the gauge links along the time direction, therefore it is a 3D complex field. Here we are interested in the real part only (the image part is qualitatively the same). 

\begin{figure}[ht!]
\begin{center}
\includegraphics[width=8cm]{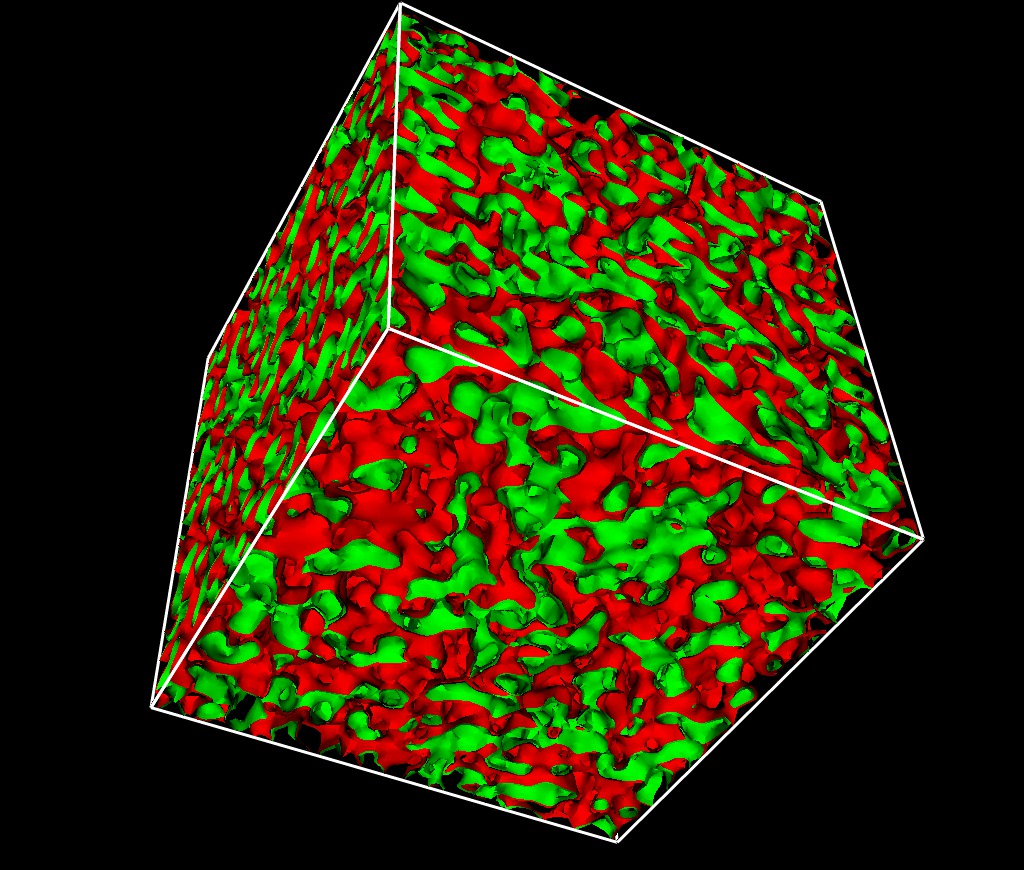}
\end{center}
\caption{Visualizaiton of contour plots for the Polyakov lines. Different colors represent positive and negative values of the real part of the Polyakov lines.\label{poly}}\end{figure}

We can visualize Polyakov lines using the {\ft -polyakov\_vtk} option:

\begin{lstlisting}
python qcdutils_run.py \
        -gauge:load=demo/demo.nersc.mdp \
        -polyakov_vtk
\end{lstlisting}

which we can convert to images with:

\begin{lstlisting}
python qcdutils_vis.py \
     -r 'scalars_t0' -i 9 -p default 'demo/*.polyakov.vtk'
\end{lstlisting}

The output is show in fig.~\ref{poly}.

Here is the relevant code in ``fermqcd.cpp'':

\begin{lstlisting}[language=C++]
void polyakov_vtk(gauge_field& U, string filename) {
  int L[3];
  L[0]=U.lattice().size(1);
  L[1]=U.lattice().size(2);
  L[2]=U.lattice().size(3);
  mdp_lattice space(3,L,
            default_partitioning<1>,
            torus_topology,
            0, 1,false);
  mdp_matrix_field V(space,U.nc,U.nc);
  mdp_field<mdp_real> Q(space,2);
  mdp_site x(U.lattice());
  mdp_site y(space);

  int k,mu=0,nu=1;
  mdp_complex s=0;

  forallsites(y) V(y)=1;
  for(int t=0; t<L[0]; t++) {
    forallsites(y) {
      x.set(t,y(0),y(1),y(2));
      V(y)=V(y)*U(x,0);
    }
  }

  forallsites(y) {
    mdp_complex z=trace(V(y));
    Q(y,0)=real(z);
    Q(y,1)=imag(z);
  }
  Q.save_vtk(filename,-1,0,0,false);
}
[...]
if (arguments.have("-polyakov_vtk")) {
   polyakov_vtk(U,newfilename+".polyakov.vtk");
}
\end{lstlisting}

This code is a little different than the previous one. It creates a 3D lattice called {\ft space} which is a time projection of the 4D space. While $x$ lives on the full lattice, $y$ leaves only on the 3D space. $q$ is a scalar field with two components (real and imaginary part of the Polyakov lines) which lives in 3D space.

\goodbreak\subsection{Quark propagator}

Given any gauge configuration we can visualize quark propagators in two ways. We can use the normal inverter and save the proprgator at the end of the inversion for each source/sink spin/color component:

\begin{lstlisting}
python qcdutils_run.py \
   -gauge:start=load:load=demo/demo.nersc.mdp \
   -quark:kappa=0.135:source_point=center:alg=bicgstab:vtk=true | run.log
\end{lstlisting}

(here using the {\ft bicgstab}, the Stabilized Bi-Conjugate Gradient).
Alternatively we can use a modified inverter
which saves the components but also VTK visualization for the field components and the residue at each step of the inversion.
\begin{lstlisting}
python qcdutils_run.py \
   -gauge:start=load:load=demo/demo.nersc.mdp \
   -quark:kappa=0.135:source_point=center:alg=bicgstab_vtk > run.log
\end{lstlisting}

\begin{figure}[ht!]
\begin{center}
\begin{tabular}{cc}
\includegraphics[width=5cm]{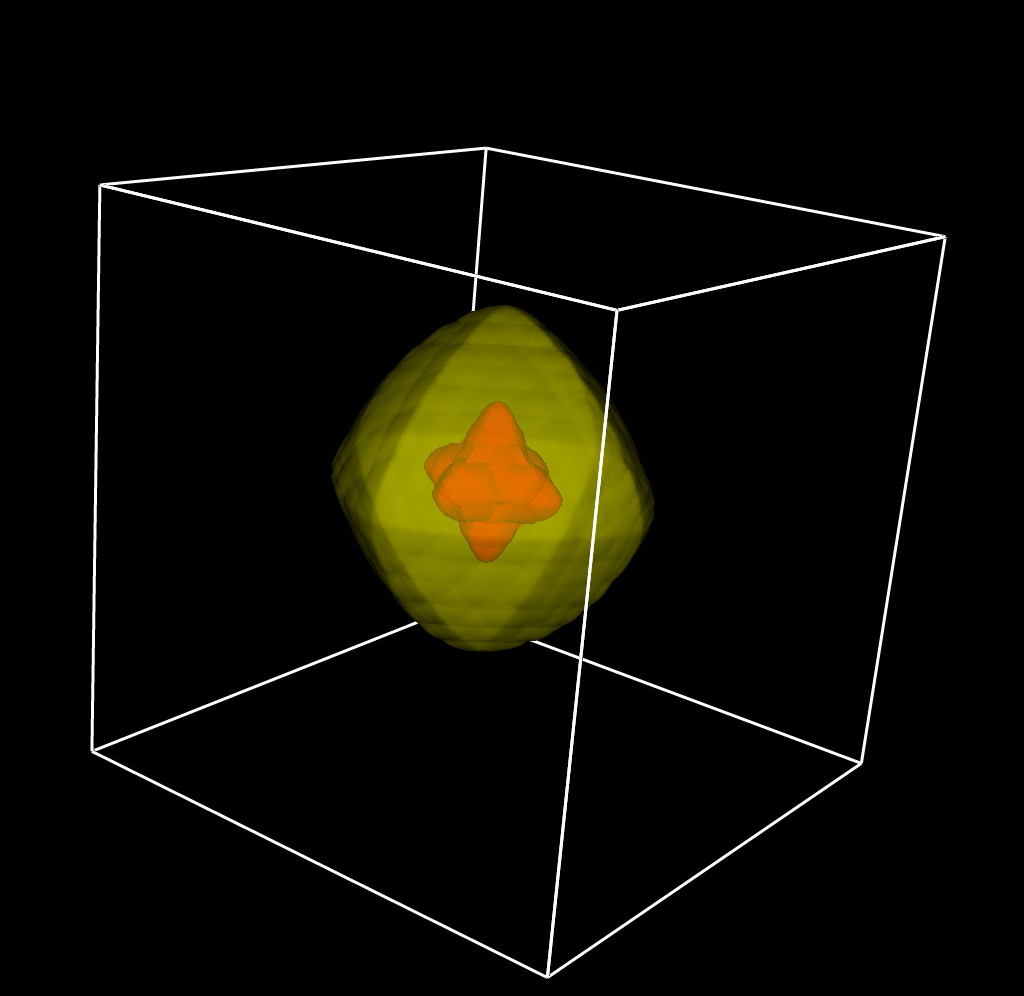} &
\includegraphics[width=5cm]{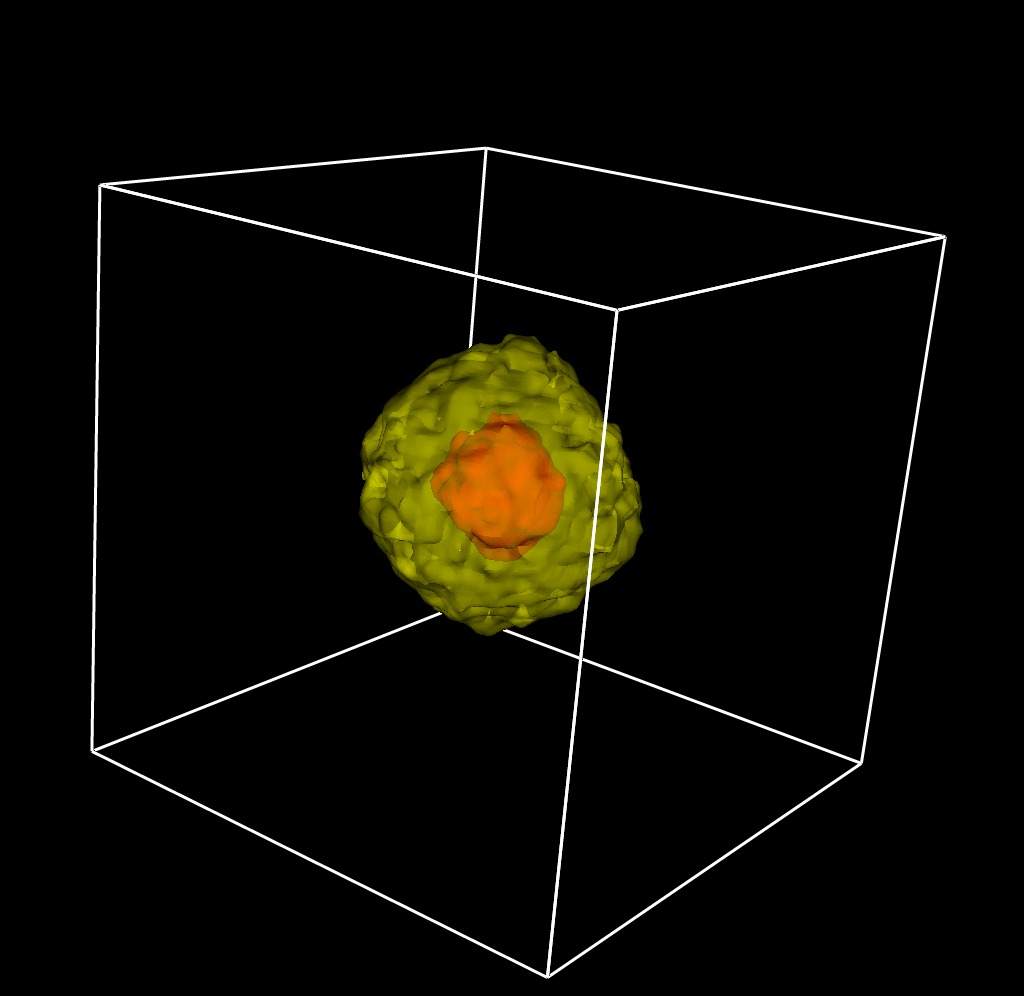} \\
\includegraphics[width=5cm]{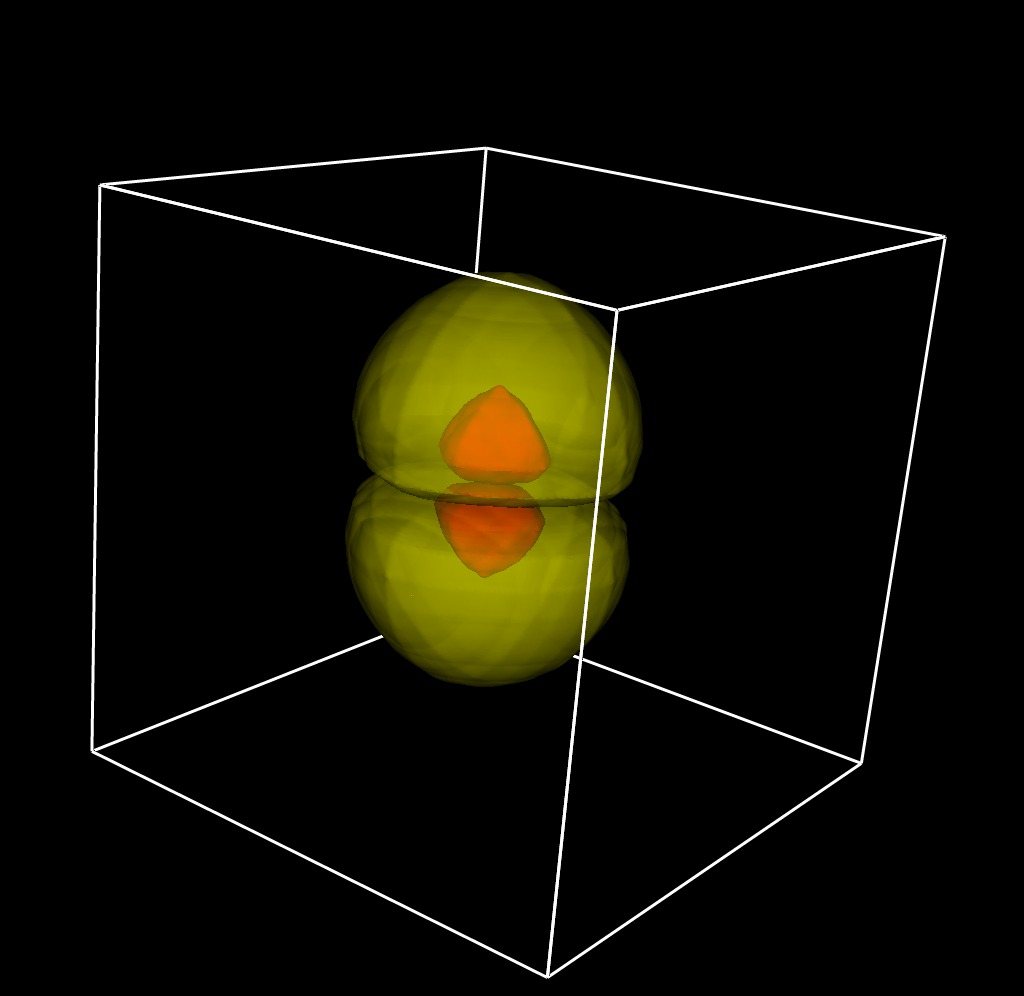} &
\includegraphics[width=5cm]{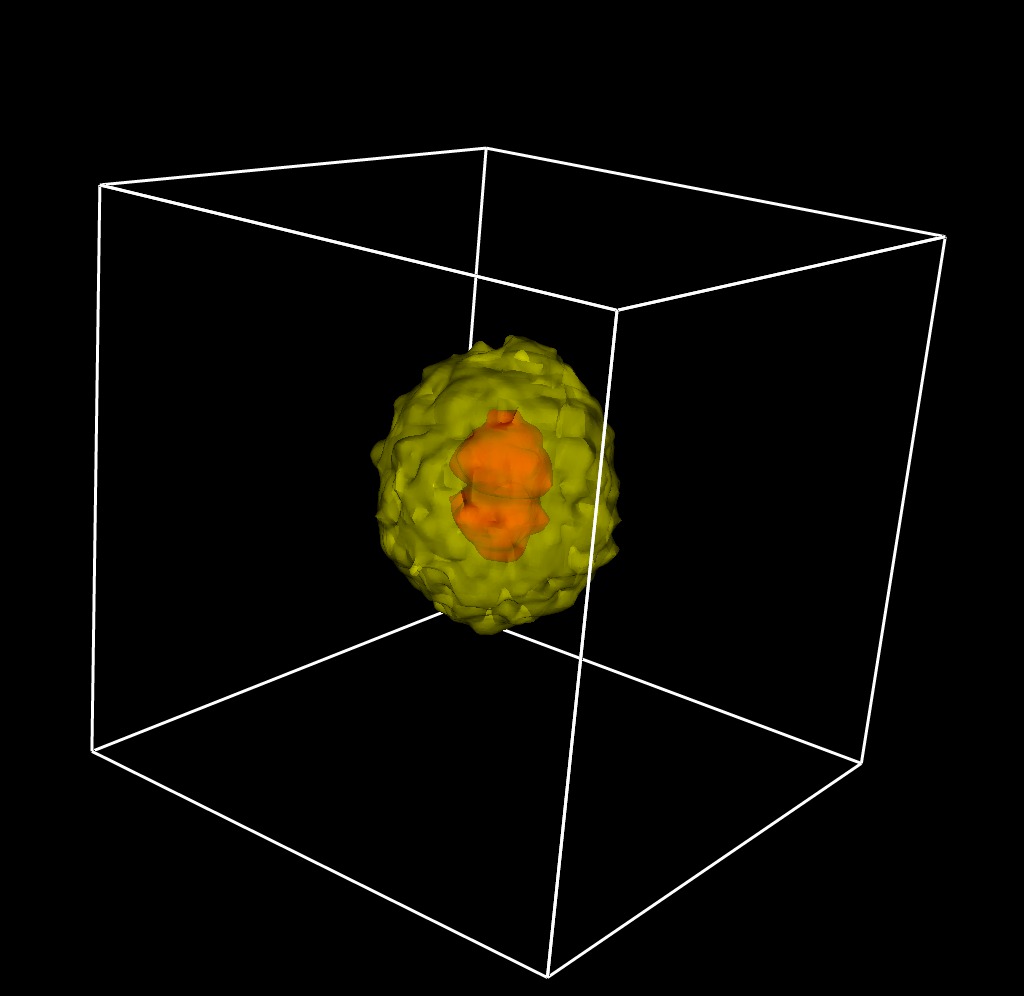} \\
\includegraphics[width=5cm]{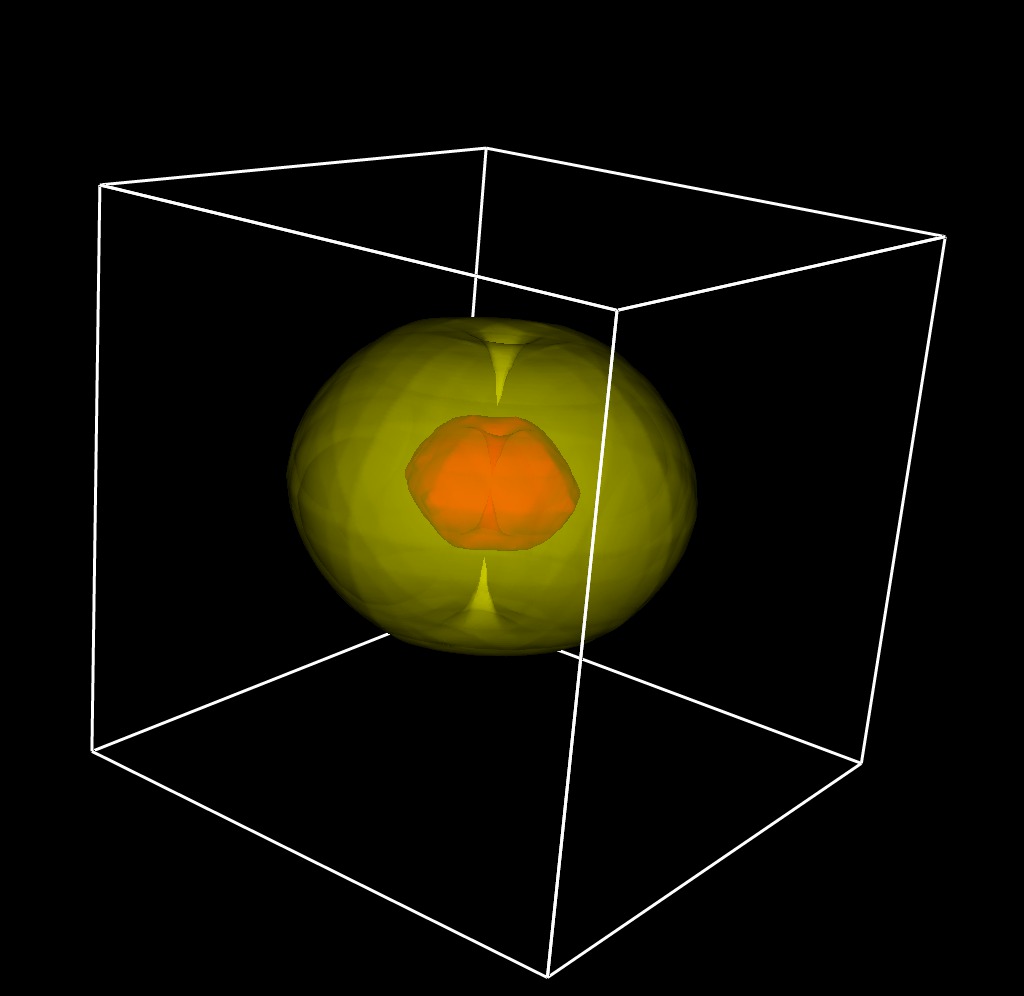} &
\includegraphics[width=5cm]{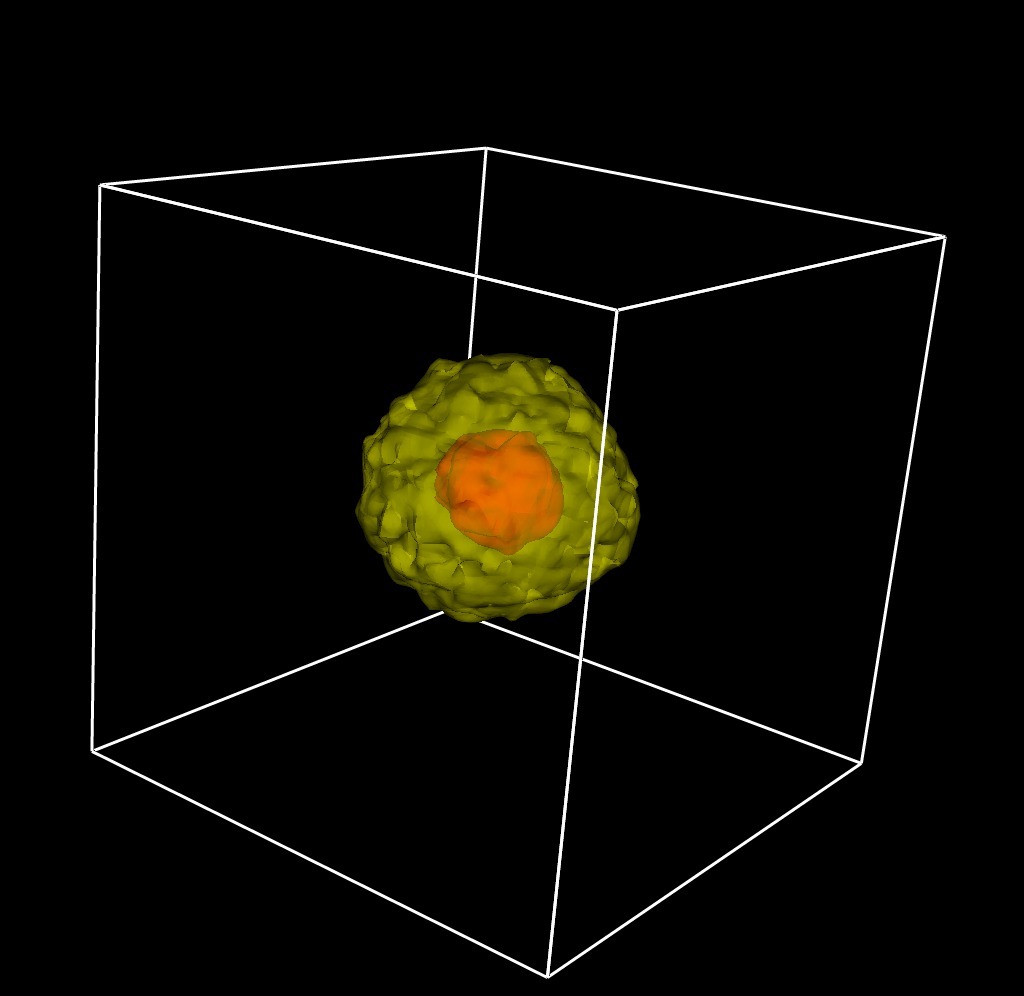} \\
\end{tabular}
\end{center}
\caption{Different components of a quark propagator on a cold gauge configuration (left) and on a thermalized gauge configuration (right).
From top to bottom, they show the magnitude of $S^{\alpha\beta ij}(t,\mathbf{x}) = S^{0000}(0,\mathbf{x)})$, $S^{0200}(0,\mathbf{x})$, and $S^{0300}(0,\mathbf{x})$.
\label{bicgstabvtk}}\end{figure}

Fig.~\ref{bicgstabvtk} shows different components of a quark propagator on a hot and a cold configuration.

From now on we assume the propagator has been computed and we reuse it.

\goodbreak\subsection{Pion propagator}

In a previous section, computed the zero momentum Fourier transform of the pion propagator. Now we want to visualize it for every point in space:

\begin{equation}
Q(t,\mathbf{x}) =
\left<\pi_{ab}(0,\mathbf{0}) | \pi_{ab}(+t,\mathbf{x})\right> = \sum_{i,\alpha}\left|S^{ii,\alpha\alpha}(t,\mathbf{x})\right|^2
\end{equation}

This can be done using the {\ft vtk=true} attribute of the {\ft -pion} option:

\begin{lstlisting}
python qcdutils_run.py \
       -gauge:start=load:load=demo/demo.nersc.mdp \
       -quark:kappa=0.135:source_point=center:load=true \
       -pion:vtk=true  > run.log
\end{lstlisting}

Notice the {\ft -quark...:load=true} which reloads the previous propagator.
We can now convert the pion VTK visualization into images using {\ft qcdutils\_vis}:

\begin{lstlisting}
python qcdutils_vtk.py -u 0.01 -l 0.00001 'demo/*.pion.vtk'
python qcdutils_vis.py -s '*' -i 9 -p default 'demo/*.pion.vtk'
\end{lstlisting}

In this case the {\ft -i 9} option is used to interpolate between time-slices in case the images are to be assembled into a movie.

\begin{figure}[ht!]
\begin{center}
\begin{tabular}{cc}
\includegraphics[width=8cm]{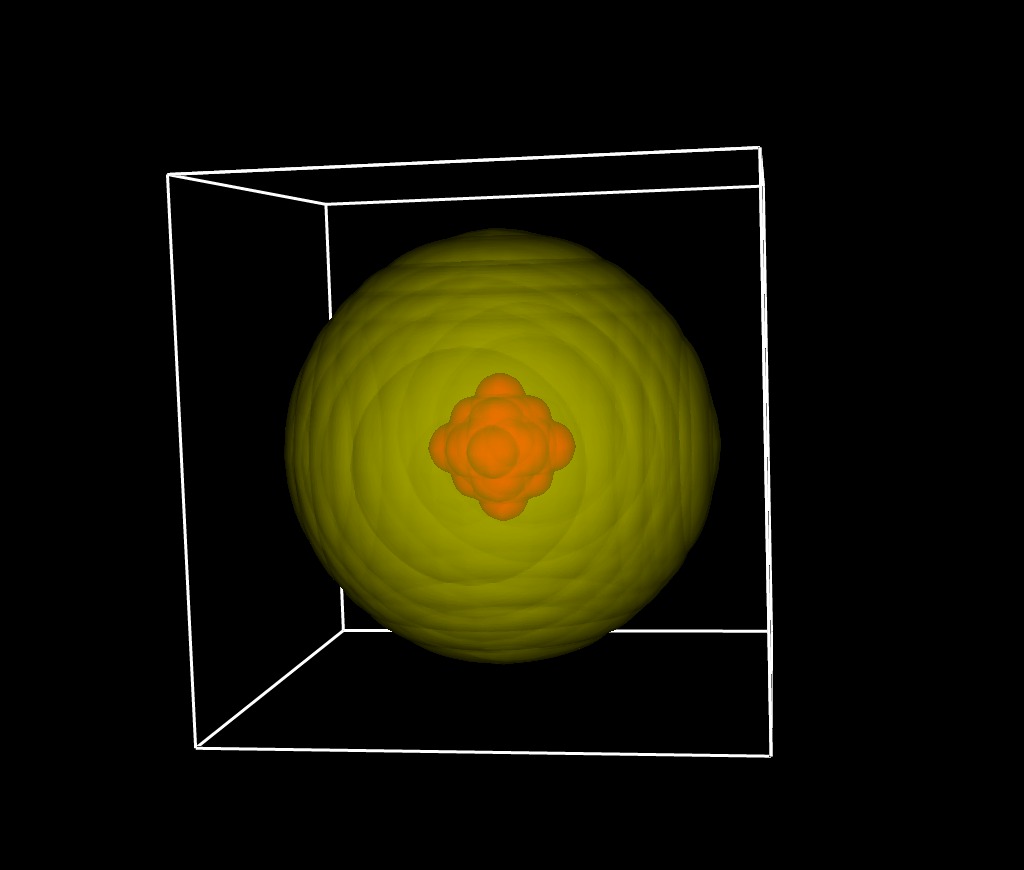} &
\includegraphics[width=8cm]{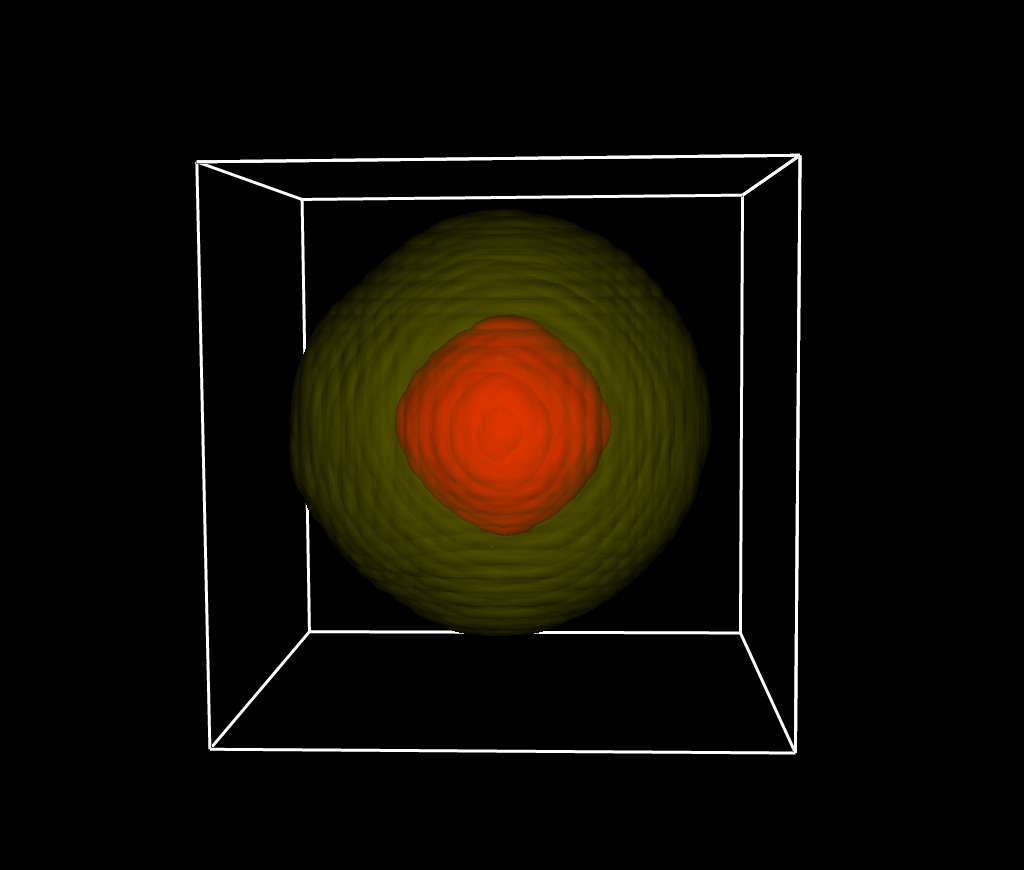}
\end{tabular}
\end{center}
\caption{Contour plot for a pion propagator.\label{pion}}\end{figure}

Examples of images are shown in fig.\ref{pion}

\goodbreak\subsection{Meson propagators}

A meson propagator is defined similarly to a pion propagator but it has a different gamma structure:

\begin{eqnarray}
Q(t,\mathbf{x}) &=& 
\left<\Gamma^{source}_{ab}(0,\mathbf{0}) | \Gamma^{sink}_{ab}(+t,\mathbf{x})\right> \\
&=& \sum_{...}
S^{ij,\beta\delta}(t,\mathbf{x})(\Gamma^{sink}\gamma^5)_{\delta\rho}
S^{\ast ij,\alpha\rho}(t,\mathbf{x})(\gamma^5\Gamma^{source})_{\alpha\beta}
\end{eqnarray}

We can visualized a Meson propagator using the following code:

\begin{lstlisting}
python qcdutils_run.py \
       -gauge:start=load:load=demo/demo.nersc.mdp \
       -quark:kappa=0.135:source_point=center:load=true \
       -meson:source_gamma=1:sink_gamma=1:vtk=true > run.log
\end{lstlisting}

and then process the VTK file as in the pion example.
In this case $\Gamma^{source} = \Gamma^{sink} = \Gamma^1$ indicates a verctor meson polarized along the $X$ axis.

\goodbreak\subsection{Current insertions}

We can also visualize the mass density and the charge density of a heavy-light meson by inserting an operator ($\bar q q$ and $\bar q \gamma^0 q$ respectively) in bewteen meson bra-kets.

\begin{eqnarray}
Q(t,\mathbf{x}) &=&
\left<\Gamma^{source}_{ha}(-t,\mathbf{x})\right|
\bar q_{a} \Gamma^{current} q_{b}
\left|\Gamma^{sink}_{bh}(+t,\mathbf{x})\right> \nonumber \\
&=& \mathrm{tr}(\Gamma^{source} \gamma5 S^\dagger(-t,\mathbf{x}) \gamma^5 \Gamma^{current} S(t,\mathbf{x}) \Gamma^{sink} H^\dagger(-t,t,\mathbf{x}))
\end{eqnarray}

\begin{figure}[ht!]
\begin{center}
\includegraphics[width=8cm]{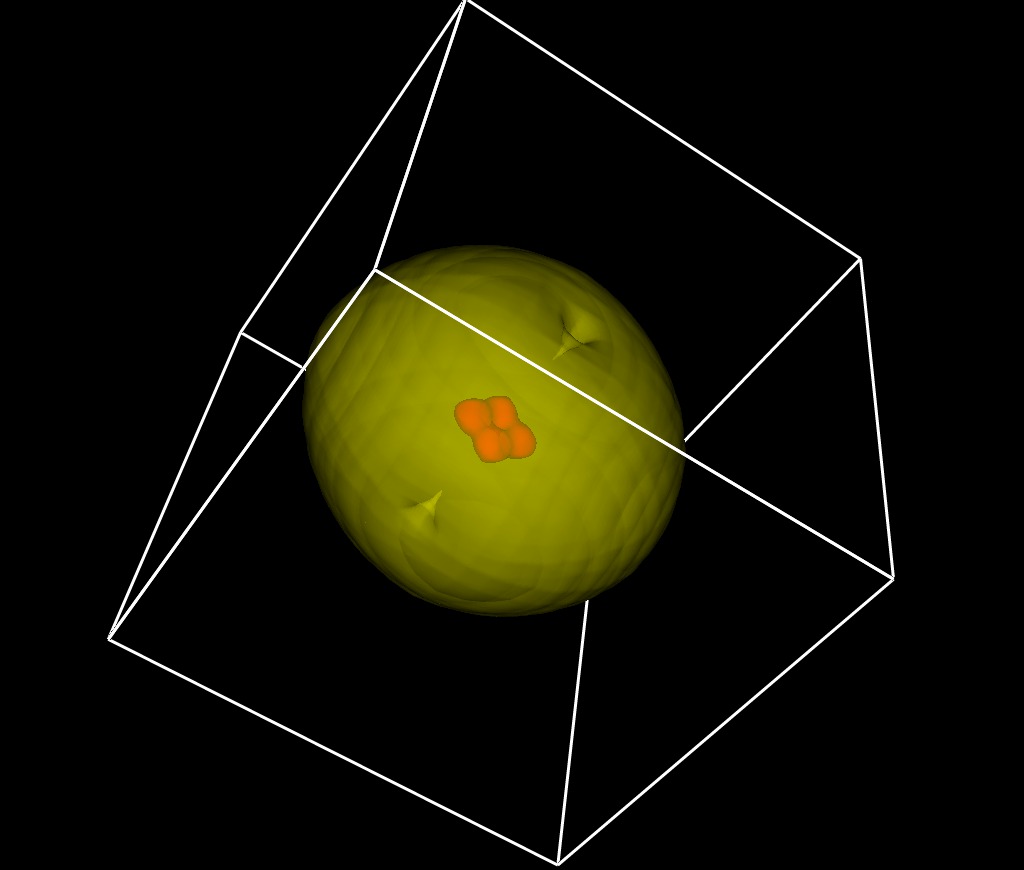}
\end{center}
\caption{Contour plot for a three points correlation function.\label{bstarbpi}}\end{figure}

Here we measure the mass distribution for a static vector meson:

\begin{lstlisting}
python qcdutils_run.py \
   -gauge:start=load:load=demo/demo.nersc.mdp \
   -quark:kappa=0.135:source_point=center:load=true  \
   -current_static:source_gamma=1:sink_gamma=1:current_gamma=I:vtk=true \
   > run.log
\end{lstlisting}

Using the same diagram we can compute the spatial distribution of $B^\ast \rightarrow B\pi$ by inserting the axial current ($\bar q \gamma^5 q$) in between a static $B$ ($\bar q \gamma^5 h$) and and a static $B^\ast$ ($\bar q \gamma^1 h$):

\begin{lstlisting}
python qcdutils_run.py \
   -gauge:start=load:load=demo/demo.nersc.mdp \
   -quark:kappa=0.135:source_point=center:load=true  \
   -current_static:source_gamma=1:sink_gamma=5:current_gamma=5:vtk=true \
   > run.log
\end{lstlisting}

A sample image is shown in fig.~\ref{bstarbpi}.

\goodbreak\subsection{Localized instantons}

FermiQCD allows the creation of custom gauge configurations with localized topological charge. Here we consider the case of a pion propagator on a single gauge configuration in presence of one t'Hooft instanton (localized lump of topological charge). Here is the code:

\begin{lstlisting}
python qcdutils_run.py  \
       -gauge:start=instantons:nt=20:nx=20:t0=0:x0=4.5:y0=10:z0=10 \
       -topcharge_vtk \
       -quark:kappa=0.120:source_point=center \
       -pion:vtk=true > run.log
\end{lstlisting}

This code places the center of the instanton at point $(t_0,x_0,y_0,z_0)=(0,4.5,10,10)$ and then computes a pion propagator with source on time slice 0 but spatial coordinates $(x,y,z)=(10,10,10)$ (center).

\begin{figure}[ht!]
\begin{center}
\begin{tabular}{cc}
\includegraphics[width=8cm]{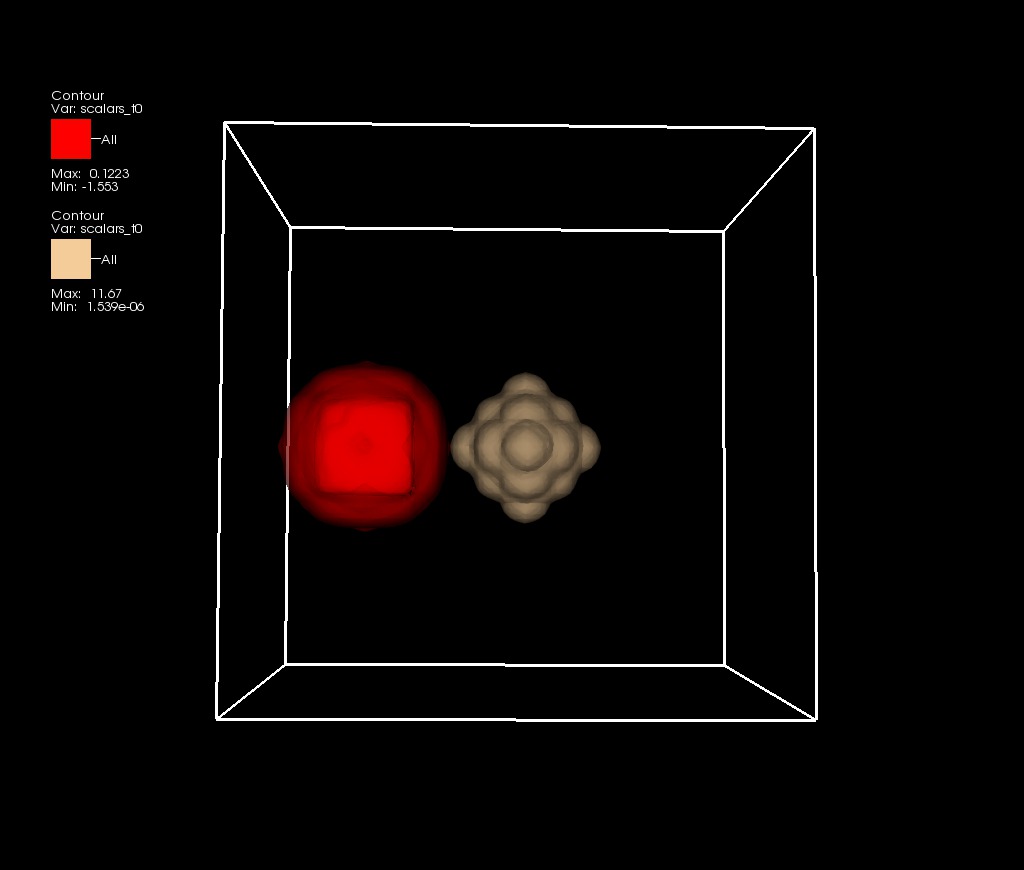} &
\includegraphics[width=8cm]{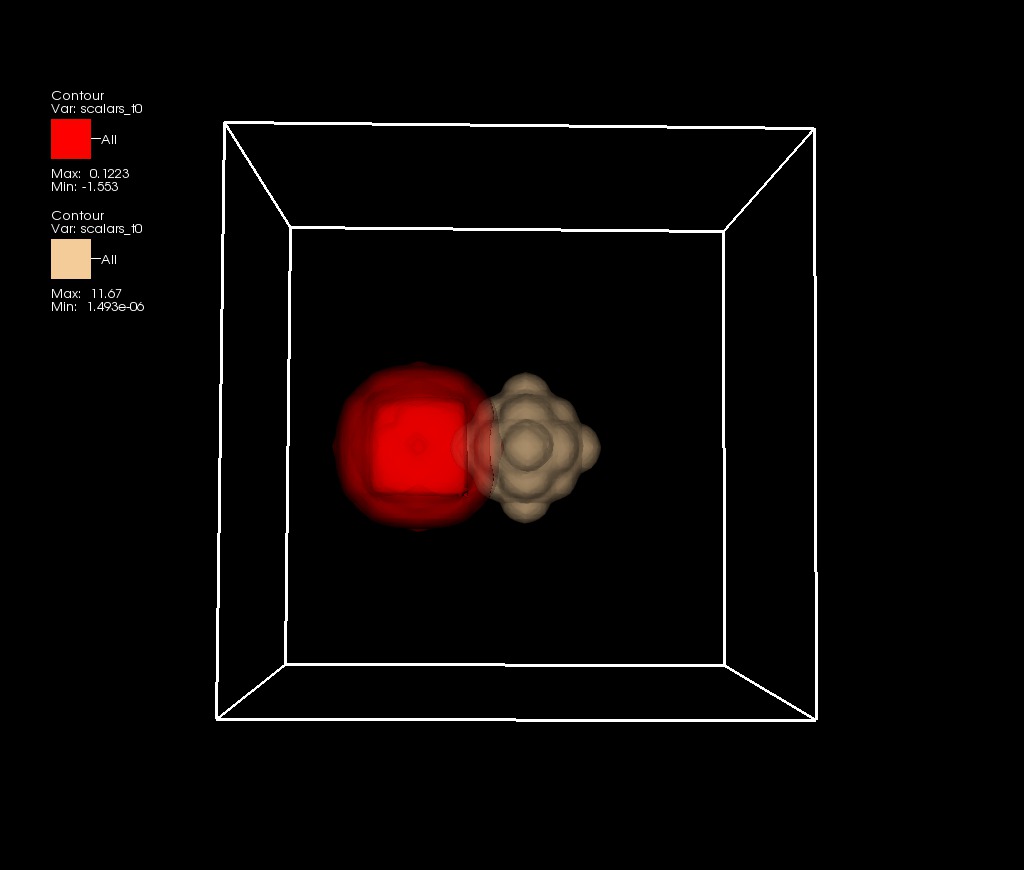} \\
\includegraphics[width=8cm]{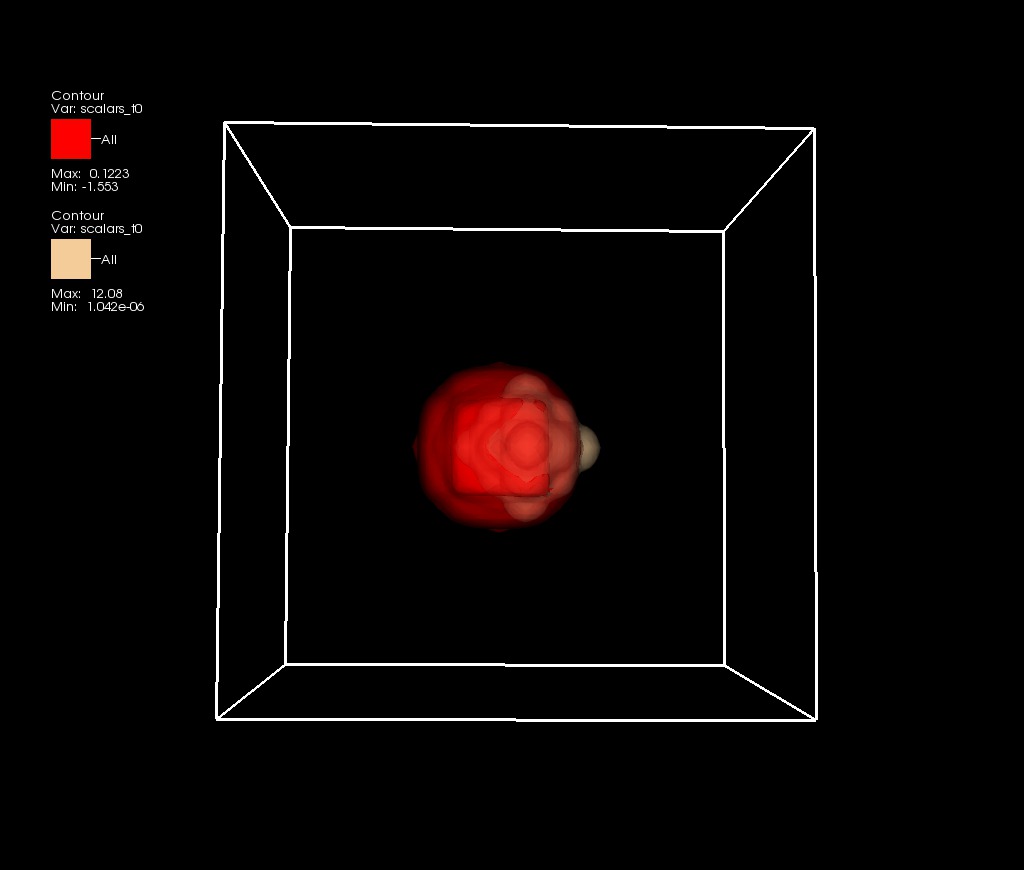} &
\includegraphics[width=8cm]{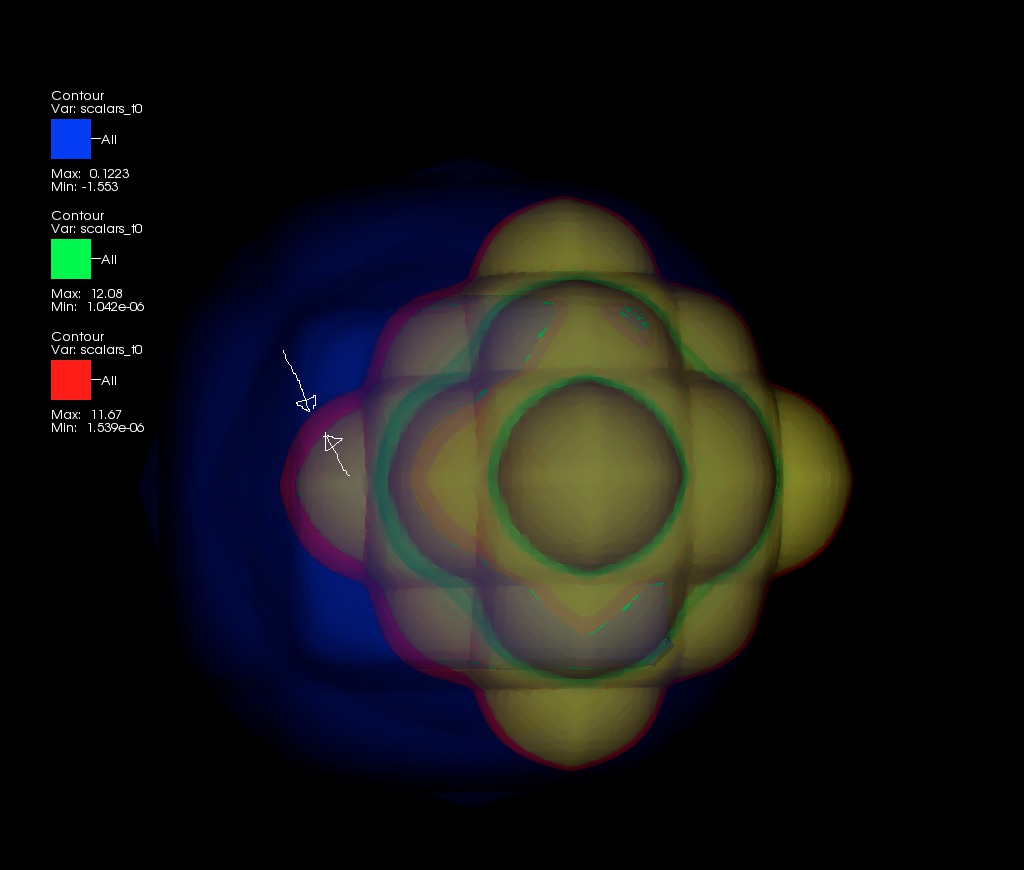}
\end{tabular}
\end{center}
\caption{Pion propagator on a semi-cold configuration in presence of one localized instanton. The bottom right image shows the instanton (in blue) and an overlay of the pion propagator with and without the instanton. The difference shows that propagator shrinks as the instanton nears.\label{insts}}\end{figure}

\begin{figure}[ht!]
\begin{center}
\begin{tabular}{cc}
\includegraphics[width=5cm]{cold.quark.00.00.jpg} &
\includegraphics[width=5cm]{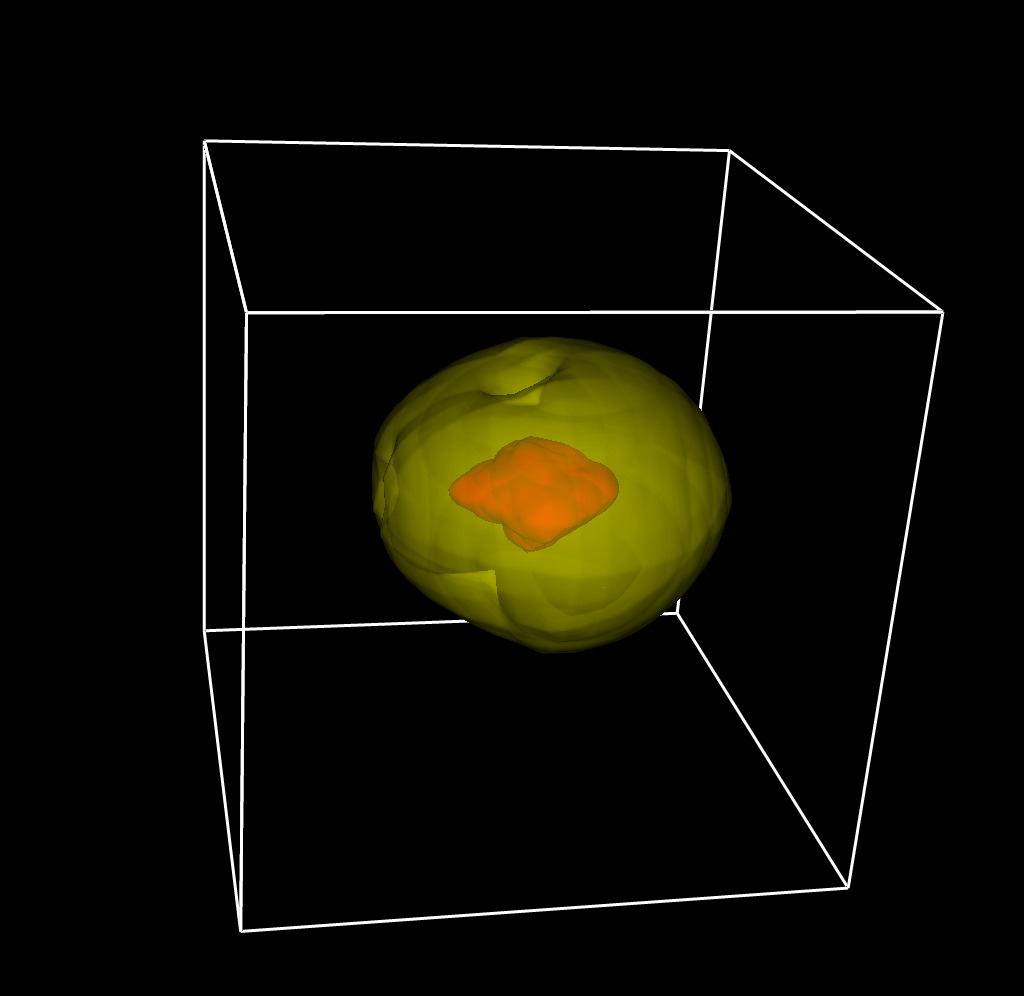} \\
\includegraphics[width=5cm]{cold.quark.00.20.jpg} &
\includegraphics[width=5cm]{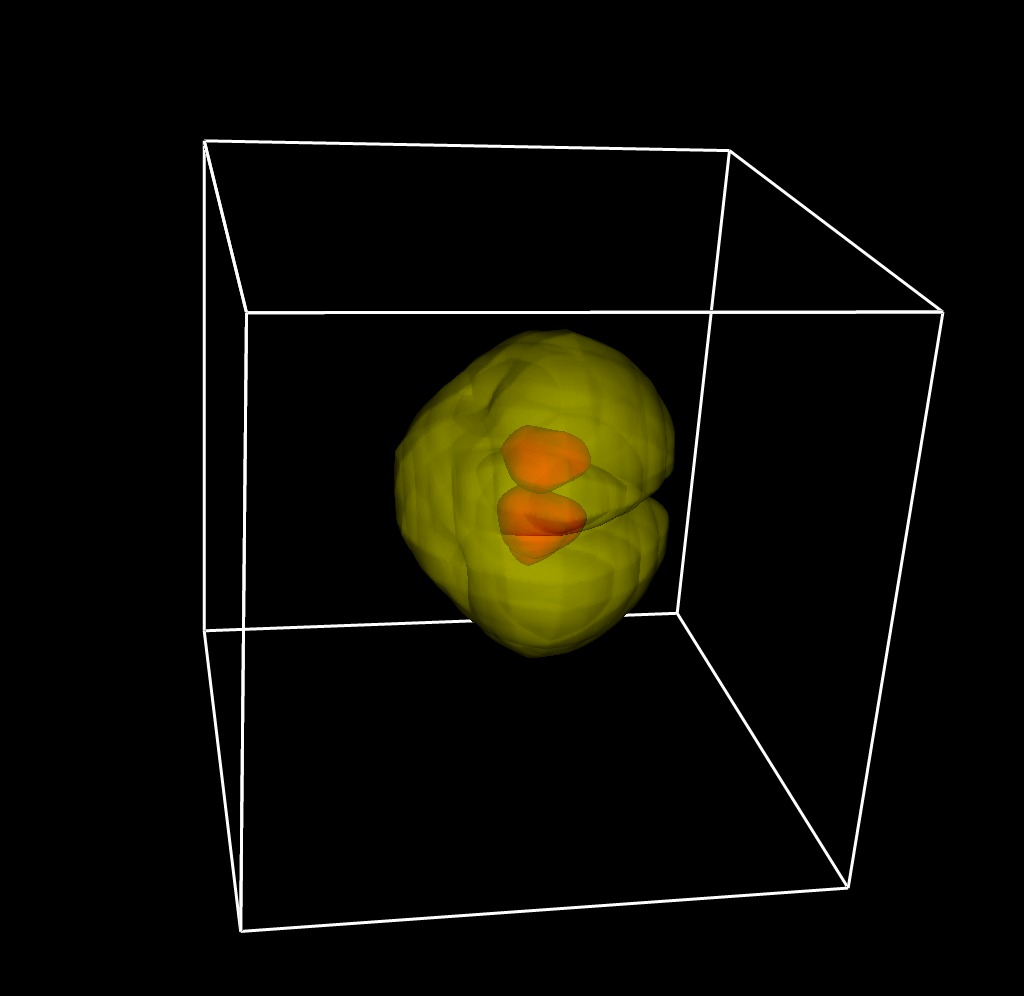} \\
\includegraphics[width=5cm]{cold.quark.00.30.jpg} &
\includegraphics[width=5cm]{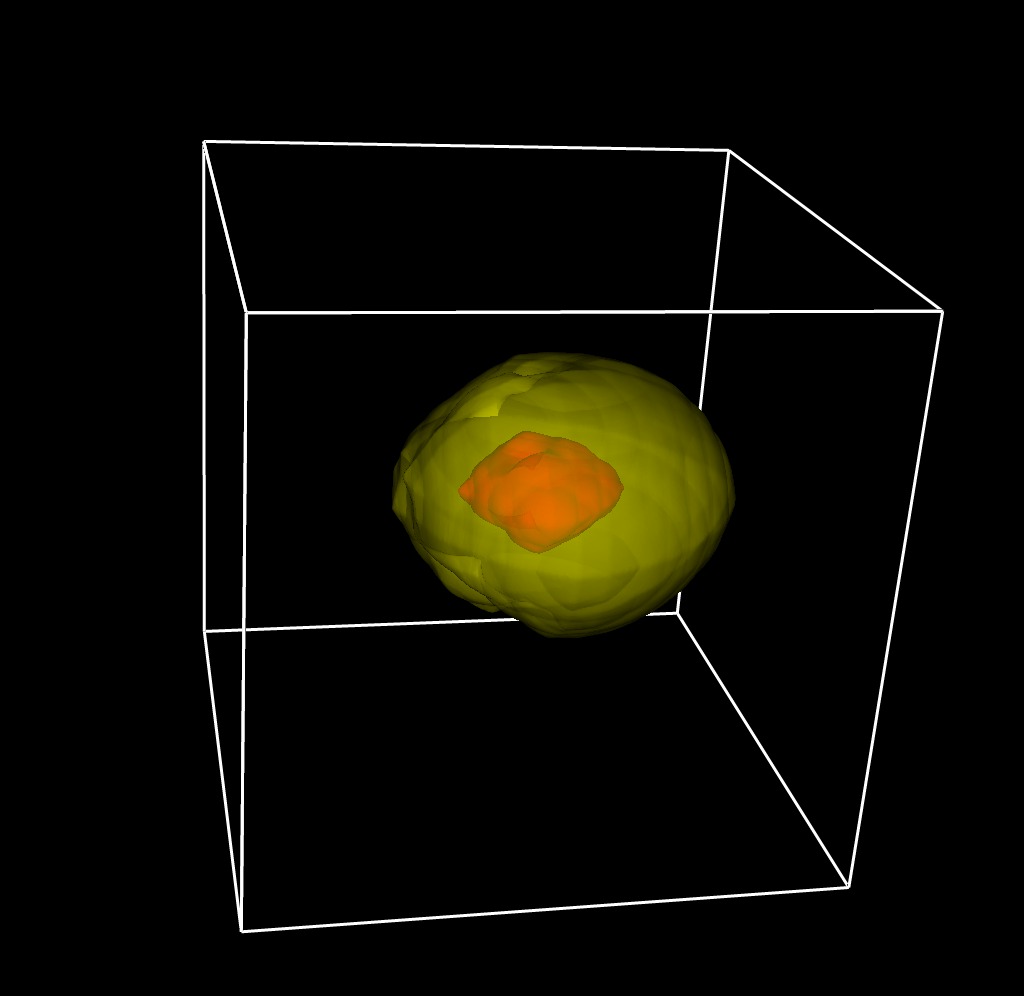} \\
\end{tabular}
\end{center}
\caption{Comparion of quark propagator components on a cold configuration (left) and in presense of a localized instanton (right). The instanton is located as in fig.~\ref{insts} (bottom-right).\label{insts2}}\end{figure}

Fig.~\ref{insts} show the pion propagator in presence of the instanton as the instanton nears the center of the propagator. Each image has been generated using the above command by placing the instanton at different locations. The last image shows a superposition of the pion propagator with and without the instanton in order to emphasize the difference. The difference is small but visible. The propagator retracts as the instanton nears. One may say that the quark interacts with the instanton and acquires mass thus making the propagator decrease faster when going through the instanton. Fig.~\ref{insts2} shows the effect of the instanton on individual components of the quark propagator.

\goodbreak\section{Analysis with {\ft qcdutils\_boot.py}, {\ft qcdutils\_plot.py}, {\ft qcdutils\_fit.py}}

The console output of the {\ft qcdutils\_run} program consists of human readble text with comments and results of measurments performed on each gauge configuration. Here are some examples of measurements logged in the output:

\begin{lstlisting}
...
plaquette = 0.654346
C2[0] = 14.5234
C3[0][0] =  1.214321
C3[0][0] =  1.123425
...
\end{lstlisting}

{\ft qcdutils\_boot.py} is a tool that can extract the values for these measurements, aggregate them and analyze them in various ways. For example it computes the average and bootstrap errors~\cite{bootstrap} of any function of the measurements. {\ft qcdutils\_plot.py} is a tool to visualize the results of the analysis. It uses the Python {\ft matplotlib} package, one of the most powerful and versatile plotting libraries available, although {\ft qcdutils\_plot.py} uses only a small subset of the available functionality.

Now, let us consider a typical Lattice QCD computation where one or more observables are measured on each Markov Chain Monte Carlo step (on each gauge configuration). We label the observables with $Y_j$ (gauge configuration, 2-point correlation function for a given value of $t$, etc.)

We also refer to each measurements with $y_{ij}$  where $i$ labels the gauge configuration and $j$ labels the observable (same index as $Y_j$). $y_{i0}$  could be, for example, the plaquette on the $i$ th gauge configuration.

The expectation value of each one observable is computed by averaging its measurements over the MCMC steps:

\begin{equation}
\bar Y_j = \left<0\right|Y_j\left|0\right> = \frac 1N \sum_i y_{ij}
\end{equation}

Here $N$  is the number of the measurements. The statistical error on the average for this simple case can be estimated using the following formula:

\begin{equation}
\delta Y_j \simeq \sqrt{ \frac 1{N(N-1)} \sum_i (y_{ij}-\bar Y_j)^2 }
\end{equation}

Usually we are interested in the expectation value of non-trivial functions of the observables:

\begin{equation}
\bar f = \left<0\right|f(Y_1,Y_2,...T_M)\left|0\right> = f(\bar Y_1,\bar Y_2, ...,\bar Y_M)
\end{equation}

Often the $y_{ij}$ are not normal distributed and may depend on each other therefore standard error analysis does not apply.

The proper technique for estimating the error on $\bar f$ is the bootstrap algorithm. It consists of the following steps:

\begin{itemize}
\item We build $K$ vectors $b^k$ of size N. The elements of these vectors $b^k_i$  are chosen at random, uniformly between $\{1,2,...N\}$.

\item For every $k$  we compute:

\begin{equation}
\bar Y^k_j = \frac 1N \sum_i y_{{b^k_i}j}
\end{equation}

\item Again for each $k$ we compute:

\begin{equation}
\bar f^k = f(\bar Y^k_1,\bar Y^k_2, ...,\bar Y^k_M)
\end{equation}

\item We then sort the resulting values for $\bar f^k$.

\item We define the $\alpha$ percent confidence interval as $[\bar f^{k^\prime},\bar f^{k^{\prime\prime}}]$  where $k^\prime = \floor{(1-\alpha)K/2}$  and $k^{\prime\prime} = \floor{1+\alpha)K/2}$.
\end{itemize}

{\ft qcdutils\_boot} is a program that takes as input $f(Y_0,Y_1,..)$ in the form of a mathematical expression where the $Y_j$ are represented by their string pattern. It locates and extracts the corresponding $y_{ij}$ values from the log files and stores them in a file called ``qcdutils\_raw.csv''. It computes the autocorrelation for each of the $y_{ij}$ and stores them in ``qcdutils\_autocorrelation.csv''. It compute the moving averages for each of the $\bar Y_j$ and stores them in ``qcdutils\_trails.csv''. It generates the $K$ bootstrap samples $\bar f^k$ and saves them in ``qcdutils\_samples.csv''. Finally it compute the mean and the 68\% confidence level intervals  $[f^{k^{\prime}}, f^{k^{\prime\prime}}]$ and stores it ``qcdutils\_results.csv''.

Mind that these files are created in the current working directory and they are overwritten every time the {\ft qcdutils\_boot} is run. Move them somewhere else to preserve them.

Moreover, if the input expression for $f$ depends on wildcards, the program repeats the analysis for all matching expressions.

{\ft qcdutils\_boot} performs this analysis without need to write any code. It only needs the input $f$ in the syntax explained below and the list of log-files to analyze for data.

\goodbreak\subsection{A simple example}

Consider the output of one of the previous {\ft qcdutils\_run}:

\begin{lstlisting}
python qcdutils_run.py -gauge:load=*.mdp -plaquette > run.log
\end{lstlisting}

In this case the observable is $Y_0$=``plaquette''. We can analyze it with

\begin{lstlisting}
python qcdutils_boot.py 'run.log' '"plaquette"'
\end{lstlisting}

This produces the following output:

\begin{lstlisting}[keywords={}]
< plaquette > = min: 0.26, mean: 0.32, max: 0.38
average trails saved in qcdutils_trails.csv
bootstrap samples saved in qcdutils_samples.csv
results saved in qcdutils_results.csv
\end{lstlisting}

Notice that {\ft qcdutils\_run} takes three arguments:
\begin{itemize}
\item A file name or file pattern (for example ``run.log'')
\item An expression (for example ``plaquette'').
\item A condition (optional)
\end{itemize}
Each of the argument must be enclosed in single quotes.

The represents $f(Y_0,Y_1,...)$ and the $Y_j$ are the names of observables in double quotes.

In {\ft '"plaquette"'} the outer single quote delimits the expression and the term plaquette between double quotes, determines the string we want to parse from the in file. 

{\ft qcdutils} uses the observable name to find all the occurrences of

\begin{lstlisting}
plaquette = ...
\end{lstlisting}
\noindent or
\begin{lstlisting}
plaquette: ...
\end{lstlisting}
\noindent in the input files and maps them into $y_{i0}$  where $i$  labels the occurrence. In this case we have a single observable (plaquette) so we use $0$ to label it.

The program opens the file or the files matching the file patterns and parses them for the values of the ``plaquette'' thus filling an internal table of $y$ s. It gives the output as the result:

\begin{lstlisting}[keywords={}]
< plaquette > = min: 0.26, mean: 0.32, max: 0.38
\end{lstlisting}

Here ``mean" is the mean of the expression ``plaquette''. $\min$ and $\max$ are the 65\% confidence intervals computed using the bootstrap.

Here is example of the content of the ``qcdutils\_results.csv'' file for the average plaquette case:

\begin{lstlisting}
"plaquette","[min]","[mean]","[max]"
"plaquette",0.26,0.32,0.38
\end{lstlisting}

In general it contains one row for each matching expression.

You can plot the content of the files generated by {\ft qcdutils\_boot} using {\ft qcdutils\_plot}:

\begin{lstlisting}
python qcdutils_plot.py -r -a -t -b
\end{lstlisting}

Here {\ft -r} indicates that we want to plot the raw data, {\ft -a} indicates we want a plot of autocorrelations, {\ft -t} is for partial averages, and {\ft -b} means we want a plot of bootstrap samples. {\ft qcdutils\_plot} loops over all the files reads the data in them and for each $Y_j$ it makes one plot with raw data ($y_{ij}$), one with autocorrelations, one with partial averages. Then for each $f$ it  makes one plot with the bootstrap samples, and one plot with the final results found in ``qcdutils\_results.csv''.

The plots are in PNG files which have a name prefix equal to the name of the data source file, followed by a serialization of the expression for $Y_j$ or $f$, depending on the case.

\begin{figure}[ht!]
\begin{center}
\begin{tabular}{cc}
\includegraphics[width=8cm]{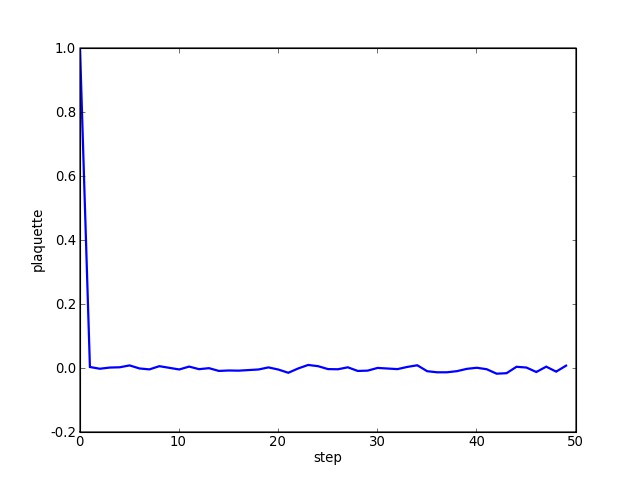} &
\includegraphics[width=8cm]{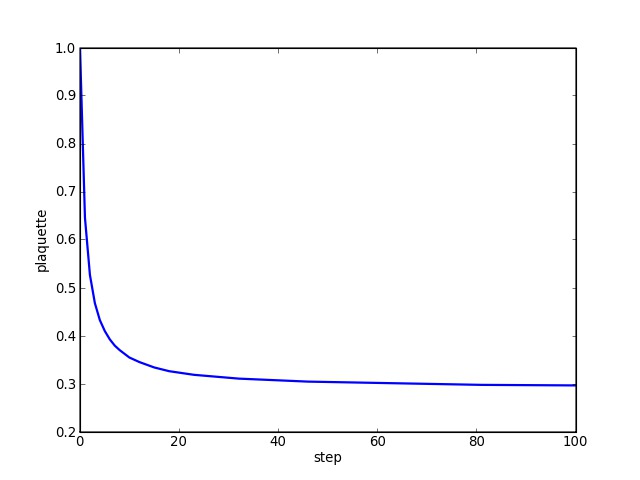}
\end{tabular}
\end{center}
\caption{Example plots of autocorrelation (left) and partial averages (right).\label{plotplaq}}
\end{figure}

For example in the case of the plaquette, the autocorrelations and the partial averages are in the files:
\begin{lstlisting}
qcdutils_autocorrelations_plaquette.png
qcdutils_trails_plaquette.png
\end{lstlisting}

and they are shown in fig.~\ref{plotplaq}.

Similarly, if you want to bootstrap $f(Y_0)=exp(Y_0/3)$ where $Y_0$ is the plaquette you would run:

\begin{lstlisting}
python qcdutils_boot.py run.log 'exp("plaquette"/3)'
\end{lstlisting}

It produces output like this:

\begin{lstlisting}[keywords={}]
< exp(plaquette/3) > = min: 1.092, mean: 1.114, max: 1.145
\end{lstlisting}

Notice that again the observable $Y_0$  is identified for convenience by {\ft "plaquette"}. The double quotes are necessary to avoid naming conflicts between patterns and functions.

Also notice that running {\ft qcdutils\_boot} twice does not guarantee generating the same exact results twice. That is because the bootstrap samples are random.

\goodbreak\subsection{2-point and 3-point correlation functions}

In order to explain more complex cases we could generate 2- and 3- points correlation functions using  something like:

\begin{lstlisting}
python qcdutils_run.py \
       -gauge:start=cold:beta=4:n=10:steps=5:therm=100 \
       -quark:kappa=0.11:c_sw=0.4:save=false -pion 
       -4quark:operator=5Ix5I > run.log
\end{lstlisting}

For testing purposes can also run:

\begin{lstlisting}
python qcdutils_boot.py -t
\end{lstlisting}

Where {\ft -t} stands for test. This creates and analyzes a file called ``test\_samples.log'' which contains {\it random} measurements for C2 and C3. Once this file is being created we can filter and study, for example, only the 2-point correlation function C2:

\begin{lstlisting}
python qcdutils_boot.py run.log '"C2[<t>]"'
\end{lstlisting}

Notice that {\ft <t>} means we wish to define a variable {\ft t} to be used internally for the analysis and whose values are to be determined by pattern-matching the data. The {\ft t} correspond to the $j$  of the previous abstract discussion.
{\ft "C2[<t>]"} matches {\ft C2[0]} with $t=0$, {\ft C2[1]} matches with $t=1$, etc.

The command above produces something like:

\begin{lstlisting}
reading file test_samples.log
C2[00] occurs 100 times
...
C2[15] occurs 100 times
raw data saved in qcdutils_raw_data.csv
autocorrelation for C2[02] and d=1 is -0.180453
...
autocorrelation for C2[06] and d=1 is -0.0436378
autocorrelations saved in qcdutils_autocorrelations.csv
< C2[00] > = min: 1.988, mean: 1.999, max: 2.008
< C2[01] > = min: 1.617, mean: 1.629, max: 1.64
< C2[02] > = min: 1.328, mean: 1.345, max: 1.359
< C2[03] > = min: 1.064, mean: 1.079, max: 1.094
< C2[04] > = min: 0.878, mean: 0.894, max: 0.908
< C2[05] > = min: 0.722, mean: 0.733, max: 0.744
< C2[06] > = min: 0.574, mean: 0.584, max: 0.597
< C2[07] > = min: 0.478, mean: 0.49, max: 0.5
< C2[08] > = min: 0.395, mean: 0.407, max: 0.419
< C2[09] > = min: 0.322, mean: 0.331, max: 0.339
< C2[10] > = min: 0.268, mean: 0.277, max: 0.286
< C2[11] > = min: 0.225, mean: 0.231, max: 0.237
< C2[12] > = min: 0.18, mean: 0.186, max: 0.192
< C2[13] > = min: 0.138, mean: 0.144, max: 0.151
< C2[14] > = min: 0.107, mean: 0.112, max: 0.118
< C2[15] > = min: 0.0883, mean: 0.0933, max: 0.0982
average trails saved in qcdutils_trails.csv
bootstrap samples saved in qcdutils_samples.csv
results saved in qcdutils_results.csv
\end{lstlisting}

which we can plot as usual with
\begin{lstlisting}
python qcdutils_plot.py -r -a -b -t
\end{lstlisting}

\begin{figure}[ht!]
\begin{tabular}{ccc}
\includegraphics[width=5cm]{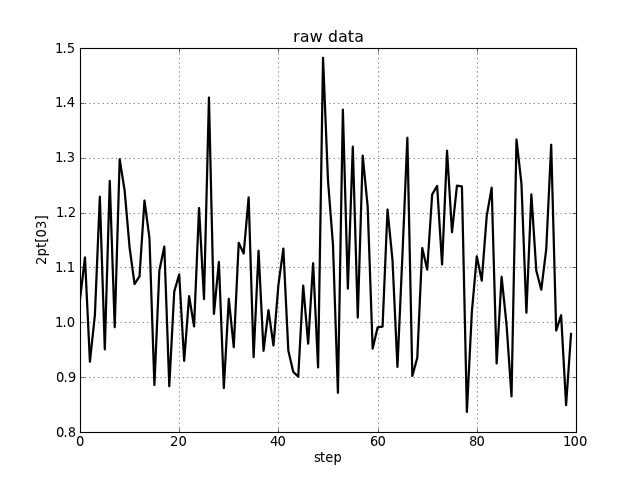} &
\includegraphics[width=5cm]{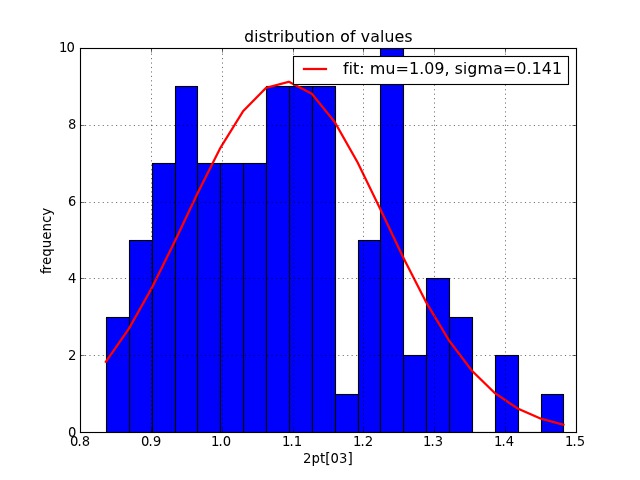} &
\includegraphics[width=5cm]{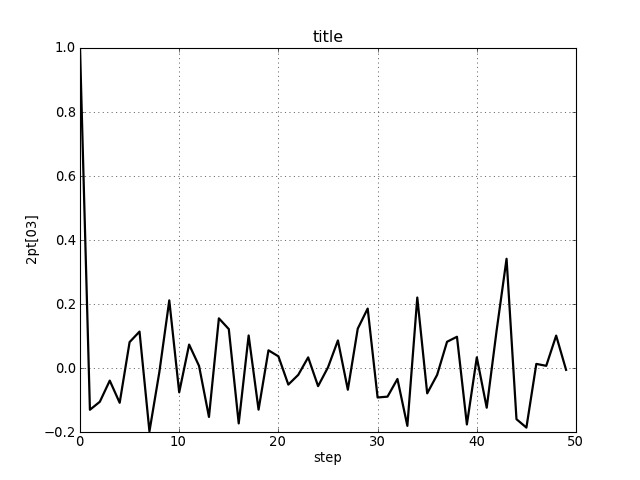} \\
\end{tabular}
\caption{Example plots for the raw data (left), the distribution of raw data (center) and autocorrelations (right) for C2.\label{plot1}}
\end{figure}

This produces about 60 plots. Some of them are shown in fig.\ref{plot1}.

We can as easily compute the log of C2 (for every {\ft t}):

\begin{lstlisting}
python qcdutils_boot.py test_samples.log 'log("C2[<t>]")'
\end{lstlisting}

or the log of the ratio between C2 at two consecutive time-slices:

\begin{lstlisting}
python qcdutils_boot.py run2.log \
           'log("C2[<t1>]"/"C2[<t2>]")' \
           't2==t1+1 if t1<8 else t2==t1-1'
\end{lstlisting}

In this case we used two implicit variables {\ft t1} and {\ft t2} but we used the third argument of {\ft qcdutils\_boot} to set a condition to link the two. This produces the following output:

\begin{lstlisting}
reading file test_samples.log
C2[00] occurs 200 times
...
C2[15] occurs 200 times
raw data saved in qcdutils_raw_data.csv
autocorrelation for C2[02] and d=1 is -0.176706
...
autocorrelation for C2[06] and d=1 is -0.0476376
autocorrelations saved in qcdutils_autocorrelations.csv
< log(C2[00]/C2[01]) > = min: 0.196, mean: 0.204, max: 0.212
< log(C2[01]/C2[02]) > = min: 0.18, mean: 0.191, max: 0.202
[...]
< log(C2[13]/C2[14]) > = min: 0.208, mean: 0.249, max: 0.291
< log(C2[14]/C2[15]) > = min: 0.147, mean: 0.189, max: 0.227
average trails saved in qcdutils_trails.csv
bootstrap samples saved in qcdutils_samples.csv
results saved in qcdutils_results.csv
\end{lstlisting}

\begin{figure}[ht!]
\begin{center}
\includegraphics[width=8cm]{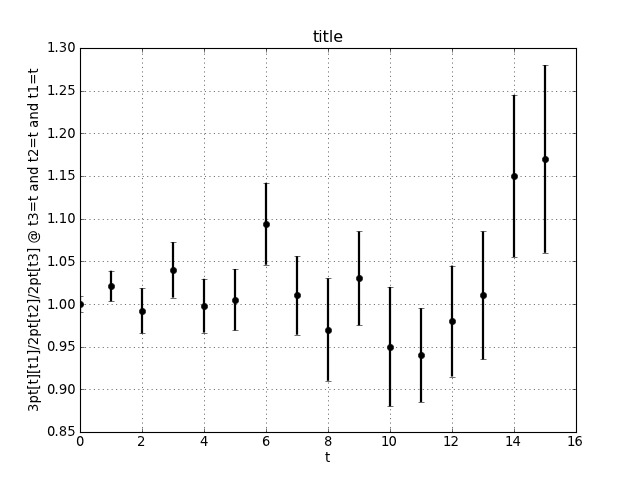}
\end{center}
\caption{Example plot showing results of the bootstrap analysis.\label{plotc3summary}}
\end{figure}

In the same fashion we can compute a matrix element as the ratio between a 3-point correlation function (C3) and a 2-point correlation function (C2):

\begin{lstlisting}
python qcdutils_boot.py test_samples.log \
       '"C3[<t>][<t1>]"/"C2[<t2>]"/"C2[<t3>]"' \
       't3==t and t2==t and t1==t' > run.log
python qcdutils_plot.py -a -t -b -r
\end{lstlisting}

Some of the generated plots can be seen in fig.\ref{plotc3}-\ref{plotc3summary}.

\begin{figure}[ht!]
\begin{tabular}{cc}
\includegraphics[width=8cm]{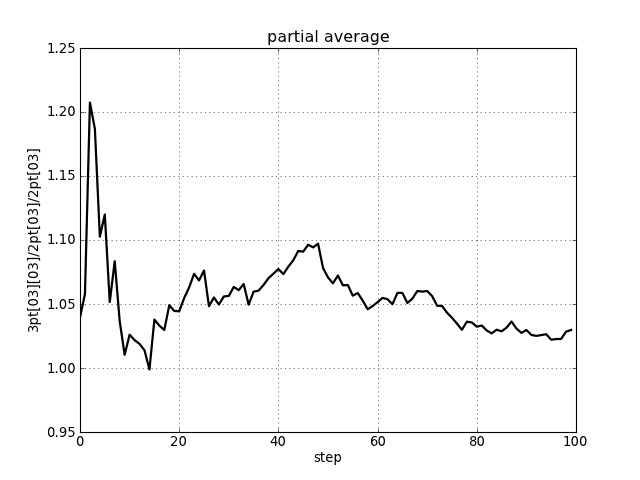} &
\includegraphics[width=8cm]{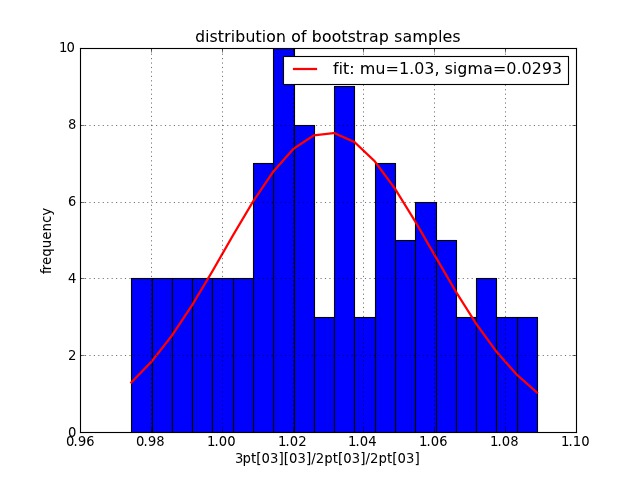}
\end{tabular}
\caption{Example plots of moving averages (left) and distribution of bootstrap samples (right) for the ratio C3/C2$^2$.\label{plotc3}}
\end{figure}

\goodbreak\subsection{Fitting data with {\ft qcdutils\_fit.py}}

{\ft qcdutils\_fit.py} is a fitting and extrapolation utility. It can read and understand the output of {\ft qcdutils\_boot.py}. Internally it uses a ``stabilized'' multidimensional Newton method to minimize $\chi^2$. It is stabilized by reverting to the steepest descent in case the Newton step fails to reduce the $\chi^2$. The length of the steepest descent step is adjusted dynamically to guarantee that each step of the algorithm reduces the $\chi^2$. The program accepts for input any function and any number of the parameters. It also accepts, optionally, Bayesian priors for those parameters and they can be used to further stabilize the fit~\cite{lepage}. A more sophisticated approach is described in ref.~\cite{fits}.

In Euclidean space C2 can be modeled by an exponential $a \exp(-bt)$ and $b$ is the mass of the lowest energy state which propagates between the source and the sink. Here is an example in which we fit C2 using a single exponential:

\begin{lstlisting}
python qcdutils_boot.py -t
python qcdutils_boot.py test_samples.log '"C2[<t>]"' > run.log
python qcdutils_fit.py 'a*exp(-b*t)@a=2,b=0.3'
\end{lstlisting}

The input data is read from the output of {\ft qcdutils\_boot}. 
The expression in quotes is the fitting formula. You can name the fitting parameters as you wish (in this case {\ft a} and {\ft b}) but the other parameters (in this case {\ft t}) must match the parameters defined in the argument of {\ft qcdutils\_boot} ({\ft <t>}). The {\ft @} symbol separates the fitting function (left) from the initial estimates for the fitting parameters (on the right, separated by commas). Every parameter to be determined by the fit must have an initial value.

\begin{figure}[ht!]
\begin{center}
\includegraphics[width=8cm]{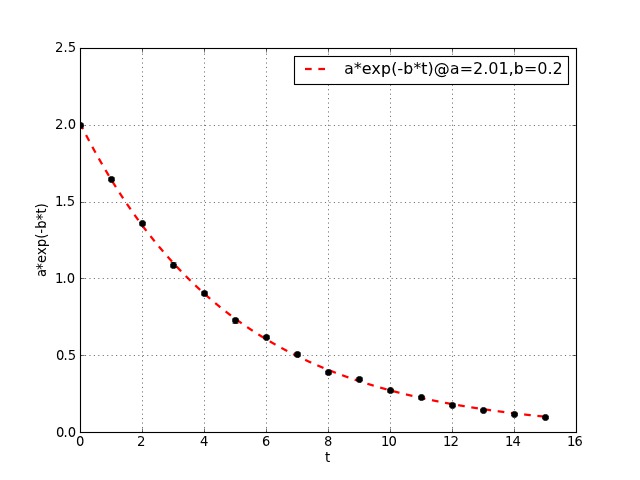}
\includegraphics[width=8cm]{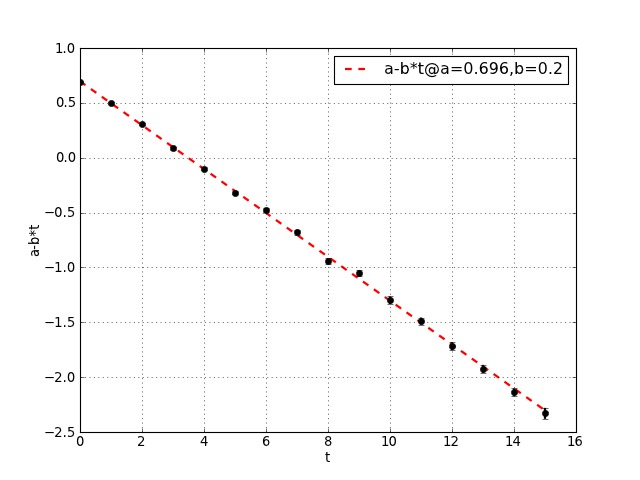}
\end{center}
\caption{Example fits for a two points correlation function (left) and its log (right).\label{fit12}}
\end{figure}

The output looks something like this:

\begin{lstlisting}
a = 1.99864
b = 0.200645
chi2= 12.8048378376
chi2/dof= 0.984987525973
\end{lstlisting}

{\ft qcdutils\_fit.py} also generates the plot of fig.~\ref{fit12} (left).

If C2 is a meson propagator, $b$ here represents the mass of the meson (of the lowest energy state with the same quantum numbers as the operator used to create the meson).

Similary we can analyze and fit the log of C2:

\begin{lstlisting}
python qcdutils_boot.py test_samples.log 'log("C2[<t>]")' > run.log
python qcdutils_fit.py 'a-b*t@a=1,b=0.3'
\end{lstlisting}

which produces something like:

\begin{lstlisting}
a = 0.69169
b = 0.200627
chi2= 12.1641201448
chi2/dof= 0.935701549598
\end{lstlisting}

and the plot of fig.~\ref{fit12} (right)

\begin{figure}[ht!]
\begin{center}
\includegraphics[width=8cm]{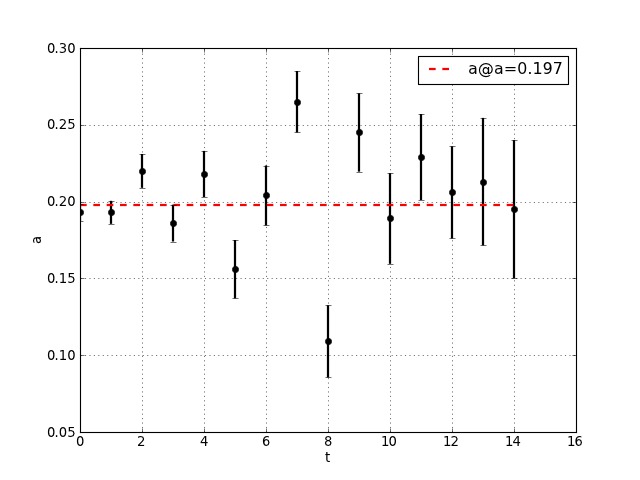}
\end{center}
\caption{Example plot of fit of $log(C2(t)/C2(t-1))$.\label{logc2c2}}
\end{figure}

If our goal is obtaining $b$ we can also cancel the $a$ dependency in the analysis:

\begin{lstlisting}
python qcdutils_boot.py test_samples.log \
       'log("C2[<t>]"/"C2[<t1>]")' 't1==t+1' > run.log
python qcdutils_fit.py 'b@b=0'
\end{lstlisting}

and obtain:

\begin{lstlisting}
b = 0.201755
chi2= 12.4502446913
chi2/dof= 0.9577111301
\end{lstlisting}

The generated plot is shown in fig.~\ref{logc2c2}.

Notice that the variable names {\ft a} and {\ft b} are arbitrary and you can choose any name.

Similarly we can fit 3-point correlation functions:

\begin{lstlisting}
python qcdutils_boot.py test_samples.log '"C3[<t1>][<t2>]"' > run.log
python qcdutils_fit.py 'a*exp(-b*(t1+t2))@a=3,b=0.3,_b=0.2'
\end{lstlisting}

In this case we have stabilized the plot with a Bayesian prior, indicated by {\ft \_b}. A variable starting with underscore indicates the uncertainty associated with our a priori knowledge about the corresponding variable without underscore. In other words {\ft b=0.3,\_b=0.2} is equivalent to {\ft b=}$0.3 \pm 0.2$. The result of this fit yields something like:

\begin{lstlisting}
a = 3.78387
b = 0.195542
chi2= 2070.73759118
chi2/dof= 8.18473356199
\end{lstlisting}

A call to {\ft qcdutils\_plot.py} generates the plot of fig.\ref{plotc3} (left)

In order to extract a matrix element (for example a 4-quark operator) we fit the ratio between C3 and C2:

\begin{lstlisting}
python qcdutils_boot.py test_samples.log \
       '"C3[<t>][<t1>]"/"C2[<t2>]"/"C2[<t3>]"' \
       't3==t and t2==t and t1==t' > run.log
python qcdutils_fit.py 'a@a=0'
\end{lstlisting}

It produces output like:

\begin{lstlisting}
a = 1.00658
chi2= 13.2075581022
chi2/dof= 0.943397007297
\end{lstlisting}

It produces the plot in fig.~\ref{plotc3} (right).

\begin{figure}[ht!]
\begin{center}
\begin{tabular}{cc}
\includegraphics[width=8cm]{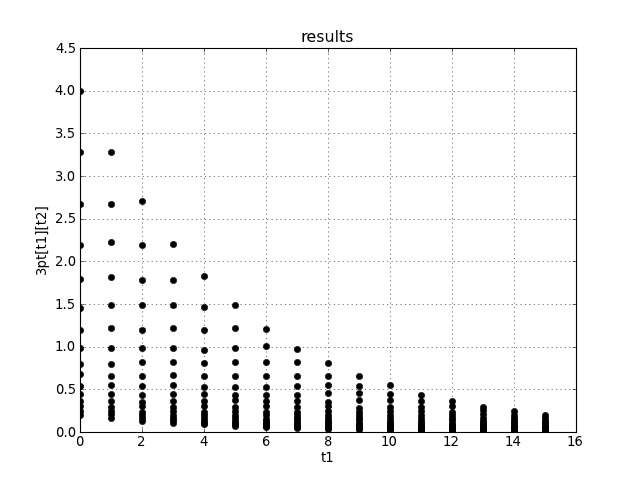} &
\includegraphics[width=8cm]{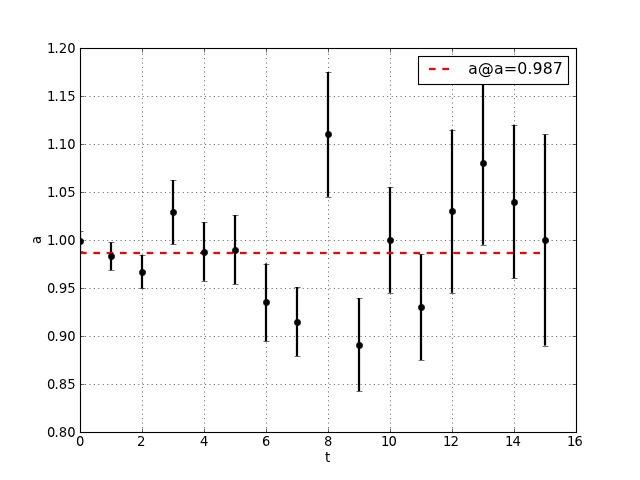}
\end{tabular}
\end{center}
\caption{Example plot showing C3[t][t1] (left) and the fit of C3[t][t1]/C2[t]c2[t1] (right).\label{plotc3}}
\end{figure}

You can use {\ft qcdutils\_fit.py} to perform exatrapolations by using the {\ft -extrapolate} command line option:

\begin{lstlisting}
python qcdutils_fit.py -extrapolate x=100 'ax+b@a=1,b=0'
\end{lstlisting}

The extrapolated point will be added to the generated plot and represented by a square.

\goodbreak\subsection{Dimensional analysis and error propagation}

In this section we did not discuss error propagation but we have developed a utility called {\it Buckingham} which is avalable from:

\url{http://code.google.com/p/buckingham/}

It provides dimensional analysis, unit conversion, and aritmetic operation with error propgation. We plan to discuss it in a separate manual but we here provide one example of usage (from inside a Python shell):

\begin{lstlisting}
>>> from buckingham import *
>>> a = Number(2.0, error=0.3, dims="fermi")
>>> b = Number(1.0, error=0.2, dims="second^2")
>>> c = a/b
>>> print c, c.units()
(2.000 pm 0.500)/10^15 meter*second^-2
>>> print c.convert('fermi*second^-2')
2.000 pm 0.500
>>> print c.convert('lightyear*day^-2')
(1.578 pm 0.395)/10^21
\end{lstlisting}

(here {\ft pm} stands for $\pm$) 
Buckingham supports 944 unit types (including {\ft eV}) and their combinations.

\appendix

\section{Filename conventions}

\begin{lstlisting}
Gauge configuration in NERSC format (3x3 or 3x2)
  *.nersc
Gauge configuration in Fermiqcd format
  *.mdp
Gauge configuration in MILC format
  *.milc
Generic LIME file 
  *.lime
Gauge configuration in ILDG format
  *.ildg
SciDAC quark propagator
  *.scidac
Quark propagator in FermiQCD format
  *.prop.mdp
Time slice for gauge configuration in FermiQCD format:
  *.t[NNNN].mdp
Time slice for propagator in FermiQCD format:
  *.t[NNNN].prop.mdp
Quark field for a given SPIN, COLOR source:
  *.s[SPIN].c[COLOR].quark
Generic log file
  *.log
VTK file containing real trace of plaquettes
  *.plaquette.vtk
VTK file containing real part of Polyakov lines
  *.polyakov.vtk
VTK file containing topological charge density
  *.topcharge.vtk
VTK file containing topological charge density for a cooled config 
  *.topcharge.cool[STEP].vtk
VTK file containt a the norm squared of a pion propagator
  *.pion.vtk
HTML file generated by qcdutils_vtk, represents a VTK file.
  *.vtk.html
VisIt visualization script generated by qcdutils_vis.py
  qcdutils_vis_[UUID].py
VisIt image generates by the previous script
  qcdutils_vis_[UUID]_[FRAME].jpeg
Raw data extract from a log file by qcdutils_boot
  qcdutils_raw_data.csv
Autocorrelations computed by qcdutils_boot
  qcdutils_autocorrelations.csv
Partial averages computed by qcdutils_boot
  qcdutils_trails.csv
Bootstrap samples generated by qcdutils_boot
  qcdutils_samples.csv
Means and bootstrap errors computed by qcdutils_boot
  qcdutils_results.csv
\end{lstlisting}

\goodbreak\section{Help Pages}

\goodbreak\subsection{\ft qcdutils\_get.py}
\begin{lstlisting}
$ qcdutils_get.py -h
Usage:

    qcdutils_get.py [options] sources

Examples:

    qcdutils_get.py --test
    qcdutils_get.py --convert ildg gauge.cold.12x8x8x8
    qcdutils_get.py --convert mdp --float *.ildg
    qcdutils_get.py --convert split.mdp *.mdp

Options:
  -h, --help            show this help message and exit
  -q, --quiet           no progress bars
  -d DESTINATION, --destination=DESTINATION
                        destination folder
  -c CONVERT, --convert=CONVERT
                        converts a field to format
                        (ildg,split.prop.mdp,prop.ildg,prop.mdp,split.mdp,mdp)
  -4, --float           converts to float precision
  -8, --double          converts to double precision
  -t, --tests           runs some tests
  -n, --noprogressbar   disable progress bar
\end{lstlisting}

\goodbreak\subsection{\ft qcdutils\_run.py}
\begin{lstlisting}
$ qcdutils_run.py -h
qcdutils_run.py is a tool to help you download and use fermiqcd from

    http://code.google.com/p/fermiqcd

When you run:
 
    python qcdutils_run.py [args]
 
It will:
- create a folder called fermiqcd/ in the current working directory
- connect to google code and download fermiqcd.cpp + required libraries
- if -mpi in [args] compile fermiqcd with mpiCC else with g++
- if -mpi in [args] run fermiqcd.exe with mpiCC else run it normally
- pass the [args] to the compiled fermiqcd.exe

Some [args] are handled by qcdutils_run.py:
-download force downloading of the libraries
-compile  force recompiling of code
-source   runs and compiles a different source file
-mpi      for use with mpi (mpiCC and mpirun but be installed)

Other [args] are handled by fermiqcd.cpp for example
-cold     make a cold gauge configuration
-load     load a gauge configuration
-quark    make a quark
-pion     make a pion
(run it with no options for a longer list of options)

You can find the source code in fermiqcd/fermiqcd.cpp

More examples:
    qcdutils_run.py -gauge:start=cold:nt=16:nx=4
    qcdutils_run.py -gauge:start=hot:nt=16:nx=4
    qcdutils_run.py -gauge:load=cold.mdp
    qcdutils_run.py -gauge:load=cold.mdp:steps=10:beta=5.7
    qcdutils_run.py -gauge:load=*.mdp -plaquette
    qcdutils_run.py -gauge:load=*.mdp -plaquette_vtk
    qcdutils_run.py -gauge:load=*.mdp -polyakov_vtk
    qcdutils_run.py -gauge:load=*.mdp -cool:steps=20 -topcharge_vtk
    qcdutils_run.py -gauge:load=*.mdp -quark:kappa=0.12:alg=minres_vtk
    qcdutils_run.py -gauge:load=*.mdp -quark:kappa=0.12 -pion
    qcdutils_run.py -gauge:load=*.mdp -quark:kappa=0.12 -pion_vtk

Options:
    -cool
        alg = ape
        alpha = 0.7
        steps = 20
        cooling = 10
    -cool_vtk
        n = 20
        alpha = 0.7
        steps = 1
        cooling = 10
    -quark
        action = clover_fast (default) or clover_slow or clover_sse2
        alg = bicgstab (default) or minres or bicgstab_vtk or minres_vtk
        abs_precision = 1e-12
        rel_precision = 1e-12
        source_t = 0
        source_x = 0
        source_y = 0
        source_z = 0
        source_point = zero (default) or center
        load = false (default) or true
        save = true (default) or false
        matrices = FERMILAB (default) or MILC or 
                   UKQCD or Minkowsy-Dirac or Minkowsy-Chiral
        kappa = 0.12
        kappa_t = quark["kappa"]
        kappa_s = quark["kappa"]
        r_t = 1.0
        r_s = 1.0
        c_sw = 0.0
        c_E = 0.0
        c_B = 0.0
    -meson
        source = 5
        sink = 5
        current = I
    -4quark
        source = 5 (default) or I or 0 or 1 or 2 or 3 or 05 or 
                 15 or 25 or 35 or 01 or 02 or 03 or 12 or 13 or 23
        operator = 5Ix5I (default) or 0Ix0I or 1Ix1I or 2Ix2I or 3Ix3I or 
                   05Ix05I or 15Ix15I or 25Ix25I or 35Ix35I or 01Ix01I or 
                   02Ix02I or 03Ix03I or 12Ix12I or 13Ix13I or 23Ix23I or 
                   5Tx5T or 0Tx0T or 1Tx1T or 2Tx2T or 3Tx3T or 05Tx05T or 
                   15Tx15T or 25Tx25T or 35Tx35T or 01Tx01T or 02Tx02T or 
                   03Tx03T or 12Tx12T or 13Tx13T or 23Tx23T
    -gauge
        nt = 16
        nx = 4
        ny = nx
        nz = ny
        start = load (default) or cold or hot or instantons
        load = demo.mdp
        n = 0
        steps = 1
        therm = 10
        beta = 0
        zeta = 1.0
        u_t = 1.0
        u_s = 1.0
        prefix = 
        action = wilson (default) or wilson_improved or wilson_sse2
        save = true
        t0 = 0
        x0 = 0
        y0 = 0
        z0 = 0
        r0 = 1.0
        t1 = 1
        x1 = 1
        y1 = 1
        z1 = 1
        r1 = 0.0
    -baryon
    -pion
    -pion_vtk
    -meson_vtk
    -current_static
    -current_static_vtk
    -plaquette
    -plaquette_vtk
    -polyakov_vtk
    -topcharge_vtk
\end{lstlisting}

\goodbreak\subsection{\ft qcdutils\_vis.py}
\begin{lstlisting}
$ qcdutils_vis.py -h
Usage: 
This is a utility script to manipulate vtk files containing scalar files.
Files can be split, interpolated, and converted to jpeg images.
The conversion to jpeg is done by dynamically generating a visit script 
that reads the files, and computes optimal contour plots.

Examples:

1) make a dummy vtk file

   qcdutils_vis.py -m 10 folder/test.vtk

2) reads fields from multiple vtk files

   qcdutils_vis.py -r field folder/*.vtk

3) extract fields as multiple files

   qcdutils_vis.py -s field folder/*.vtk

(fields in files will be renamed as "slice")
4) interpolate vtk files

   qcdutils_vis.py -i 9 folder/*.vtk

tricubic Resample/Interpolate individual vtk files

   visit -v 10x10x10 folder/*.vtk 

6) render a vtk file as a jpeg image

   qcdutils_vis.py -p 'AnnotationAttributes[axes3D.bboxFlag=0];
      ResampleAttributes[samplesX=160;samplesY=160;samplesZ=160];
      ContourAttributes[SetMultiColor(9,$orange)]' 'folder/*.vtk'
  
   or simply

   qcdutils_vis.py -p default 'folder/*.vtk'

Options:
  -h, --help            show this help message and exit
  -r READ, --read=READ  name of the field to read from the vtk file
  -s SPLIT, --split=SPLIT
                        name of the field to split from the vtk file
  -i INTERPOLATE, --interpolate=INTERPOLATE
                        name of the vtk files to add/interpolate
  -c CUBIC, --cubic-interpolate=CUBIC
                        new size for the lattice 10x10x10
  -m MAKE, --make=MAKE  make a dummy vtk file with size^3 whete size if arg of
                        make
  -p PIPELINE, --pipeline=PIPELINE
                        visualizaiton pipeline instructions
\end{lstlisting}

\goodbreak\subsection{\ft qcdutils\_vtk.py}
\begin{lstlisting}
$ qcdutils_vtk.py -h
Usage: qcdutils_vtk.py filename.vtk

Options:
  -h, --help            show this help message and exit
  -u UPPER, --upper-threshold=UPPER
                        treshold for isosurface
  -l LOWER, --lower-threshold=LOWER
                        treshold for isosurface
  -R UPPER_RED, --upper-red=UPPER_RED
                        color component for upper isosurface
  -G UPPER_GREEN, --upper-green=UPPER_GREEN
                        color component for upper isosurface
  -B UPPER_BLUE, --upper-blue=UPPER_BLUE
                        color component for upper isosurface
  -r LOWER_RED, --lower-red=LOWER_RED
                        color component for lower isosurface
  -g LOWER_GREEN, --lower-green=LOWER_GREEN
                        color component for lower isosurface
  -b LOWER_BLUE, --lower-blue=LOWER_BLUE
                        color component for lower isosurface
\end{lstlisting}

\goodbreak\subsection{\ft qcdutils\_boot.py}
\begin{lstlisting}
$ qcdutils_boot.py -h
Usage: qcdutils_boot.py *.log 'x[<a>]/y[<b>]' 'abs(a-b)==1'
  scans all files *.log for expressions of the form
    x[<a>]=<value> and y[<b>]=<value>
  and computes the average and bootstrap errors of x[<a>]/y[<b>]
  where <a> and <b> satisfy the condition abs(a-b)==1.

This is program to scan the log files of a Markov Chain Monte Carlo Algorithm,
parse for expressions and compute the average and bootstrap errors of any
function of those expressions.It also compute the convergence trails of the
averages.

Options:
  --version             show program's version number and exit
  -h, --help            show this help message and exit
  -b MIN, --minimum_index=MIN
                        the first occurrence of expression to be considered
  -e MAX, --maxmium_index=MAX
                        the last occurrence +1 of expression to be considered
  -n NSAMPLES, --number_of_samples=NSAMPLES
                        number of required bootstrap samples
  -p PERCENT, --percentage=PERCENT
                        percentage in the lower and upper tails
  -t, --test            make a test!
  -r, --raw             Load raw data instead of parsing input
  -a, --advanced        In advanced mode use regular expressions for variable
                        patterns
  -i IMPORT_MODULE, --import_module=IMPORT_MODULE
                        import a python module for expression evaluation
  -o OUTPUT_PREFIX, --output_prefix=OUTPUT_PREFIX
                        path+prefix used to build output files
\end{lstlisting}

\goodbreak\subsection{\ft qcdutils\_plot.py}
\begin{lstlisting}
$ qcdutils_plot.py -h
Usage: python qcdutils_plot.py

plot the output of qcdutils.py

Options:
  --version             show program's version number and exit
  -h, --help            show this help message and exit
  -i INPUT_PREFIX, --input_prefix=INPUT_PREFIX
                        the prefix used to build input filenames
  -r, --raw             make raw data plots
  -a, --autocorrelations
                        make autocorrelation plots
  -t, --trails          make trails plots
  -b, --bootstrap-samples
                        make bootstrap samples plots
  -v PLOT_VARIABLES, --plot_variables=PLOT_VARIABLES
                        plotting variables
  -R RANGE, --range=RANGE
                        range as in 0:1000
\end{lstlisting}

\goodbreak\subsection{\ft qcdutils\_fit.py}
\begin{lstlisting}
$ qcdutils_fit.py -h
Usage: qcdutils_fit.py [OPTIONS] 'expression@values'
  Example: qcdutils-fit.py 'a*x+b@a=3,b=0'
  default filename is qcdutils_results.csv
  ...., 'x', 'min', 'mean', 'max'
  ...., 23, 10, 11, 12
  ...., etc etc etc

Options:
  --version             show program's version number and exit
  -h, --help            show this help message and exit
  -i INPUT, --input=INPUT
                        input file (default qcdutils_results.csv)
  -c CONDITION, --condition=CONDITION
                        sets a filter on the points to be fitted
  -p PLOT, --plot=PLOT  plots the hessian (not implemented yet)
  -t, --test            test a fit
  -e EXTRAPOLATIONS, --extrapolate=EXTRAPOLATIONS
                        extrpolation point
  -a AP, --absolute_precision=AP
                        absolute precision
  -r RP, --relative_precision=RP
                        relative precision
  -n NS, --number_steps=NS
                        number of steps
\end{lstlisting}

\end{document}